\documentclass[aps,prb,showpacs,showkeys,twocolumn,superscriptaddress,floatfix]{revtex4}
\usepackage{epsfig}
\usepackage{graphics}
\usepackage{bm}
\usepackage[ansinew]{inputenc}
\usepackage{amsmath}
\newcommand{\un}{\bm}
\newcommand{\ul}[1]{\underline{#1}} 
\begin{document}
\title{Stochastic Foundation of Quantum Mechanics and the Origin of Particle Spin} 
\author{L.~Fritsche}
\thanks{Corresponding author}
\email{lfritsche@t-online.de} \affiliation{
Institut f\"ur Theoretische Physik der Technischen Universit\"at
Clausthal, D-38678 Clausthal-Zellerfeld, Germany}
\author{M.~Haugk} \affiliation{Hewlett-Packard GmbH, Schickardtstr.~25\\
D-71034 B\"oblingen, Germany}
\begin{abstract}
The present contribution is aimed at removing most of the obstacles in understanding the quantum mechanics of massive particles. We advance the opinion that the probabilistic character of quantum mechanics does not originate from uncertainties caused by the process of measurement or observation, but rather reflects the presence of objectively existing vacuum fluctuations whose action on massive particles is calibrated by Planck's constant and effects an additional irregular motion. As in the theory of diffusion the behavior of a single particle will be described by an ensemble of identically prepared but statistically independent one-particle systems. Energy conservation despite the occurrence of a Brownian-type additional motion is achieved by subdividing the ensemble into two equally large sub-ensembles for each of which one obtains an equation of motion that has the form of a Navier-Stokes- or ``anti''-Navier-Stokes-type equation, respectively. By averaging over the total ensemble one obtains a new equation of motion which can be converted into the time-dependent Schr\"odinger equation.  We clarify the problem of the uniqueness of the wave function and the quantization of orbital momentum. The concept allows the inclusion of electromagnetic fields and can be extended to interacting N-particle systems. We analyze the problem of how an experimental setup can consistently be decomposed into the quantum system under study and the residual quantum system ``apparatus''. The irregular extra motion of the particle under study allows a decomposition of the associated ensemble into two subensembles the members of which perform, respectively, a right-handed or left-handed irregular circular motion about a given axis which becomes physically relevant in the presence of a magnetic field.  We demonstrate that this orientation-decomposed ``Zitterbewegung'' behaves - in accordance with Schr\"odinger's original idea - as a spin-type angular momentum which appears in addition to a possible orbital angular moment of the particle. We derive the non-relativistic time-dependent Pauli equation and propose a theory of the Stern-Gerlach experiment. The Dirac equation proves to be derivable by drawing on similar arguments used in obtaining the Pauli equation. We, further, attempt to put Bell's theorem and the Kochen-Specker theorem into perspective.
\end{abstract}
\maketitle
\begin{center}
{\bf {\Large Contents}}
\end{center} 
\begin{enumerate}
\item Introduction
\item Origin of quantum mechanical randomness
\item Defining ensembles and averages
\item Navier-Stokes equations 
\item The time-independent Schrödinger equation
\item Including currents
\item The velocity potential and phase uniqueness 
\item Quantization of angular momentum
\item An instructive objection and quantum beats
\item The time-dependent Schrödinger equation
\item The uncertainty relation and the issue of ``measurement''
\item Averaging over the total ensemble
\item Conservative diffusion 
\item The time-dependent Schrödinger equation in the presence of an electromagnetic field
\item A model for ``non-Markovian'' diffusion \\
      illustrating the origin of non-locality
\item Operators and commutators
\item Collaps of the wave function and the node problem
\item The Feynman path integral
\item The time-dependent N-particle Schr\"odinger equation
\item States of identical particles and entanglement
\item A borderline case of entanglement
\item Decomposing an experimental setup into the quantum system under study and a remainder. Schrödinger's cat
\item The origin of particle spin
\item Generalizing one-particle quantum mechanics by including particle spin
\item The time-dependent non-relativistic Pauli equation
\item The Cayley-Klein parameters and Pauli spin matrices
\item Spin precession in a magnetic field
\item A theory of the Stern-Gerlach experiment
\item The time-dependent Dirac equation
\item Spatial particle correlation beyond the limit of entanglement. Spooky action at a distance
\item Concluding remarks
\item Appendix: Derivation of the Navier-Stokes equation
\end{enumerate}
\vspace{1.0cm}
``... to skeptics, heretics and na\"ive realists everywhere.\\
Keep doubting; let others keep the faith.''
\\[0.2cm]
David Wick in: {\it The Infamous Boundary} \cite{Wick}
\\[0.5cm]
``I have never been able to discover any well-founded reasons
as to why there exists so high a degree of confidence in the .....current form of quantum theory.'' 
\\[0.2cm]
David Bohm in: {\it Wholeness and the Implicate Order} \cite{DavidB}

\section{Introduction}\label{kapitel1}   
There exists a rich literature on attempts that have been made to derive non-relativistic quantum mechanics from a concept of dissipationless stochastic point mechanics. A precursor of the idea of correlating the probabilistic character of quantum mechanics with the action of a stochastic background field may be seen in the paper by Bohm and Vigier \cite{Bohm_1}. The present contribution draws on later  work on this subject but avoids certain implications that have often been criticized during the past 20 years. (S. e.g. W.~Weizel \cite{Weizel}, E.~Nelson \cite{Nelson_1}, \cite{Nelson_2}, Guerra and Morato, \cite{Guerra}, M.~Baublitz \cite{Baublitz}, L.~de la Pe\~na and A.~H.~Cetto \cite{Pena}, Petroni and Morato \cite{Petroni}, T.~C.~Wallstrom \cite{Wallstrom} and numerous references therein. For a rather complete review see R.F. Streater \cite{Streater}. An earlier review covering work up to 1986 is given in a book by Namsrai \cite{Namsrai}.) A more recent contribution is due to Fritsche and Haugk \cite{FritscheHaugk}.\\
The derivation of the time-dependent Schr\"odinger equation constitutes the focus of the following considerations. A very interesting alternative to our approach that also relates to vacuum fluctuations, but draws on non-equilibrium thermodynamics has recently been put forward by Grössing \cite{Groessing_1}$^{,}$\cite{Groessing_2}. Based on the concept of our derivation one is led to conclude that every conceivable situation of a physical system is exhaustively described by the respective solution to the Schr\"odinger equation, and that there can be no independent measurement problem. As for this point we side with J. Bell \cite{Bell_1} who argues that the attempt to base the interpretation of quantum mechanics on some notion of ``measurement'' has raised more problems than it has solved. As we shall outline in Section \ref{kapitel15.1} ``measurements'' relate outcomes, e.~g.~detector readings, to characteristic properties of a quantum system by using solutions to the Schrödinger (or Pauli) equation as primordeal  information. Without these equations and their solutions ``measurements'', i.~e.~in general, detector or ``pointer'' readings, constitute a set of worthless data.

\section{Origin of quantum mechanical randomness} \label{kapitel1.1}

We interpret the fact that microscopic particles move and behave differently from macroscopic objects as reflecting the active role of the vacuum providing a background for energy fluctuations. The latter will henceforth be referred to as vacuum fluctuations. The consequences of their existence  have  already been discussed quite some time ago, s. e.$\,$g.~Bess \cite{Bess}, Puthoff \cite{Puthoff1}, \cite{Puthoff2}, Boyer \cite{Boyer}, Calogero \cite{Calogero}, Carati and Calgani \cite{Carati}. Present day quantum mechanics is strongly shaped by historical contingencies in its development, and it has become almost impossible to tell fiction from facts. Statements on ``the measurement of positions at different times'' and ``there is no momentum of a particle in advance of its measurement'' are typical of this school of thought (s. e.$\,$g.~Streater \cite{Streater}), yet they are definitely void of meaning. What kind of experimental setup should allow a perfectly accurate position measurement at a perfectly accurate time point? And how does the setup look like that allows the measurement of a particle momentum in the spirit of orthodox quantum mechanics; i.~e. with zero variance? It is totally impossible to perform non-fictional measurements on quantities that would conform to their quantum mechanical definition, e.$\,$g. measuring commuting observables like energy and angular momentum at the same time. The ``observables'' around which a substantial portion of quantum mechanical literature revolves are in reality non-observables. Further, there is simply no evidence of a causal interrelation between the probabilistic character of quantum mechanics and indeterminacies introduced by ``the observer''.\\
By contrast, there is every reason to believe that vacuum fluctuations are real and constitute an objective property of nature. Zero point motion of particles constitutes the most obvious evidence of their existence. It is this zero point motion which, for example, keeps liquid $^4$He ``molten'' down to the very lowest temperatures and explains this extraordinary material property. \\
One could view vacuum fluctuations as caused by an exchange of energy between the mechanical system in question and the embedding vacuum that serves as an energy reservoir in terms of virtual particles: if that reservoir reduces its content of virtual particles, the energy of the system under study increases so that the energy of the entire system comprising this ``vacuum reservoir'' is conserved. Considerations of Calogero \cite{Calogero} point in a similar direction. In that sense quantum mechanical systems may be viewed as open systems like classical point mass systems in contact with a heat bath. This analogy will become particularly visible in our treatment. We shall use the terms ``point mass'' and ``point charge'' with the reservation that the actual size of the particles in question might well be finite of the order 10$^{-13}\,$cm, but very small compared to atomic diameters of the order $10^{-8}\,$cm. Occasionally ``point mass'' will stand for the centroid of an atom or some composite system.
\\[0.2cm]
In the following we shall focus on the description of the subsystem ``point mass in real-space'' which is open toward the active vacuum and  whose energy is therefore conserved  only on average. \\
An implication of this concept is that charged point masses, despite their irregular motion, do not emit or absorb radiation on the average. In stationary situations a charged point mass will exchange photons with the vacuum in a way that does not change its average energy and momentum. 
\\[0.2cm]
{\it Radiation only occurs when the probability density of the point mass, being at its various positions in space, or the associated current density becomes time-dependent.} 
\\[0.2cm]
This is analogous to a system kept by non-heat conducting fibers in a vacuum chamber whose walls serve as a heat bath. In a stationary state situation the system exchanges constantly photons with the heat bath without changing its average energy. However, if its temperature is, for example, higher than that of the wall, the system starts radiating, that is, there is now a net flow of photons leaving the system.
\\[0.2cm]
If one disregards the details of the energy transfer between the two systems, vacuum fluctuations appear as an irregular temporary departure of the particle in question from its energy conserving trajectory in that it changes its energy by an average amount $\Delta E$ for an average time interval $\Delta t$ so that $\Delta E\,\Delta\,t =f\,\hbar$ where $h=2\,\pi \hbar$ denotes Planck's constant and the factor $f$ is about unity. It is this departure from classical energy conservation which explains, as already alluded to, why a harmonic oscillator in its state of lowest energy is irregularly driven out of the position where it would be classically at rest. Furthermore, it explains the stability of a hydrogen atom in its ground-state (which applies quite generally to all atoms and their compounds), the zero-point motion of atoms in molecules and solids and the ``tunneling'' of particles through a potential wall which actually amounts to overcoming that wall.\\
Zero-point motion is commonly associated with the uncertainty relation which, however, merely shifts the problem of understanding a non-classical phenome\-non to understanding the origin of a non-classical relation. Moreover, it amounts to keeping a blind eye on the fact that one is dealing here with a ground-state phenomenon which is certainly not observer-induced. Only if the contrary would apply, one would be justified in referring to the uncertainty relation.
\\[0.2 cm]
A Boltzmann distribution over the energy levels of some system is completely independent of the details and the kind of the energy exchange between the heat bath and the system. The distribution contains only one universal parameter, viz. Boltzmann's constant. Similarly, a system's stationary zero-temperature states that emerge from exchanging energy with the vacuum do not depend on the details of this exchange and on the kind of particles involved, but only depend on another universal constant, viz.~Planck's constant. \\
The envisaged derivation implies that particle trajectories persist under the influence of the stochastic vacuum forces. Their existence becomes particularly obvious with tracks of $\alpha$-particles in a track chamber, but also with the trajectories of electrons in a field electron microscope. Their property of forming straight lines from the field-emission tip (assumed semi-spherical) to the monitoring screen is actually presupposed in calculating the magnification of the microscope. Conversely, purely quantum mechanical behavior occurs at lowest energies when the trajectories do no longer possess a classical reference in the limit $\hbar\rightarrow 0$. Trajectories still persist in that case, but the respective particle now performs a purely irregular motion.\\
The existence of particle trajectories is denied by the Copenhagen school of thought because ``things that cannot be observed do not exist''. Supporter of this view have to live with the conflict that a complex-valued wavefunction or its associated state vector, which constitutes the center of quantum mechanics, cannot be observed as well. By contrast, we believe that the validity of assumptions can only be scrutinized by checking the consistency of the resulting theory against experimentally accessible quantities and laws. We are here in complete accord with Ballentine who states in his seminal article \cite{Ballentine}:
\\[0.1cm]
{\it ``...quantum theory is not inconsistent with the supposition that a particle has at any instant both a definite position and a definite momentum, although there is a widespread folklore to the contrary.''}
\\[0.2cm]
In Section \ref{kapitel2} we briefly discuss the construction of ensemble averages of quantities that appear in the Navier-Stokes equation given in Section \ref{kapitel3}. We regard this equation as a mathematical object that derives entirely from classical concepts. Details of its derivation, which goes essentially back to Gebelein \cite{Gebelein}, will be relegated to the Appendix, Section  \ref{kapitel22}. We discuss the construction of a ``Brownian'' and an ``anti-Brownian'' sub-ensemble. The motional behavior of the latter is governed by an ``anti-Navier-Stokes'' equation. We explain why a system of statistically independent particles moves according to the arithmetic mean of these two equations when their motion is governed by classical mechanics plus ``conservative'' stochastic forces. On forming this arithmetic mean we arrive at an equation that can be converted into the Schr\"odinger equation. We demonstrate that Wallstrom's objection \cite{Wallstrom} against the legitimacy of this conversion and his arguments in favor of the standard approach to the quantization of orbital momentum are based on a misunderstanding and ignore fundamental considerations of Pauli \cite{Pauli_1} and Born and Jordan \cite{BornJordan} in the early days of ``conventional'' quantum mechanics. In Section \ref{kapitel10} we show how the derivation of the time-dependent Schr\"odinger equation can be extended by including electromagnetic fields. The derivation can be extended further to interacting many-particle systems. 
\section{Defining ensembles and averages}\label{kapitel2} 
As in the theory of diffusion we start with considering a point-like particle that is driven by an external conservative force ${\un F}({\un r})$ and moves in an environment where it is exposed to additional stochastic forces. To gain access to quantities that are commonly discussed within this framework we construct a sufficiently large set of $N$ identical systems (an ensemble of systems) under the supposition that there is no correlation between the stochastic forces of different systems. As a fundamental consequence, one is led then, as in the theory of diffusion, to a form of quantum mechanics that merely describes ensemble behavior. But this, again, is in accord with Ballentine's view \cite{Ballentine}: \textit{``..in general, quantum theory predicts nothing which is relevant to a single measurement (excluding strict conservation laws like those of charge, energy or momentum).''}
\\[0.2cm]
The relative freqency with which the particle appears at the time $t$ in an elementary volume $\Delta^3 {\un r}$ around the point ${\un r}$ is given by 
\begin{eqnarray}
\label{eqn:def_density}
\frac{n({\un r},t)}{N}=\rho({\un r},t)\,\Delta^3 {\un r}
\end{eqnarray}
where $n({\un r},t)$ is the number of particles in $\Delta^3 {\un r}$, and $\rho({\un r},t)$ denotes the probability density. We, furthermore, introduce $N_{{\un r}}$ for the number of elementary volumes into which the total volume ${\cal{V}}$ is thought to be subdivided. Since the sum over all elementary cells yields $N$ particles we have
\begin{eqnarray}
\label{eqn:norm_integral}
\sum_{\un r}^{N_{{\un r}}}\frac{n({\un r},t)}{N}=1\quad \mbox{that is} \quad \int_{{\cal{V}}}\rho({\un r},t)\,d^3 {\un r}=1\,.
\end{eqnarray}
We refrain here from discussing the proper limiting case $N\rightarrow \infty$ and relating relative frequencies to probabilities, as this matter has extensively been analyzed elsewhere (s. e.$\,$g. Streater \cite{Streater}). We assume that there will always be a smooth function $\rho({\un r},t)$ for any finite $N$ that provides a least mean square fit to the actually histogram-type function $\frac{n({\un r}_j,t)}{N} $ in real-space where $j$ numbers the cubes into which the normalization volume $\cal{V}$ is thought to be subdivided, and ${\un r}_j$ denotes the centroid of the particle positions in the respective cube.
\\[0.2cm]
The relative frequency $\frac{n({\un r},t)}{N}$ which we shall below express as the mod squared of some wave function $\psi({\un r},t)$, refers - when multiplied by $\Delta^3r$ - to the subset of identically prepared systems where the particle appears at ${\un r}$ and nowhere else simultaneously, otherwise the term ``particle'' would be meaningless. We think that the commonly used phraseology ``probability of {\bf finding} the particle at ${\un r}$'' is inappropriate because it suggests that one would have placed a detector at ${\un r}$ monitoring the occurrence of that particle. However, a detector would - apart from causing various uncontrollable perturbations -  terminate the motion of the particle on impact, and hence there would be a shadow area behind the detector where $\frac{n({\un r},t)}{N}\approx 0$, different from the original unperturbed situation. Wherever in the following the quantity $\frac{n({\un r},t)}{N}\,\Delta^3r$ or $\rho({\un r},t)\,\Delta^3r$ will appear it is clearly to be understood as the probability of the particle {\bf being} in $\Delta^3r$ around ${\un r}$.
\\[0.2cm]
We temporarily number the particles in $\Delta^3 {\un r}$ at time $t$ by an index $i$, ($i=1,2\ldots n({\un r},t)$).  The particles move, in general, at different velocities ${\un v}_i(t)$. We define the ensemble average of the latter as
\begin{eqnarray}
\label{eqn:average_v}  
{\un v}({\un r},t)=\frac{1}{n({\un r},t)}\,\sum_{i=1}^{n({\un r},t)} {\un v}_i(t)\,.
\end{eqnarray}
As is familiar from the theory of diffusion, the individual velocities ${\un v}_i(t)$ in $\Delta^3 {\un r}$ will in general be quite different from ${\un v}({\un r},t)$ which we shall come back to later in Section \ref{kapitel6}. By contrast, in Bohm's version of quantum mechanics \cite{Bohm1} the true particle trajectories are, for no obvious reason, identified with the streamlines of the velocity field ${\un v}({\un r},t)$. This is one of the points where our approach differs fundamentally from Bohm's and reflects a concomitant feature of our definition of ${\un v}({\un r},t)$: \\
In performing the average according to Eq.(\ref{eqn:average_v}) one sums over velocities ${\un v}_i(t)$ of different trajectories that run through sometimes very different regions of the available space of the one-particle system. Hence, they are influenced by the classical field ${\un F}({\un r})$ in those regions. This carries over to the ensemble average ${\un v}({\un r},t)$. That means: if one places a diaphragm somewhere so that a continuous subset of trajectories is blocked out, ${\un v}({\un r},t)$ changes. That kind of non-local sensitivity explains why the streamlines of the field ${\un v}({\un r},t)$ are affected by portions of the space which may be far away. The unfamiliar feature of non-locality will be illustrated by a particularly surprising example in Section \ref{kapitel5.1}.
\\[0.2cm]
As a general property of the stochastic forces that act on the respective particle in each system, we require them to ensure ergodicity in the following sense:\\
If the system is not explicitly time-dependent, that is, when it is in a bound stationary state and if one would follow the particle on its trajectory within the range it is bound to, one would see it successively occur in all the cubes  $\Delta^3r$ over which - in the ensemble average - all particles of the ensemble are distributed at a certain instant $t$. Thus, instead of forming the ensemble average according to Eq.(\ref{eqn:def_density}) it can for a single particle just as well be defined as
\begin{eqnarray}
\label{eqn:time_average_rho}
\lim_{T\to\infty} 
\frac{\overline{\Delta t}({\un r})}{T}=\rho({\un r})\,\Delta^3r
\end{eqnarray}
where $\overline{\Delta t}({\un r})$ denotes the overall time which the particle has spent occurring repeatedly in $\Delta^3r$
around ${\un r}$ within the total time span $T$. \\
The velocity ${\un v}({\un r})$ can be defined analogously 
\begin{eqnarray}
\label{eqn:time-average_v}  
{\un v}({\un r})=\frac{1}{\hat{n}({\un r})}\,\sum_{i=1}^{\hat{n}({\un r})} {\un v}(t_i)
\end{eqnarray}
where $\hat{n}({\un r})$ is the number of times the particle has occurred in $\Delta^3r$ around ${\un r}$, and $t_i$ denotes some point within the time span the particle has spent there the $i^{th}$ time. In realistic cases in which the system under study undergoes transitions between quasi-stationary states, one has to allow $T$ to be  finite, and quasi-stationarity can only be ensured if the changes are sufficiently slow on a time scale of unit length $T$. Practical experience shows, that this applies to the majority of cases. However, in Section \ref{kapitel15} we shall give an example where $T$ must be expected to be far too long to justify a classification of the states in a photo emission transition as quasi-stationary.\\
Yet, the bulk of this article will deal with ensemble averages.  
\\[0.2cm]
\section{Navier-Stokes equations}\label{kapitel3}
If a particle of mass $m_0$ moves in an environment of kinematic viscosity $\nu$ the resulting ensemble average of its velocity ${\un v}({\un r},t)$ is just the sum of the so-called ``convective velocity'' ${\un v}_c({\un r},t)$ and a ``diffusive velocity'' ${\un u}({\un r},t)$ driven by the stochastic forces of the embedding medium:
\begin{eqnarray}
\label{eqn:convectivevelocity}
{\un v}({\un r},t)={\un v}_c({\un r},t)+{\un u}({\un r},t) 
\end{eqnarray}
Employing the Smoluchowski equation (s. Section \ref{kapitel22}) for the probability density $\rho({\un r},t)$, similarly for the probability current density ${\un j}_c({\un r},t)=\rho({\un r},t)\,{\un v}_c({\un r},t)$  and invoking Einstein's law \cite{Einstein1} for the mean square displacement we obtain a Navier-Stokes-type equation of the form 
\begin{eqnarray}
\label{eqn:Navier_Stokes}
\frac{\partial}{\partial t}\,({\un v}-{\un
u})+\left[({\un v}+{\un u})\cdot\nabla ({\un v}- {\un u})\right]-\nu\,\Delta
({\un v}-{\un u}) \nonumber \\
=\frac{1}{m_0}\,{\un F}({\un r})\,.
\end{eqnarray}
with ${\un F}({\un r})=-\nabla V({\un r})$ denoting the external conservative force acting on the particle. The ``osmotic'' or ``diffusive'' velocity  ${\un u}({\un r},t)$ is defined by
\begin{eqnarray}
\label{eqn:osmotic_velocity}
{\un u}({\un r},t)=-\nu\,\frac{\nabla \rho({\un r},t)}{\rho({\un r},t)}\,,
\end{eqnarray}
or equivalently in terms of the diffusion current density ${\un j}_D$
\begin{eqnarray}
\label{eqn:diffusion_currentdensity}
{\un j}_D({\un r},t)=-\nu\,\nabla \rho({\un r},t) \;\; \mbox{``Fick's law''}
\end{eqnarray}
where
\begin{eqnarray}
\label{eqn:diffusion_currentdensity_2}
{\un j}_D({\un r},t)=\rho({\un r},t)\,{\un u}({\un r},t)\,.
\end{eqnarray}
In the special case when ${\un v}_c\equiv 0$ the equation of continuity reduces to
\begin{eqnarray}
\label{equation_continuity}
\frac{\partial}{\partial t}\,\rho + \nabla \cdot\rho\,{\un u}=0\,,
\end{eqnarray}
which on insertion of ${\un u}({\un r},t)$ from Eq.(\ref{eqn:osmotic_velocity}) attains the form of the diffusion equation
\begin{eqnarray}
\label{diffusion_equation}
\frac{\partial}{\partial t}\,\rho =\nu\,\Delta \rho\,.
\end{eqnarray}
On the other hand, when $\nu=0$ one has ${\un u}({\un r},t)\equiv 0$, and hence all particles move now along smooth trajectories ${\un r}(t)$ so that the various velocities ${\un v}_i(t)$ under the sum in Eq.(\ref{eqn:average_v}) become equal: ${\un v}_i(t)={\un v}({\un r}(t))$. Thus
$$
\frac{\partial}{\partial x_k}\,{\un v}({\un r}(t))\equiv 0\quad (k=1,2,3) \;\;\rightarrow {\un v}\cdot \nabla\,{\un v}\equiv 0\,,
$$
and consequently Eq.(\ref{eqn:Navier_Stokes}) reduces to Newton's second law. 
\\[0.2cm]  
The set of equations (\ref{eqn:Navier_Stokes}) to (\ref{eqn:diffusion_currentdensity}) will be derived in Section \ref{kapitel22}.\\
Eq.(\ref{eqn:osmotic_velocity}) may be rewritten
\begin{eqnarray}
\label{eqn:osmotic-velocity1}
 {\un u}({\un r},t)=-\nu \frac{\nabla \rho({\un
r},t)}{\rho({\un r},t)}=-\nu \,\nabla \,\ln[\rho({\un r},t)/\rho_0]
\end{eqnarray}
where $\rho_0$ denotes a constant density that has merely been inserted for dimensional reasons. As ${\un u}({\un r},t)$ can be expressed as a gradient of a function, we have
\begin{eqnarray}
\label{eqn:curlfree_u}
\nabla\times {\un u}({\un r},t)=0\,,
\end{eqnarray}
and hence
\begin{eqnarray}
\label{eqn:u_dot_nabla_u} ({\un u}\cdot \nabla){\un u}=\nabla\,\frac{{\un
u}^2}{2}\,.
\end{eqnarray}
If we, further, make use of the identity
\begin{eqnarray}
\label{eqn:identity}
\nabla\times(\nabla\times {\un a})=\nabla(\nabla\cdot {\un a})-\Delta {\un a}
\end{eqnarray}
and observe Eq.(\ref{eqn:curlfree_u}) we obtain $\Delta {\un u}({\un r},t)=\nabla(\nabla\cdot {\un u}({\un r},t))$. Thus, Eq.(\ref{eqn:Navier_Stokes})  in conjunction with Eq.(\ref{eqn:osmotic_velocity}) may be cast as
\begin{eqnarray}
\label{eqn:Stoch_equation1}
m_0\frac{d}{dt}{\un v({\un r},t)}={\un F}({\un r})-\nabla
V_{stoch}({\un r},t)+\vec{\Omega}({\un r},t)\,,
\end{eqnarray}
where $V_{stoch}({\un r},t)$ and $\vec{\Omega}({\un r},t)$ are abbreviations which stand for
\begin{eqnarray}
\label{eqn:Stochastic_potential}
V_{stoch}=\nu^2\, \left[\frac{1}{2}
\left(\frac{\nabla \rho }{\rho}\right)^2-\frac{\nabla^2\rho }{\rho}\right]\,,
\end{eqnarray}
and
\begin{eqnarray*}
\label{eqn:Omega} \vec{\Omega}=\frac{\partial{\un u}}{\partial t}+({\un
v}\cdot\nabla)\,{\un u}-({\un u}\cdot\nabla)\, {\un v}+\nu\,\Delta\,{\un v}\,.
\end{eqnarray*}
In deriving (\ref{eqn:Stochastic_potential}) we have observed that $\frac{1}{\nu}\,\nabla {\un u}=-\frac{\Delta \rho}{\rho}+(\frac{\nabla \rho}{\rho})^2$. 
Furthermore, we have introduced $\frac{d{\un v}}{dt}$ as the ``convective (or hydrodynamic) acceleration'' which in the present context merely represents an abbreviation
\begin{eqnarray}
\label{eqn:convective_acceleration}
\frac{d{\un v}({\un r},t)}{dt}=\frac{\partial\,{\un v}}{\partial t}\,
+{\un v}\cdot\nabla {\un v}\,.
\end{eqnarray}
The ``stochastic potential'' $V_{stoch}({\un r},t)$ depends on $\nu^2$ whereas $\vec{\Omega}({\un r},t)$ is proportional to $\nu$.\\
The latter constant is associated with the occurrence of the stochastic forces  which - in the absence of an external force ${\un F}({\un r})$ - would slow down the particle within a characteristic time $\tau$.\\
Since the physical vacuum does not represent an embedding medium whose stochastic forces can cause a particle to slow down completely, we modify the character of the stochastic forces by assuming that they change periodically after a laps of $\approx \tau$ sec from down-slowing ``Brownian'' to motion enhancing ``anti-Brownian'' and vice versa. The ``anti-Brownian'' forces act as if the kinematic viscosity would have a negative sign. Hence, the corresponding equation of motion has the form
\begin{eqnarray}
\label{eqn:Stoch_equation2}
m_0\frac{d}{dt}{\un v({\un r},t)}={\un F}({\un r})-\nabla
V_{stoch}({\un r},t)-\vec{\Omega}({\un r},t)\,.
\end{eqnarray}
In Section \ref{kapitel11} we give an example of an embedding medium that acts on a test particle by alternating Brownian/anti-Brownian forces. \\
If we now additionally assume that the temporal changes that occur with all quantities in Eqs.(\ref{eqn:Stoch_equation1}) and (\ref{eqn:Stoch_equation2}) are slow on a scale of unit length $\tau$ - which is the standard requirement also in diffusion theory - the motion of the ensemble will be governed by the arithmetic mean of these equations , that is by
\begin{eqnarray}
\label{eqn:Bohm_equation1}
m_0\frac{d}{dt}{\un v({\un r},t)}={\un F}({\un r})-\nabla V_{stoch}({\un r},t)\,.
\end{eqnarray}
A more detailed definition of the stochastic forces that ensure ``conservative diffusion'' will be given in Section \ref{kapitel8}.
One might suspect that our subdivision into a Brownian ``B``-ensemble and an anti-Brownian ``A''-ensemble is unnecessarily clumsy and could be avoided at the outset by assuming vacuum forces that neither possess down-slowing components nor counterparts that effect motion enhancement, but rather consist of random (Gaussian) forces whose components form a normal distribution. However, from Einstein's theory of Brownian motion the kinematic viscosity (or ``diffusion constant'') emerges as
\begin{eqnarray}
\label{eqn:EinsteinLaw}
\nu=\frac{k_B\,T\,\tau}{m_0} \quad \mbox{(Einstein:}\: \overline{\Delta x_i\,\Delta x_j}=2\,\delta_{ij}\,\nu\,\Delta t\,; \\
\quad i,j=1,2,3\nonumber \,)
\end{eqnarray}
where $m_0$ is the mass of the particle under study, $\Delta x_i\,,\Delta x_j$ are displacements of its position and $T$ is the effective temperature of the embedding medium. This temperature enters into the derivation as the width of the distribution of the random (Gaussian) forces that act on the particle apart from the {\bf directional} down-slowing force. Because of the latter there is a down-slowing motion that we have already alluded to. The associated time constant is denoted by $\tau$. Equating the down-slowing forces to zero amounts to $\tau \rightarrow \infty$ which would yield infinite kinematic viscosity. Hence, there is no alternative to our approach.\\
Obviously, the physical dimension of the numerator of the above fraction in Eq.(\ref{eqn:EinsteinLaw}) is that of an action, i.e. energy$\times$time. As $\nu$ appears via $V_{stoch}({\un r},t)$ in Eq.(\ref{eqn:Bohm_equation1}) which is constructed to describe dissipationless motion in a ``stochastic vacuum'' whose effect on a particle can only be associated with a new constant of nature, one is justified in equating $k_B\,T\,\tau$ with $\frac{1}{2}\,\hbar$ where $h=2\pi\,\hbar$ is Planck's constant. Of course, instead of $1/2$ there could be any other dimensionless prefactor in front of $\hbar$, but it turns out that the numerical results of all quantum mechanical calculations that follow from Eq.(\ref{eqn:Bohm_equation1}) are only consistent with the above choice. {\it Clearly, that choice has to be made only once and for all.}\\
Having thus calibrated the ``vacuum-$\nu$''we rewrite Eq.(\ref{eqn:Bohm_equation1}) in the form
\begin{eqnarray}
\label{eqn:Bohm_equation2}
m_0\frac{d}{dt}{\un v({\un r},t)}={\un F}({\un r})-\nabla
V_{QP}({\un r},t)\,,
\end{eqnarray}
where we have substituted the subscript of $V_{stoch}({\un r},t)$ by ``$QP$''
\begin{eqnarray}
\label{eqn:Quantum_potential}
V_{QP}={\textstyle \frac{\hbar^2}{4\,m_0}}\, \left[\frac{1}{2}
\left(\frac{\nabla \rho }{\rho}\right)^2-\frac{\nabla^2\rho }{\rho}\right]\; \mbox{{\small``quantum potential''}}\,,
\end{eqnarray}
and we have set 
\begin{eqnarray}
\label{eqn:Definition_hbar}
\frac{\hbar}{2m_0}=\nu=\frac{k_B\,T\,\tau}{m_0}\,.
\end{eqnarray}
The ``quantum potential'' has first been introduced by de Broglie \cite{Broglie} and later been taken up again by David Bohm \cite{Bohm1}. Obviously Eq.(\ref{eqn:Bohm_equation2}) may be viewed as a modification of Newton's second law.\\
The assumption made above, viz. that all changes of the ensemble properties have to be sufficiently slow on a time scale of unit length $\tau$ may raise questions about the validity of such a constraint. Eqs.({\ref{eqn:Bohm_equation2}) and (\ref{eqn:Quantum_potential}) will prove equivalent to the time-dependent Schr\"odinger equation whose validity is unquestioned at the non-relativistic level. Hence, $\tau$ is obviously sufficiently small within the experimentally tested range of the Schr\"odinger equation. Conversely, as one may conclude then from Eq.(\ref{eqn:Definition_hbar}) the ``effective temperature'' of the vacuum must be very high compared to those temperatures commonly considered in applied thermodynamics and astrophysics. 
\\[0.2cm]
Fundamentally different from our approach Bohm \cite{Bohm1} derives Eqs.({\ref{eqn:Bohm_equation2}) and (\ref{eqn:Quantum_potential}) by choosing the opposite direction starting from the time-dependent Schr\"odinger equation which he just considers given. Hence, he does not offer any new insight into what makes the motion of a microscopic particle different from what classical mechanics predicts. In the context of Bohm's mechanics Eq.({\ref{eqn:Bohm_equation2}) is frequently cast such that it resembles the Hamilton-Jacobi equation. To this end one sets
$$
{\un v}({\un r},t)=\frac{1}{m_0}\,\nabla S({\un r},t)\quad \mbox{and} \quad \rho({\un r},t)=R^2({\un r},t)
$$
which implies, again for no obvious reason, that ${\un v}({\un r},t)$ is curl-free. \\
Eq.({\ref{eqn:Bohm_equation2}) in conjunction with (\ref{eqn:Quantum_potential}) then attains the form
$$
\frac{1}{m_0}\,\nabla\left[\frac{\partial S}{\partial t}+\frac{(\nabla S)^2}{2\,m_0}+V({\un r})-\frac{\hbar}{2\,m_0}\,\frac{\Delta R}{R}\right]=0\,.
$$
This is equivalent to
$$
\frac{\partial S}{\partial t}+\frac{(\nabla S)^2}{2\,m_0}+V({\un r})-\frac{\hbar}{2\,m_0}\,\frac{\Delta R}{R}=0\,,
$$
and becomes identical with the Hamilton-Jacobi equation in the limit $\hbar\rightarrow 0$. However, the connection to classical mechanics is far more evident from Eq.(\ref{eqn:Bohm_equation2}), which reduces to Newton's second law
$$
{\un F}=m_0\,\frac{d}{dt}{\un v}
$$
as $\hbar$ tends to zero. In addition, Eq.(\ref{eqn:Bohm_equation2}) lends itself to a thought-experiment that is particularly illustrative of the quantum character of particle motion.\\
One starts with setting $\hbar=0$ and assumes that all particles of the ensemble commence their motion under identical initial conditions. Their positions and trajectories will coincide then at any later time. One now lets $\hbar$ take on a finite value. As a consequence of the now occurring stochastic forces whose action on some particle is statistically independent from that on any other particle, the particle positions start diverging and form a cloud around the formerly common position along the trajectory. The particles of the ensemble now reach positions that are not accessible under energy conservation. It is hence obvious that the vacuum provides an embedding medium of a ``universal noise''  consisting of energy fluctuations which cause shifts of the individual particle trajectories such that the classical momentum and the energy are conserved on the average. This is reflected in the expectation value of the ``vacuum force'' ${\un F}_{QP}=-\nabla V_{QP}({\un r},t)$ which equals zero:
\begin{eqnarray}
\label{eqn:Expect_value_vacuum_force}
\int \rho({\un r},t)\,{\un F}_{QP}({\un r},t)\,d^3r=0\,.
\end{eqnarray}
We shift the proof of this equation to Section \ref{kapitel9}. Eq.(\ref{eqn:Expect_value_vacuum_force}) may be interpreted in the sense that the particles undergo only reversible scatterings. Figuratively speaking, the vacuum keeps track of the energy balance and remembers at later positions of a particle departures from its classical momentum and energy that occurred at previous positions. The undulatory properties of the \textbf{probability density}  reside in this memory effect which gives rise to an unfamiliar non-locality.  Hence, from our point of view it is illegitimate to correlate these properties with a \textbf{wave-like character of the particle}. We definitely side with Nevill Mott (1964) who argues:
\\[0.2cm]
 {\it ``Students should not be taught to doubt that electrons, protons and the like are 
particles....The waves cannot be observed in any way than by observing particles.''}

\section{The time-independent Schr\"odinger equation}\label{kapitel4} 

As a first application we discuss the stationary state of a particle that is bound to a potential without symmetry elements. Hence, the real-space dependence of the potential does not display any distinct direction. That means, when a particle of the ensemble appears with a velocity ${\un v}_i(t)$ in the elementary volume $\Delta^3 r$ around ${\un r}$ there will always be another particle in that volume with approximately the opposite velocity, so that 
\begin{eqnarray}
\label{eqn:zero_velocity}
{\un v}({\un r},t)=\frac{1}{n({\un r},t)}\,\sum_i^{n({\un r},\,t)} {\un v}_i(t)\equiv 0\,.
\end{eqnarray}
Hence, Eq.(\ref{eqn:Bohm_equation2}) reduces to
\begin{eqnarray*}
\nabla\left(\frac{\hbar^2}{4\,m_0}\left[-\frac{1}{\rho}\nabla^{2}\rho+
\frac{1}{2}\left(\frac{\nabla\rho}{\rho}\right)^{2}\right]+V({\un r})\right)=0\,.
\end{eqnarray*}
This is equivalent to:
\begin{eqnarray}
\label{eqn:DglgRho}
\frac{\hbar^2}{4\,m_0}\left[-\frac{1}{\rho}\nabla^{2}\rho+\frac{1}{2}\left(
\frac{\nabla\rho}{\rho}\right)^{2}\right]+V({\un r})=E\,,
\end{eqnarray}
where $E$ denotes a constant. Eq.(\ref{eqn:DglgRho}) represents a \textbf{non-linear} partial differential equation in $\rho({\un r})$.\\
On substituting $\rho({\un r})$ by a function $\psi({\un r})$ defined through
\begin{eqnarray}
\label{eqn:Madelung_transform1}
\rho({\un r})=\psi^2({\un r})
\end{eqnarray}
one obtains because of
$$
\nabla \rho=2\,\psi\,\nabla\psi\;;\quad \frac{1}{2}\,\left(\frac{\nabla
\rho}{\rho}\right)^2=2\,\left(\frac{\nabla\psi}{\psi}\right)^2
$$
and
\begin{eqnarray*}
\nabla^2\rho=2\,\psi\,\nabla^2\psi+2\,(\nabla\psi)^2 \qquad\\
-\frac{1}{\rho}\,\nabla^2 \rho=-2\,\frac{\nabla^2
\psi}{\psi}-2\left(\frac{\nabla^2 \psi}{\psi}\right)^2
\end{eqnarray*}
a \textbf{linear} differential equation 
\begin{eqnarray}
\label{eqn:DglgPhi}
\frac{\hbar^2}{2\,m_0}\left[-\frac{1}{\psi}\nabla^{2}\psi\right]+V({\un r})=E \quad \mbox{that is}\: \nonumber\\
-\frac{\hbar^2}{2\,m_0}\,\nabla^{2}\psi+V({\un r})\,\psi=E\;\psi\qquad \quad
\end{eqnarray}
which constitutes the time-independent Schr\"odinger equation.

\section{Including currents}\label{kapitel4.1}

For the familiar problem of a particle in a box Eq.(\ref{eqn:DglgPhi}) reduces in the one-dimensional case to
\begin{eqnarray}
\label{eqn:particle_box}
\left[\frac{d^2}{dx^2}+k^2\right]\psi(x)=0
\end{eqnarray} 
where we have set 
$$
k^2=\textstyle\frac{2m_0}{\hbar^2}\,E\,; \quad E={\textstyle \frac{m_0}{2}}\,v^2  
$$
and
\begin{eqnarray*}
V(x)=\left\{\begin{array}{r@{\quad \quad}l}
\,0  & \mbox{for}\;\; 0 \leq x \leq a \\
\infty & \mbox{else}
\end{array} \right. 
\end{eqnarray*}
The solutions 
\begin{eqnarray}
\label{eqn:momentum_quantization}
\psi(x)={\textstyle \frac{1}{\sqrt{a/2}}}\,\sin k_{n}x \;\; \mbox{where}\: k_n=\frac{\pi}{a}\,n\,;\;\; n=1,2,3..
\end{eqnarray}
may be recast as
$$
\psi(x)={\textstyle \frac{1}{\sqrt{2}}}\,[\psi_{+}(x)+\psi_{-}(x)]
$$
where
$$
\psi_{\pm}(x)={\textstyle \frac{1}{\sqrt{a}}}\,e^{\pm \,i\varphi(x)}\;;\quad \varphi(x)=k_{n}x+{\textstyle \frac{\pi}{2}}\,.
$$
In the spirit of our approach the two independent solutions to the differential equation (\ref{eqn:particle_box}), $\psi_{\pm}(x)$,  refer to the particle moving at a velocity $v_n=\frac{\hbar\,k_n}{m_0}$ either to the right or (after reflection at $x=a$) to the left where it is reflected again at $x=0$.\\
We are thus led to surmise that we have in the general case of a freely moving particle
\begin{eqnarray}
\label{eqn:velocity_plus_wavef}
\psi({\un r})=|\psi({\un r})|\,e^{i\varphi({\un r})} \quad \mbox{and} \quad   {\un v}({\un r})=\frac{\hbar}{m_0}\,\nabla \varphi({\un r})\,.
\end{eqnarray}
The validity of this conjecture will be shown in Section \ref{kapitel5}.\\
In a stationary state of the one-particle system in which $\frac{\partial}{\partial t}{\un v}=0$ but ${\un v}({\un r})\not=0$ we have according to Eq.(\ref{eqn:convective_acceleration}) $\frac{d}{dt}{\un v}={\un v}\cdot\nabla {\un v}=\frac{1}{2}\nabla{\un v}^2$ where we have exploited in advance that, according to Eq.(\ref{eqn:velocity_plus_wavef}), ${\un v}({\un r})$ is curl-free. Hence, in the presence of a stationary current Eq.(\ref{eqn:DglgRho}) contains the kinetic energy $\frac{m_0}{2}\,{\un v}^2$ as an additional term, that is
\begin{eqnarray}
\label{eqn:DglgRhoKinetic} 
{\textstyle \frac{\hbar^2}{4\,m_0}}\left[-\frac{1}{\rho}\nabla^{2}\rho+\frac{1}{2}\left(
\frac{\nabla\rho}{\rho}\right)^{2}\right]+V({\un r})+\frac{m_0}{2}\,{\un v}^2=E\,.
\end{eqnarray}
If one now makes use of Eq.(\ref{eqn:velocity_plus_wavef}) instead of Eq.(\ref{eqn:Madelung_transform1})
\begin{eqnarray}
\label{eqn:Madelung_1}
\rho({\un r})=|\psi({\un r})|^2=\left(\psi({\un r})\,e^{-i\varphi({\un r})}\right)^2
\end{eqnarray}
and substitutes ${\un v}({\un r})$ by $\frac{\hbar}{m_0}\,\nabla \varphi({\un r})$ the bracketed term in Eq.(\ref{eqn:DglgRhoKinetic}) becomes
\begin{eqnarray*}
\frac{\hbar^2}{4\,m_0}\left[-\frac{1}{\rho}\nabla^{2}\rho+\frac{1}{2}\left(\frac{\nabla\rho}{\rho}\right)^{2}\right]= \qquad \qquad \qquad \qquad \\
-\frac{\hbar^2}{2\,m_0}\,\frac{1}{\psi}\nabla^{2}\psi+\underbrace{\frac{\hbar^2}{2m_0}\,(\nabla \varphi)^2}_{=\frac{m_0}{2}\,{\un v}^2} \qquad \qquad\\
+\underbrace{i\left[\frac{\hbar^2}{2\,m_0}\,\nabla^2 \varphi + \frac{\hbar^2}{2\,m_0}\,\left(2\,\nabla \varphi\cdot \frac{\nabla \psi}{\psi}\right)\right]}_{=\frac{i\hbar}{2}\left[\nabla \cdot\,{\un v}+2{\un v}\cdot \frac{\nabla \psi}{\psi}\right]}\,.
\end{eqnarray*}
Invoking the equation of continuity in the form
$$
\nabla\cdot{\un j}=\nabla \cdot\rho\,{\un v}=\rho\,\nabla\cdot{\un v}+{\un v}\cdot\nabla\rho=0
$$
it can readily be shown that the term $i [...]$ on the right-hand side equals $-m_0\,{\un v}^2$. Hence we have from Eq.(\ref{eqn:DglgRhoKinetic})
\begin{eqnarray*}
-\frac{\hbar^2}{2\,m_0}\,\frac{1}{\psi}\nabla^{2}\psi +V({\un r})=E\,,
\end{eqnarray*}
that is 
\begin{eqnarray}
\label{eqn:Schroedinger_Equation_2}
-\frac{\hbar^2}{2\,m_0}\,\nabla^{2}\psi +V({\un r})\,\psi=E\,\psi
\end{eqnarray} 
as before without a current.\\
It should be noticed that $\varphi$ may well be time-dependent even when $\nabla \varphi$ is not, that is, we have in general
$$
\varphi({\un r},t)=\varphi_0({\un r})+f(t)
$$
where $f(t)$ is a real-valued function. In this case the wave function $\psi({\un r},t)$ attains the form 
\begin{eqnarray}
\label{eqn:time_dependent_wave function_1}
\psi({\un r},t)=\hat{\psi}({\un r})\,e^{i\,f(t)} \quad \mbox{where} \quad \hat{\psi}({\un r})=|\hat{\psi}({\un r})|\,e^{i\,\varphi_0({\un r})}
\end{eqnarray}
and hence, its time-derivative may be cast as
\begin{eqnarray}
\label{eqn:time_derivative_wave function_1}
i\hbar\,\frac{\partial}{\partial t}\psi({\un r},t)=-\hbar\,\dot{f}\,\psi({\un r},t)\,.
\end{eqnarray}
Since $f(t)$ is primarily unspecified and $-\hbar\dot{f}$ possesses the dimension of an energy the latter may justifiably be identified with the energy $E$ which is the only energy-related constant characterizing the wave function of the system:
\begin{eqnarray}
\label{eqn:time_derivative_wave function_2}
-\hbar\,\dot{f}=E\,; \quad \mbox{that is} \quad if(t)=-\frac{i}{\hbar}E\,t\,.
\end{eqnarray}
As a result, we have from Eq.(\ref{eqn:time_dependent_wave function_1}) 
\begin{eqnarray}
\label{eqn:time_dependent_wave function_2}
\psi({\un r},t)=\hat{\psi}({\un r})\,e^{-\frac{i}{\hbar}E\,t}
\end{eqnarray}
for a wave function in a stationary state. Furthermore, we have from Eqs.(\ref{eqn:Schroedinger_Equation_2}), (\ref{eqn:time_derivative_wave function_1}) and (\ref{eqn:time_derivative_wave function_2}) 
\begin{eqnarray}
\label{eqn:Schroedinger_Equation_3} 
-\frac{\hbar^2}{2\,m_0}\,\nabla^{2}\psi({\un r},t) +V({\un r})\,\psi({\un r},t)=i\hbar\,\frac{\partial}{\partial t}\psi({\un r},t)
\end{eqnarray}
which constitutes the time-dependent Schr\"odinger equation. Its validity is here still restricted to stationary systems, but it will be shown in Section \ref{kapitel6} that it retains this form also for non-stationary systems. However, in order to achieve this consistency, one has to introduce the negative sign in Eq.(\ref{eqn:time_derivative_wave function_2}) which seems to lack reason and can actually not be justified without reference to Section \ref{kapitel6}.

\section{The velocity potential and phase uniqueness}\label{kapitel5}

We rewrite Eq.(\ref{eqn:Bohm_equation2}) in the form
\begin{eqnarray}
\label{eqn:Helmholtz_equation}
m_0\frac{d}{dt}{\un v({\un r},t)}=-\nabla P({\un r},t) 
\end{eqnarray}
where
$$
P({\un r},t)=\frac{1}{m_0}\,[V({\un r})+V_{PQ}({\un r},t)]\,,
$$
and we have made use of Eq.(\ref{eqn:convective_acceleration}) defining the ``hydrodynamic'' or convective acceleration
\begin{eqnarray*}
\frac{d}{d\,t}{\un v}({\un r},t)=\frac{\partial}{\partial\,t}{\un v}+ ({\un v}\cdot\nabla)\,{\un v}\,.
\end{eqnarray*}
In hydrodynamics Eq.(\ref{eqn:Helmholtz_equation}) corresponds to the Euler equation of perfect (frictionless) fluids and constitutes the starting point of Helmholtz's theory of vortices. Thomson's more elaborate analysis on vortices \cite{Thomson} builds on Helmholtz's considerations. We confine ourselves here to reporting only the general ideas as far as they directly concern the present theory. \\
If we set $\vec{\omega}=\nabla\times {\un v}$ for the curl of the ensemble average of the particle velocity,
we have from Eq.(\ref{eqn:identity})
$$
({\un v}\cdot\nabla)\,{\un v}=\nabla\frac{{\un v}^2}{2}-{\un v}\times\vec{\omega}\,.
$$
We now form the curl of Eq.(\ref{eqn:Helmholtz_equation}) and use this expression together with Eq.(\ref{eqn:convective_acceleration}).  The result may be cast as
\begin{eqnarray}
\label{eqn:ZeitablOmega} \frac{\partial}{\partial t}\,\vec{\omega}({\un
r},t)-\nabla \times [{\un v}({\un r},t)\times \vec{\omega}({\un r},t)]=0\,,
\end{eqnarray}
where we have used $\nabla \times \nabla P=0\quad\mbox{and} \quad \nabla\times\nabla {\un v}^2=0$. 
One recognizes from Eq.(\ref{eqn:ZeitablOmega}) that $\frac{\partial}{\partial t}\,\vec{\omega}({\un
r},t)|_{t=0}$ becomes zero for some chosen time, which we here equate to zero for convenience, if $\vec{\omega}({\un r},t)|_{t=0}=0$ at that time. Forming the time derivative of Eq.(\ref{eqn:ZeitablOmega}) and setting again $t=0$ we see that the second time derivative of $\vec{\omega}({\un r},t)$ vanishes as well. This can be carried further to any higher order of the time derivative. Hence, the system stays curl-free if it is curl-free at $t=0$. We now consider an ensemble of free particles $({\un F}({\un r})\equiv 0)$ when $\hbar=0$. They may start their motion at $t=0$ at the same point in real-space and with the same momentum ${\un p}_0=m_0\,{\un v}_0$. If one allows $\hbar$ to attain its natural value, the particle positions diverge and form a point cloud. Outside this cloud there are no particles and therefore ${\un v}({\un r},t)\equiv 0$. Since the ensemble does not exchange momentum with the vacuum on the average and consequently no angular momentum, we have everywhere within the space of normalization
\begin{eqnarray}
\label{eqn:curl_zero} \nabla\times {\un v}({\un r},t)=\vec{\omega}({\un
r},t)\equiv 0\quad \forall\,{\un r}, t\,.
\end{eqnarray}
If one now turns on some (physically realistic) potential $V({\un r})$, weighting it with a smooth switch function from zero to one, starting at $t=t_0$, the velocity distribution ${\un v}({\un r},t)$ for $t>t_0$ will now change differently, of course, but because of Eqs.(\ref{eqn:ZeitablOmega}) and (\ref{eqn:curl_zero}) for $t=t_0$, we have as before $\vec{\omega}({\un r},t_0)\equiv 0$ and $\frac{\partial}{\partial t}\,\vec{\omega}({\un r},t)|_{t=t_0}\equiv 0$ which again applies to any higher order time-derivative at $t=t_0$. We thus arrive at the conclusion that an ensemble whose equation of motion is given by Eq.(\ref{eqn:Helmholtz_equation}) is curl-free. In other words, ${\un v}({\un r},t)$ possesses a potential $\varphi({\un r},t)$ which we express in the form 
\begin{eqnarray}
\label{eqn:Gradphi}
{\un v}({\un r},t)=\frac{\hbar}{m_0}\,\nabla \varphi({\un r},t)\,.
\end{eqnarray}
Because of the prefactor $\hbar/m_0$ the function $\varphi({\un r},t)$ becomes dimensionless. Eq.(\ref{eqn:Gradphi}) may equivalently be cast as
\begin{eqnarray}
\label{eqn:Gradphi_2}
\varphi({\un r})=\frac{m_0}{\hbar}\,\int_{{\un r}_0}^{{\un r}} {\un v}({\un r}')\cdot d{\un r}'
\end{eqnarray}
where we have omitted the time-dependence in confining ourselves to a stationary state situation.
As in the theory of perfect fluids there may be singular vortex lines which occur if $V({\un r})$ possesses axial or spherical symmetry. A vortex line then defines an axis of quantization. The latter may be regarded as the boundary line of a semi-plane. Even in the presence of a vortex line,    can $\varphi({\un r})$ be defined such that it remains unique if one only stipulates that the starting point of the line integral in Eq.(\ref{eqn:Gradphi_2}), ${\un r}_0$,  lies on one side of this semi-plane and that the path along which the integral is performed never crosses that semi-plane. The point ${\un r}_0$ may be chosen at will. In general, $\varphi({\un r})$ will now be discontinuous at the semi-plane. The ensuing section deals with this particular problem.

\section{Quantization of angular momentum}\label{kapitel5.1}

The primary objective of this section is to disprove Wallstrom's notable objection \cite{Wallstrom} against Madelung's conviction, also held by other theorists of this school of thought, that Newton's modified second law (\ref{eqn:Bohm_equation2}) is equivalent to the time-dependent Schr\"odinger equation which we shall derive below. In so doing we have to exploit the uniqueness of the velocity potential shown in the preceding section. By contrast, in standard quantum mechanics the time-dependent Schr\"odinger equation is regarded as given. It is customarily converted into the equation of continuity
$$
\dot{\rho}+\nabla\cdot[{\textstyle\frac{\hbar}{2i\,m_0}}\{\psi^{*}\nabla\,\psi-\psi\,\nabla\,\psi^{*}\}]=0
$$
to show that the bracketed expression has to be interpreted as the current density ${\un j}({\un r},t)$. This conclusion is only legitimate if 
${\un j}({\un r},t)$ has been proven to be curl-free which, however, is only tacitly presupposed. Inserting
\begin{eqnarray}
\label{eqn:psi_polar_represent}
\psi({\un r},t)=|\psi({\un r},t)|\,e^{i\varphi({\un r},\,t)}
\end{eqnarray}
into the bracketed expression yields
$$
{\un j}({\un r},t)=|\psi({\un r},t)|^2\,\underbrace{\frac{\hbar}{m_0}\,\nabla \varphi({\un r},t)}_{={\un v}({\un r},t)}\,,
$$
as a consequence of which one obtains Eq.(\ref{eqn:Gradphi}). If one is dealing with a stationary state whose velocity field contains a vortex line, e.$\,$g. an excited state of a hydrogen electron possessing an orbital momentum, we have
\begin{eqnarray}
\label{eqn:angular_momentum_integral_1}
\oint{\un v}({\un r})\cdot d{\un r}\not=0
\end{eqnarray}
for any path encircling the vortex line (=quantization axis). On inserting here ${\un v}={\textstyle \frac{\hbar}{m_0}}\,\nabla \varphi$ one obtains
\begin{eqnarray}
\label{eqn:angular_momentum_integral_2}
\int_{{\un r}_0}^{{\un r}} \nabla \varphi({\un r})\cdot d{\un r}=\varphi({\un r})-\varphi({\un r}_0) \not=0
\end{eqnarray}
where ${\un r}$ and ${\un r}_0$ are two points facing each other across the semi-plane, introduced in Section \ref{kapitel5}, at an infinitesimal distance.
Thus, in general the phase of the wave function, and consequently the wave function itself, will be discontinuous at the semi-plane as opposed to  $\rho({\un r})$ and ${\un j}({\un r})$ which may be presupposed to be smooth functions everywhere.\\
Clearly, as follows from Eq.(\ref{eqn:psi_polar_represent}), $\psi({\un r})$ remains continuous at the semi-plane if
\begin{eqnarray}
\label{eqn:angular_momentum_quantization}
\varphi({\un r})-\varphi({\un r}_0)=2m\,\pi\quad \mbox{where} \quad m=\mbox{integer}\,.
\end{eqnarray}
But there is no immediately obvious reason why one should require $\psi({\un r})$ to be continuous because only $\rho({\un r})$ and ${\un j}({\un r})$ can be regarded as reflecting physical properties of the system. We are hence led to conclude that without an additional argument {\bf neither} our derivation {\bf nor} standard quantum mechanics yields a justification of the proven relation
\begin{eqnarray}
\label{eqn:angular_momentum_integral_3}
m_0 \oint{\un v}({\un r})\cdot d{\un r}=2m\pi\,\hbar=m\,h\; \mbox{where} \; m=\mbox{integer}
\end{eqnarray}
which comprises Eqs.(\ref{eqn:Gradphi}), (\ref{eqn:angular_momentum_integral_1}) to (\ref{eqn:angular_momentum_quantization}). This has already been pointed out more than 75 years ago by Pauli \cite{Pauli_1} and Born and Jordan \cite{BornJordan}. As opposed to these considerations Wallstrom states in his paper \cite{Wallstrom}: {\it ''To the best of my knowledge, this condition} (Eq.(\ref{eqn:angular_momentum_integral_3})) {\it has not yet found any convincing explanation {\bf  outside the context of the Schr\"odinger equation}''}. \\
This is definitely incorrect: within that context the assumption of continuity (Eq.(\ref{eqn:angular_momentum_quantization})) has to be justified by an additional argument as well. 
\\[0.2cm]
What else necessitates then the continuity of $\psi({\un r})$ everywhere? \\
We consider two states, $\hat{\psi}_{m_1}({\un r})$ and $\hat{\psi}_{m_2}({\un r})$, which are solutions to the time-independent Schr\"odinger equation for a spherically symmetric potential. The associated energies may be denoted by $E_{m_1}$ and $E_{m_2}$, and the spherical coordinates by $r,\theta,\alpha$ with $\theta=0,\pi$ defining the quantization axis (=vortex line). We assume that there is a weak magnetic field along this axis to lift a possible degeneracy. The solutions have the general form
$$
\hat{\psi}({\un r})=\frac{1}{\sqrt{2\pi}}\,F_{k_{1}\,k_{2}\,k_{3}}(r,\theta)\,e^{i\,k_{3}\, \alpha} \,.
$$ 
where $k_1,k_2,k_3$ are real numbers and 
\begin{eqnarray}
\label{eqn:alpha_dependence_of_varphi}
\varphi({\un r})=\varphi(\alpha)=k_3\,\alpha=m_{1/2}\,\alpha\,.
\end{eqnarray}
Moreover, $F_{k_{1}\,k_{2}\,m_{1/2}}(r,\theta)$ denotes real-valued functions whose square is normalized to unity. Specifically we distinguish two solutions which differ in their $k_3$-values and solve the {\it time-dependent} Schr\"odinger equation
$$
\psi_{m_{1/2}}({\un r},t)=\frac{1}{\sqrt{2\pi}}\,F_{k_{1}\,k_{2}\,m_{1/2}}(r,\theta)\,e^{i(\,m_{1/2}\,\alpha-\omega_{m_{1/2}}\,t)}
$$
where
$$
\omega_{m_{1/2}}=E_{m_{1/2}}/\hbar\,.
$$
Since the time-dependent Schr\"odinger equation is linear, the function
\begin{eqnarray*}
\psi({\un r},t)= {\textstyle  \frac{1}{\sqrt{2\pi}}}[c_{m_1}\,F_{k_{1}\,k_{2}\,m_{1}}(r,\theta)\,e^{i(\,m_{1}\,\alpha-\omega_{m_1}\,t)}\\
+c_{m_2}\,F_{k_{1}\,k_{2}\,m_{2}}(r,\theta)\,
e^{i(\,m_{2}\,\alpha-\omega_{m_2}\,t)}]
\end{eqnarray*}
constitutes a solution as well. The constants $c_{m_{1/2}}$ may be assumed real-valued. We now form
$$
\int \underbrace{|\psi({\un r},t)|^2}_{=\rho({\un r},t)}\,d^3r=c^2_{m_{1}}+c^2_{m_{2}}+ c_{m_{1}}\,c_{m_{2}}\,I_{m_{1}\,m_{2}}
$$
$$ 
\times\frac{1}{2\pi}\left[\int_{0}^{2\pi} e^{i(m_2-m_1)\alpha}\,d\alpha \: \:e^{i(\omega_{m_1}-\omega_{m_2})\,t}+ c.c.\right]
$$
where
$$
I_{m_{1}\,m_{2}}=\int F_{k_{1}\,k_{2}\,m_{1}}(r,\theta)\,
F_{k_{1}\,k_{2}\,m_{2}}(r,\theta)\,r^2\,dr\,\sin \theta \,d \theta \,. 
$$
According to Eq.(\ref{eqn:norm_integral}) the real-space integral of $\rho({\un r},t)$ must be unity at any time. This is obviously only possible if $m_2-m_1$ is an integer. Since $m_1$ may attain the particular value zero, corresponding to a state without circular current, it follows then that $m_2$ must be an integer itself and hence, if one invokes Eq.(\ref{eqn:alpha_dependence_of_varphi})
$$
m_0\oint{\un v}({\un r})\cdot d{\un r}=\hbar\oint \nabla \varphi({\un r})\cdot d{\un r}
$$
$$
=\hbar\int_0^{2\pi}\frac{\partial}{\partial \alpha}\,(m_2\,\alpha)\,d\alpha=m_2\,h 
$$
where $m_2$ is now proven to be an integer number, in agreement with the conjecture (\ref{eqn:angular_momentum_integral_3}).

\section{An instructive objection and quantum beats}\label{kapitel5.1}

An apparently serious objection against a stochastic foundation of quantum mechanics along the lines of the preceding sections goes back to Mielnik and Tengstrand \cite{Mielnik}. The authors refer to an experimental setup as sketched in Figure \ref{Interference} where the test particle enters from a distant source on the left-hand side and is kept within a tube that extends up to a screen on the right. The tube contains an impermeable partition that completely seals off the upper part (A) from the lower part (B). It possesses a limited, but macroscopic length of, say, 10$\,$cm. The authors argue that according to conventional quantum mechanics the incoming wave would split up into an upper and totally independent lower portion. Yet both portions retain their capability of interfering with each other when they merge again within the area C and beyond. However, if the wave portions are replaced by the set of irregular trajectories which stochastic quantum mechanics claims to be an equivalent of, it seems to be very unlikely that stochastic-force controlled trajectories can preserve information over so long a distance as well as waves. This criticism amounts to perceiving the preceding derivation of the Schr\"odinger equation from Eq.(\ref{eqn:Bohm_equation2}) as ill-founded or even erroneous. It is just the solution to the Schr\"odinger equation for the particular setup around which the authors' consideration revolve. On the other hand, it is easy to verify the validity of the derivation. There is simply no step where one may be in doubt. But one has to keep in mind that the solutions $\psi({\un r})=|\psi({\un r})|\,e^{i\varphi({\un r})}$ to the Schr\"odinger equation provide only information on ensemble properties and not on a particular trajectory that is a member of the ensemble under study. For example, the velocity ${\un v}({\un r})=\frac{\hbar}{m_0}\,\nabla \varphi({\un r})$ at some point in the area marked C represents such an average over all trajectories of the ensemble running through that point. 
\begin{figure}[ht]    
\epsfig{file=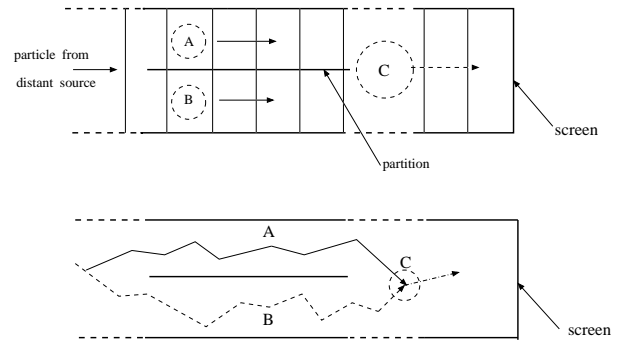,width=8cm}
\caption[Interference of trajectories] {\label{Interference} Interference of trajectories}
\end{figure}
This ensemble defines the probability in which direction a particular particle that has arrived at C, e.~g.  along the ``A''-trajectory, will move further. (S. Figure \ref{Interference}, lower panel.) This is analogous to considerations we shall discuss in the context of the Smoluchowski equation (Section \ref{kapitel22}). The properties of the ensemble are just an image of the property of the vacuum fluctuations to ensure the absence of dissipation. This manifests itself in the fact that ${\un v}({\un r})$ is curl-free as in ideal fluids. An individual particle that has moved along the ``A''-trajectory and arrives at C ``feels'', so to speak, the possibility of a ``B''-trajectory.  As stated above, it continues its trajectory depending also on the family of ``B''-trajectories running through C. If the partition in the tube would be elongated and the point C correspondingly shifted to the right, irrespective of how much, the ``A''- and ``B''subset of trajectories would now be different, but the scattering probability at C of  a particle that has moved along an ``A''(or ``B``)-trajectory would still be influenced by possible ``B'' (or ``A'')-trajectories. Furthermore, if one would place some electrostatic array into the upper part of the setup which would cause a spatially varying electrostatic potential, the ``A''-trajectories would change accordingly and give rise to a different interference pattern within the ``C''-range. To make the surprising content of this observation even more striking we consider a situation where one has particles enter the setup one by one from the left so that only one particle traverses the setup at a time. First, we switch the electrostatic array off so that there is no  extra potential along the ``A''-trajectory. If one has placed a detector, an electron multiplier, for example, at some position ${\un r}_{screen}$ on the screen, it would monitor the incoming electrons at a certain rate. These electrons come either along an `A''- or a ``B``-trajectory. Once the extra potential has been turned on, the count rate at ${\un r}_{screen}$ changes. Although an electron may have moved along the unmodified ``B``-portion of the setup, it feels the modification of the ``A''-portion when it arrives in ``C''. As explained above, this is due to the change of the vacuum scattering probability at C. Electrons that have arrived at some elementary volume within C and have so far preferentially been scattered into ${\un r}_{screen}$ are now also scattered to other positions on the screen, thereby changing the count rate at ${\un r}_{screen}$. \\
If the electrostatic array in the ``A''-portion would simply consist of two planar parallel grids perpendicular to the average particle motion, and if one applies an accelerating  voltage $V$ between the grids, the particles' kinetic energy $\epsilon_0$ increases by an amount $\Delta \epsilon=e\,V$ where $e$ denotes the particle charge. The wave function $\psi({\un r}_{screen},t)$ at the screen is the sum of the ``A''- and ``B''-related contributions: 
\begin{eqnarray}
\label{eqn:oscillating_interference}
={\textstyle \frac{1}{\sqrt{2}}}\left[\hat{\psi}_A({\un r}_{screen})\,e^{-i\omega_A\,t}+\hat{\psi}_B({\un r}_{screen})\,e^{-i\omega_B\,t}\right]
\end{eqnarray}
where 
$$
\hat{\psi}_{A/B}({\un r}_{screen})={\textstyle \frac{1}{\sqrt{N}}}\,e^{ik_{A/B}\,r_{screen}}
$$
and
$$
\hbar\,\omega_A=\epsilon_0+\Delta \epsilon\,;\,\hbar\,\omega_B=\epsilon_0\,,
$$
and with $1/\sqrt{N}$ denoting an appropriate normalization factor. The function $\psi({\un r},t)$ solves the time-dependent Schrödinger equation (\ref{eqn:Schroedinger_Equation_3}) for the particular array under study. If we introduce $\overline{\epsilon}=\epsilon_0+\frac{1}{2}\,\Delta\epsilon$ we may cast $\hbar\,k_{A/B}$ as $\hbar\,k_{A/B}\approx \sqrt{2m_0\,\overline{\epsilon}}\left(1\pm \textstyle{ \frac{1}{2}\,\frac{\Delta \epsilon}{\overline{\epsilon}}}\right)$ if $\frac{\Delta \epsilon}{\overline{\epsilon}}<<1$ where $m_0$ denotes the rest mass of the particle. Eq.(\ref{eqn:oscillating_interference}) can then be rewritten
\begin{eqnarray*}
\psi_{A/B}({\un r}_{screen},t)={\textstyle \frac{1}{\sqrt{N}}}\,e^{i(\overline{k}\,r_{screen}-\overline{\omega}\,t)}\,\\
\times{\textstyle \frac{1}{\sqrt{2}}}\,\left[e^{i(\Delta k\,r_{screen}-\Delta \omega\,t)}+e^{-i(\Delta k\,r_{screen}-\Delta \omega\,t)}\right]
\end{eqnarray*}
where $\hbar\,\overline {k}=\sqrt{2m_0\,\overline{\epsilon}}\,,\, \Delta k=k_A-k_B$ and $\hbar\,\Delta \omega={\textstyle \frac{1}{2}}\,\Delta \epsilon$. Hence we have for the current density ${\un j}({\un r},t)\propto$ count rate at ${\un r}_{screen}$
\begin{eqnarray*}
{\un j}({\un r}_{screen},t)={\textstyle \frac{1}{N}\,\frac{\hbar\,\overline{{\un k}}}{m_0}}\,|\psi_{A/B}({\un r}_{screen},t)|^2=\\
{\textstyle \frac{1}{N}\,\frac{\hbar\,\overline{{\un k}}}{m_0}}\,[1+\cos(2\Delta k\,r_{screen}-\Delta \epsilon\,t)]\,.
\end{eqnarray*} 
That means: the count rate oscillates at a period of $T=\frac{2\pi\,\hbar}{\Delta \epsilon}$. This most surprising effect of ``quantum beats'' has clearly been observed by Rauch and collaborators (s. Badurek et al. \cite{Badurek}) who used spin polarized neutrons instead of electrons. The energy change $\Delta \epsilon$ in the ``A''-section of the setup was in that case imparted to the respective neutron by flipping its spin within a spatially confined magnetic field along the ``A''-trajectory. (In practice one used a spin flipper also in the ``B''-portion of the setup where the corresponding magnetic field was slightly lower than in the ``A''-portion so that $\Delta \epsilon$ referred to the difference of two spin flip energies in this case.)

\section{The time-dependent Schr\"odinger equation}\label{kapitel6}

In the most general case ${\un v}$ and hence $\varphi$ are time-dependent. As already pointed out in Section \ref{kapitel4.1} the substitution of $\rho({\un r})$ has to be modified then in the form 
\begin{eqnarray}
\label{eqn:Madelung_transform2}
\psi({\un r},t)=\pm \sqrt{\rho({\un r},t)}\,e^{i\varphi\,({\un r},\,t)} 
\end{eqnarray}
which was introduced by Madelung in 1926 \cite{Madelung}
The $\pm$-sign requires a comment. As discussed in Section 2, $\rho({\un r},t)$ will generally be presupposed as a smooth function. The zeros of $\rho({\un r},t)$ pose a particular problem that occurred already in Section 3, but was not explicitly mentioned. The admissible type of zeros limits the set of functions $\rho({\un r})$ that can be mapped onto $\psi({\un r})$ according to Eq.(\ref{eqn:Madelung_1}). For simplicity we confine ourselves to the time-independent case and assume that the zeros of $\rho({\un r})$ lie on the faces of a rectangular parallelepiped defined by the equations $x_{\nu}=x_{\nu 0}$ with $x_{\nu}$ and $\nu=1,2,3$ denoting Cartesian coordinates. Hence close to $x_{\nu}=x_{\nu 0}$ and perpendicular to the respective face the density varies as $(x_{\nu}-x_{\nu 0})^2$. Since we have everywhere $\rho({\un r})\geq 0$ its square root varies as $|x_{\nu}-x_{\nu 0}|$ and thus would not be differentiable at $x_{\nu}=x_{\nu 0}$. In defining the map $\rho({\un r}) \rightarrow \psi({\un r})$ one is forced hence to choose the positive sign in front of $\sqrt{\rho({\un r})}$ outside the rectangular parallelepiped if one has chosen the minus sign inside (or vice versa) to ensure that $\psi({\un r})$ stays differentiable across the face of the rectangular parallelepiped. Hence,  mapping functions $\rho({\un r})$ onto differentiable functions $\psi({\un r})$ is only possible if the zeros of $\rho({\un r})$ subdivide the space of volume $\cal{V}$ into cells without leaving empty space. At first sight it appears that this limitation in the set of admissible functions $\rho({\un r})$ constitutes a serious drawback of the entire concept. One has to bear in mind, however, that the functions $\psi({\un r})$ are not determined as a map of $\rho({\un r})$ but rather by solving the Schr\"odinger equation (\ref{eqn:DglgPhi}) which has been the objective of the derivation. Physical meaningful solutions to Eq.(\ref{eqn:DglgPhi}) have automatically the required spatial structure of their zeros.
\\[0.2cm]
We now move on to derive the time-dependent Schr\"odinger equation under the supposition that the above considerations apply to the time-dependent case as well.\\
If one uses instead of Eq.(\ref{eqn:Bohm_equation2}) the arithmetic mean of the original Eq.(\ref{eqn:Navier_Stokes}) and its ``anti-Brownian'' analogue where the sign of $\nu$ and ${\un u}({\un r},t)$ is reversed, one obtains
\begin{eqnarray}
\label{eqn:local_acceleration0}
\frac{\partial}{\partial t}\,{\un v}+({\un v}\cdot\nabla)\,{\un v}-({\un u}\cdot\nabla)\,{\un u}+\frac{\hbar}{2\,m_0}\,\Delta {\un u}=\frac{1}{m_0}\,{\un F}({\un r})\,.
\end{eqnarray}
This can be simplified in the form:
\begin{eqnarray}
\label{eqn:local_acceleration}
\frac{\partial}{\partial\,t}{\un v}=-\frac{1}{m_0}\,\nabla V-\frac{1}{2}\,\nabla {\un v}^2+\frac{1}{2}\,\nabla {\un u}^2-\frac{\hbar}{2\,m_0}\,\Delta {\un u}\,,
\end{eqnarray}
where we have made use of the relations
$$
{\un v}\cdot\nabla {\un v}=\frac{1}{2}\nabla {\un v}^2\,; \quad {\un u}\cdot\nabla {\un u}=\frac{1}{2}\nabla {\un u}^2 \quad \mbox{and}\quad \nu=\frac{\hbar}{2m_0}\,.
$$
With the first two equations it has been observed that ${\un v}$ and ${\un u}$ are curl-free.
On differentiating ${\un u}$ with respect to time and using Eq.(\ref{eqn:osmotic_velocity}) one obtains
\begin{eqnarray}
\frac{\partial}{\partial\,t}\,{\un u}=-\frac{\hbar}{2\,m_0}\,\nabla\,(\frac{\partial\,\rho}{\partial\,t}/\rho)\,,
\end{eqnarray}
Invoking the equation of continuity
\begin{eqnarray}
\label{eqn:equation_continuity}
\partial\,\rho/\partial\,t + \nabla \cdot (\rho\,{\un v})=0
\end{eqnarray}
that is
$$
\partial\,\rho/\partial\,t + \rho\,\nabla \cdot {\un v} +{\un v}\cdot \nabla \rho =0
$$
$\frac{\partial\,\rho}{\partial\,t}/\rho$ can be replaced by $-\nabla \cdot {\un v}-{\un v}\cdot \frac{1}{\rho}\,\nabla \rho $ which yields 
\begin{eqnarray}
\label{eqn:Zeitableitung_u1}
-\frac{\hbar}{2\,m_0}\,\nabla\,(\frac{\partial\,\rho}{\partial\,t}/\rho)=\frac{\hbar}{2\,m_0}\,
\nabla\, (\nabla \cdot{\un v})\nonumber \\
-\nabla\left[{\un v}\cdot\left(-\frac{\hbar}{2\,m_0}\,\frac{1}{\rho}\,\nabla \rho\right)\right]\,.
\end{eqnarray}
Using Eq.(\ref{eqn:osmotic_velocity}) we may substitute the expression in the $[...]$-brackets on the right-hand side by ${\un v}\cdot {\un u}$. Hence Eq.(\ref{eqn:Zeitableitung_u1}) takes the form
\begin{eqnarray}
\label{eqn:Zeitableitung_u2}
\frac{\partial}{\partial\,t}\,{\un u}=\frac{\hbar}{2\,m_0}\,\nabla (\nabla \cdot{\un v})-\nabla({\un u}\cdot {\un v})\,.
\end{eqnarray}
On multiplying the equation of motion (\ref{eqn:local_acceleration}) by the imaginary unit $i$ and 
subtracting Eq.(\ref{eqn:Zeitableitung_u2}) we obtain
\begin{eqnarray*}
\frac{\partial}{\partial\,t}\,(-{\un u}+i\,{\un v})=-\frac{i}{m_0}\,\nabla V-\frac{i}{2}\,\nabla 
{\un v}^2+\frac{i}{2}\,\nabla {\un u}^2\\
-i\,\frac{\hbar}{2\,m_0}\,\Delta {\un u}-\frac{\hbar}{2\,m_0}\,\nabla (\nabla \cdot{\un v})+\nabla({\un u}\cdot {\un v})\,,
\end{eqnarray*} 
which after reordering the terms on the right-hand side becomes
\begin{eqnarray}
\label{eqn:komplexeBeschleunigung}
\frac{\partial}{\partial\,t}\,(-{\un u}+i\,{\un v})=\frac{i}{2}\,\nabla (-{\un u}+i\,{\un v})^2 \nonumber \\
+\frac{i\,\hbar}{2\,m_0}\nabla\left[\nabla \cdot(-{\un u}+i\,{\un v})\right]-\frac{i}{m_0}\,\nabla V\,.
\end{eqnarray}
Here we insert Eqs.(\ref{eqn:osmotic-velocity1}), (\ref{eqn:Gradphi}) and (\ref{eqn:Madelung_transform2}) in the form
\begin{eqnarray}
\label{eqn:komplexeGeschwindigkeit}
-{\un u}+i\,{\un v}=\frac{\hbar}{m_0}\,\nabla\ln \left[\psi/\sqrt{\rho_0}\right] \,.
\end{eqnarray}
After interchanging the operators $\partial/\partial\,t$ and $\nabla$ one obtains
\begin{eqnarray*}
\nabla\left(\frac{\hbar}{m_0}\frac{1}{\psi}\,\frac{\partial\,\psi}{\partial\,t}\right)=\qquad \qquad \qquad \\
\nabla \left[\frac{i}{2}\,\frac{\hbar^2}{m_{0}^2}\left\{\left(\frac{1}{\psi}\,\nabla 
\psi\right)^2+\nabla\cdot\left(\frac{1}{\psi}\,\nabla \psi\right)\right\}-\frac{i}{m_0}\,V \right]\,.
\end{eqnarray*}
If the gradient of some function equals that of another function the two functions can only differ by a real-space independent function of time which we denote by $\beta(t)$. Hence, if one divides the above equation by the imaginary unit the result may be cast as 
\begin{eqnarray}
\label{eqn:Schroedingerglg0}
-i\,\frac{\hbar}{m_0}\frac{1}{\psi}\,\frac{\partial\,\psi}{\partial\,t}=\frac{1}{2}\,
\frac{\hbar^2}{m_{0}^2}\left[\left(\frac{1}{\psi}\,\nabla \psi\right)^2+\nabla\cdot\left(\frac{1}{\psi}\,\nabla \psi\right)\right]\nonumber \\
-\frac{1}{m_0}\,V-i\,\beta(t)\,.
\end{eqnarray} 
One can now make use of the identity
\begin{eqnarray*}
\nabla \cdot\left(\frac{1}{\psi}\,\nabla\psi\right)=-\left(\frac{1}{\psi}\,\nabla \psi \right)^2+\frac{1}{\psi}\,\nabla^{2}\psi
\end{eqnarray*}
and multiply Eq.(\ref{eqn:Schroedingerglg0}) by $-m_{0}\,\psi$. This yields
\begin{eqnarray}
\label{eqn:Schroedingerglg1}
i\hbar\,\frac{\partial\,\psi}{\partial\,t}=-\frac{\hbar^{2}\nabla^2}{2\,m_0}\,\psi
+V\,\psi+\gamma(t)\,\psi
\end{eqnarray}
where
\begin{eqnarray*}
\gamma(t)=i\,m_{0}\,\beta(t)\,.
\end{eqnarray*} 
If $\psi({\un r},t)$ is replaced by $\widehat{\psi}({\un r},t)$ defined through 
\begin{eqnarray*}
\label{eqn:DefinitionDachpsi}
\psi({\un r},t)=\widehat{\psi}({\un r},t)\,\exp\,\left[-\frac{i}{\hbar}\int_{t_0}^{t}\gamma(t')\,dt'\right]\,,
\end{eqnarray*} 
Eq.(\ref{eqn:Schroedingerglg1}) becomes an equation for $\widehat{\psi}({\un r},t)$:
\begin{eqnarray}
\label{eqn:Schroedingerglg2}
i\hbar\,\frac{\partial\,\widehat{\psi}({\un r},t)}{\partial\,t}=\left[\frac{\widehat{{\un p}}^2}{2\,m_0}+V({\un r})\right]\,\widehat{\psi}({\un r},t)\,,
\end{eqnarray}
where
\begin{eqnarray}
\label{eqn:Momentum_operator}
\widehat{{\un p}}\equiv -i\hbar \nabla\,.
\end{eqnarray}
The two functions $\psi({\un r},t)$ and $\widehat{\psi}({\un r},t)$ differ only by a time-dependent phase factor without physical relevance. Only the functions
\begin{eqnarray}
\label{eqn:Definition_rho}
\quad \qquad  \rho({\un r},t)=\psi^{*}({\un r},t)\,\psi({\un r},t) \qquad \mbox{(density)}
\end{eqnarray}
and the current density:
\begin{eqnarray}
\label{eqn:Definition_j}\quad {\un j}({\un r},t)=\rho({\un r},t)\,\frac{\hbar}{m_0}\,\nabla\,\varphi({\un r},t)\,,
\end{eqnarray}
refer to relevant quantities of the system which obviously do not depend on this phase factor.
For this reason we may set $\gamma(t)\equiv 0$, that is replace $\widehat{\psi}({\un r},t)$ in Eq.(\ref{eqn:Schroedingerglg2}) by $\psi({\un r},t)$
without loss of generality. To simplify the notation we introduce the so-called Hamiltonian defined by
\begin{eqnarray}
\label{eqn:Hamilton_Operator}
\widehat{H}\equiv\frac{\widehat{{\un p}}^2}{2\,m_0}+V({\un r})\,.
\end{eqnarray}
Eq.(\ref{eqn:Schroedingerglg2}) then takes the familiar form of the Schr\"odinger equation
\begin{eqnarray}
\label{eqn:Schroedingerglg3}
i\hbar\,\frac{\partial\,\psi({\un r},t)}{\partial\,t}=\widehat{H}({\un r})\,\psi({\un r},t)\;.
\end{eqnarray}
The first order time derivative on the left-hand side can be tracked down to the acceleration 
$(\partial/\partial t)\,{\un v}$ in Newton's modified second law (\ref{eqn:local_acceleration0}). \\
Using
$$
\psi({\un r},t)=|\psi({\un r},t)|\,e^{i\,\varphi({\un r},t)}
$$
and inserting this into Eqs.(\ref{eqn:Definition_rho}) and (\ref{eqn:Definition_j}) one obtains the familiar expression
\begin{eqnarray}
\label{eqn:Definition2_j}
{\un j}({\un r},t)=\rho({\un r},t)\,{\un v}({\un r},t)=\qquad \quad \nonumber \\
\frac{\hbar}{2i\,m_0}\,\left[\psi^*({\un r},t)\nabla\,\psi({\un r},t)-\psi({\un r},t)\nabla\,\psi^*
({\un r},t)\right]
\end{eqnarray}
which on real-space integration and multiplication by $m_0$ yields
\begin{eqnarray}
\label{eqn:Definition_p_expect}
m_0\,\langle {\un v}(t)\rangle=\int\psi^*({\un r},t)\,\widehat{{\un p}}\;\psi({\un r},t)\,d^3r\equiv \langle \widehat{{\un p}} \rangle
\end{eqnarray}
where $\psi({\un r},t)$ has been required to satisfy the usual boundary conditions at the surface of the normalization volume. Because of Eq.(\ref{eqn:Definition_p_expect}) one is justified in terming $\widehat{{\un p}}\,$ ``momentum operator''.\\
In Bohm's version of quantum mechanics \cite{Bohm1} Eq.(\ref{eqn:Definition2_j}) is recast to define the velocity field
$$
{\un v}({\un r},t)=\frac{\hbar}{m_0}\,\Im\left(\frac{\nabla \psi({\un r},t)}{\psi({\un r},t)}\right)\,.
$$
The streamlines of this field are interpreted as true particle trajectories. From our point of view this appears to be rather absurd because the explicit ${\un r}$-dependence of ${\un v}$ comes about by forming the ensemble average over the (in principle infinite) family of true trajectories as defined in Eq.(\ref{eqn:average_v}). Bohm's definition of ${\un v}$ as describing the true velocity of the particle leads inescapably to strange results, notably with stationary real-valued wave functions $\psi({\un r})$ for which ${\un v}({\un r})\equiv 0$. Hence, the particle appears to be at rest although the kinetic energy of the particle
\begin{eqnarray}
\label{eqn:kinetic_energy_1Definition2_j}
\langle \widehat{T}\rangle=\int \psi^{*}({\un r})\,\frac{{\widehat{\un p}}^2}{2\,m_0}\,\psi({\un r})\,d^3r\equiv \frac{\langle \widehat{{\un p}}^2\rangle}{2\,m_0}
\end{eqnarray}
is definitely different from zero.\\
The time-dependent Schr\"odinger equation represents the center of non-relativi\-stic quantum mechanics. Fundamentally different from the present approach where it is derived from a new vacuum concept, in conventional quantum mechanics it falls out of the blue, and this applies to  Bohm's theory as well. As the latter associates the pattern of smooth stream lines with the set of true particle trajectories, it is forced to explain the probabilistic character of the information contained in $\psi({\un r},t)$ by an additional ``quantum equilibrium''- hypothesis. It is therefore hard to see that anything can be gained by ``going Bohmian''. The ``process of measurement'' in which a particle moves from a source to the detector where it fires a counter, is in our view described by one of the irregular trajectories which is terminated at the detector. Due to the stochastic forces that cause this irregularity, the information on the ensemble properties is naturally probabilistic.\\ 
A frequently raised objection against Bohm's theory concerns the asymmetric way in which it deals with the particle's real-space position and its momentum. In fact, the real-space position ${\un r}$ plays a pivotal role in Bohm's theory compared to the other observables which are ``contextualized'' by resorting to the wave function $\psi({\un r},t)$ that solves the Schr\"odinger equation for the system under study. By contrast, in our approach  the ensemble's $i$-th particle position ${\un r}_i$ and its velocity ${\un v}_i(t)$ enter into the theory as autonomous quantities. This is reflected in the occurrence of two independent functions $\rho({\un r},t)$ and  ${\un v}
({\un r},t)=\frac{\hbar}{2\,m_0}\,\nabla \varphi({\un r}\,t)$. It is this pair of information $\rho({\un r},t)\,;\varphi({\un r},t)$ that necessitates the description of the one-particle system by a {\bf complex}-valued function
\begin{eqnarray*}
\psi({\un r},t)=\pm \sqrt{\rho({\un r},t)}\,e^{i\,\varphi({\un r},t)}\,.
\end{eqnarray*}

\section{The uncertainty relation and the issue of ``measurement''}\label{kapitel7}

By performing a Fourier transform on $\psi({\un r},t)$ 
\begin{eqnarray}
\label{eqn:FourierTransform}
\psi({\un r},t)=\frac{1}{(2\,\pi)^{3/2}}\int C({\un k},t)\,e^{i\,{\un k}\cdot{\un r}}\,d^3k
\end{eqnarray}
Eqs.(\ref{eqn:Definition_p_expect}) and (\ref{eqn:kinetic_energy_1Definition2_j})
may alternatively be written
\begin{eqnarray}
\label{eqn:FT_KineticEnergy}
\langle \widehat{{\un p}}\rangle=\int\psi^*({\un r},t)\,\widehat{{\un p}}\;\psi({\un r},t)\,d^3r=\nonumber\\
\int C^{*}({\un k},t)\,
(\hbar\,{\un k})\,C({\un k},t)\,d^3k  \nonumber \\ 
\langle \widehat{{\un p}}^2\rangle=\int\psi^{*}({\un r},t)\,\widehat{{\un p}}^2\,\psi({\un r},t)\,d^3r= \nonumber\\
\int C^{*}({\un k},t)\,
(\hbar\,{\un k})^2\,C({\un k},t)\,d^3k
\end{eqnarray}
where
\begin{eqnarray}
\label{eqn:probability_k}
C^{*}({\un k},t)\,C({\un k},t)\,\Delta^3 k=P({\un k},t)\,\Delta^3 k
\end{eqnarray}
describes the probability of the particle possessing a momentum that lies within $\Delta^3 k$ around ${\un k}$ in the 
${\un k}$-space. We temporarily label the coordinate-components of the particle in the two spaces by an index $\nu\,;\;\nu=1,2,3$. The mean square departures of the position coordinates $x_{\nu}$ and $k_{\nu}$, respectively, from their arithmetic means $\bar{x}_{\nu}$ and $\bar{k}_{\nu}$ are given by
$$
\langle (x_{\nu}-\bar{x}_{\nu})^2\rangle_t=\int \psi^{*}({\un r},t)\,(x_{\nu}-\bar{x}_{\nu})^2\,\psi({\un r},t)\,d^3r
$$
and
$$
\langle (k_{\nu}-\bar{k}_{\nu})^2\rangle_t=\int C^{*}({\un k},t)\,(k_{\nu}-\bar{k}_{\nu})^2\,C({\un k},t)\,d^3k\,.
$$
Since $C({\un k},t)$ is the Fourier transform of $\psi({\un r},t)$ we have as a fundamental mathematical theorem 
$$
\langle (x_{\nu}-\bar{x}_{\nu})^2\rangle_t\,\langle(k_{\nu}-\bar{k}_{\nu})^2\rangle_t \ge \frac{1}{4}
$$
that is
\begin{eqnarray}
\label{eqn:FourierUncertainty}
\langle (x_{\nu}-\bar{x}_{\nu})^2\rangle_t\,\langle(\hbar\,k_{\nu}-\hbar\,\bar{k}_{\nu})^2\rangle_t \ge \frac{\hbar^2}{4}\,.
\end{eqnarray}
Following the standard notation by setting $\Delta x_{\nu}=\sqrt{\langle (x_{\nu}-\bar{x}_{\nu})^2\rangle_t}$ and $\Delta p_{\nu}=\sqrt{\langle(\hbar\,k_{\nu}-\hbar\,\bar{k}_{\nu})^2\rangle_t}=\sqrt{\langle(\widehat{{\un p}}-\langle\widehat{{\un p}}\rangle)^2\rangle_t}$ the latter relation may be cast as
\begin{eqnarray}
\label{eqn:UncertaintyRelation}
\Delta x_{\nu}\,\Delta p_{\nu}\ge \frac{\hbar}{2}
\end{eqnarray}
which constitutes the celebrated uncertainty relation. It is commonplace to interpret this relation, loosely speaking, by saying: ``momentum and position of a particle cannot be measured simultaneously with any desirable precision''.\\
From our point of view it does in no ways refer to any measurement on the position or momentum of the particle in question. It is nothing more than the theorem Eq.(\ref{eqn:FourierUncertainty}) on the product of two quantities that are interconnected by a Fourier transform. Furthermore, 
since this relation is - besides the Schr\"odinger equation - just another consequence of our concept, it cannot possibly conflict with the existence of trajectories which constitute a fundamental element of that concept. \\
Eq.(\ref{eqn:UncertaintyRelation}) is considered ground-laying for the Copenhagen interpretation of quantum mechanics. The latter is based on the conviction that it is the measurement that causes the indeterminacy in quantum mechanics and necessitates a probabilistic description of microscopic mechanical systems. In a highly respected article \cite{Heisenberg3} Heisenberg gives a revealing example of such a measurement. To pinpoint an electron moving along the $x$-axis within an experimental setup he considers a $\gamma$-ray source, that illuminates the electron beam, and a hypothetical $\gamma$-ray microscope that possesses a sufficiently high resolution in detecting the position of that electron up to an error of $\Delta x$. He demonstrates that the $\gamma$-ray photon that ``hits the electron'' and is subsequently scattered into the microscope, transfers a momentum $\Delta p_x$ to the electron so that
\begin{eqnarray}
\label{eqn:HeisenbergUncertainty}
\Delta x\,\Delta p_x \approx\hbar\,.
\end{eqnarray}
The above result reflects only a property of the microscope  
\begin{eqnarray*}
\Delta x\,\Delta k_x\approx 2\,\pi
\end{eqnarray*}
which interrelates the resolved linear dimensions $\Delta x=\lambda/\sin \alpha$ of an object and the admissible maximum angle $\alpha$ required to ensure that the scattered wave (of wavelength $\lambda$) is still captured by the front lens of the microscope, and $\Delta k_x= k\,\sin \alpha$ which describes the $k_x$-change of the wave vector of the scattered wave. But this interrelation expresses only the content of Eq.(\ref{eqn:FourierUncertainty}) in a different form. The measurement, however, is completely fictional for two reasons. Firstly, imaging systems within that regime of wavelength are for fundamental reasons unfeasible. Secondly, different from the picture insinuated by Heisenberg's phrasing, the interaction does not take place as an instantaneous collision process where a point-like particle (the photon) hits another point-like particle, the electron. Instead the transition probability of the electron for attaining a different momentum is given by the mod squared of the transition matrix element $M_{opt}$, a real-space integral that extends over a range of many light wave lengths in diameter. Moreover, the transition is not instantaneous but rather takes some time of the order $\hbar/|M_{opt}|$. Within this transition time the electron covers a distance $\Delta x'$ which has nothing to do with $\Delta x$ in Eq.(\ref{eqn:HeisenbergUncertainty}). Other examples of ``measurement'', e.$\,$g. diffraction at slits of a certain width $\Delta x$ show even more directly that the probabilistic information on the (non-relativistic) motion of a particle is exhaustively described by the Schr\"odinger equation and boundary conditions for $\psi({\un r})$, and hence this information merely reflects our vacuum concept, irrespective of whether or not results on the diffraction are verified by measurements.\\
The host of considerations invoking the uncertainty relation (\ref{eqn:UncertaintyRelation}) refer to situations where a particle is located within an interval $\Delta x$ and one interprets this confinement of the particle indiscriminately in terms of a ``measurement'' of its coordinate $x$ with limited accuracy. One concludes then from the uncertainty relation that $\Delta x$ correlates unavoidably with a variance $\overline{\Delta p_x^2}$ of its momentum such that $\Delta x\,\Delta p_x \stackrel{>}{\approx}\frac{\hbar}{2}$ where $\Delta p_x \stackrel{\mbox{{\tiny def}}}{=} \sqrt{\overline{\Delta p_x^2}} $. In reality neither a measurement on $\Delta x$ nor on $\overline{\Delta p^2}$ is truly executable. The uncertainty relation merely states that a solution of the one-dimensional Schr\"odinger equation for a particle in a box of length $\Delta x$ yields a ground state energy $\Delta E=\frac{\Delta p^2}{2\,m_0}$ where $\Delta p^2=\left(\hbar\,\frac{\pi}{\Delta x}\right)^2$. Hence one obtains simply as a consequence of solving the Schr\"odinger equation for that case ``without observer''(!) $\Delta x\, \Delta p=\pi\,\hbar$. One cannot help but quote John Bell's question phrased in his stirring article ``Against Measurement''\cite{Bell1}
\\[0.2cm]
{\it ``What exactly qualifies some physical systems to play the role of 'measurer'?''}
\\[0.2cm]
The above considerations are in line with a discussion of Heisenberg's paper by Wigner \cite{Wigner0}. 
\\[0.2cm]
\section{Averaging over the total ensemble}\label{kapitel8}

In forming the arithmetic mean of the two equations (\ref{eqn:Stoch_equation1}) and (\ref{eqn:Stoch_equation2}) we omitted to mention a problem that we wish to discuss here in more detail. \\
We temporarily decompose the entire ensemble considered so far into a ``Brownian'' and ``anti-Brownian'' sub-ensemble, each characterized by the associated stochastic forces and comprising an equally large number of members. Accordingly we distinguish the velocities ${\un v}({\un r},t)$ and the densities $\rho({\un r},t)$ in the respective sub-ensembles by subscripts $B$ (for ``Brownian'') and $A$ (for ``Anti-Brownian''). If the velocities in these two equations agree at a time $t$, they are definitely different at a later time $t+\Delta t$. Yet forming the arithmetic mean of the two equations can only lead to the same average - which we could recast as ``Newton's modified second law'', Eq.(\ref{eqn:Bohm_equation2}) - if the two velocities ${\un v}_B, {\un v}_A$ and the densities $\rho_B, \rho_A$ agree also at $t+\Delta t$ and any later time. At first sight the latter appears to be irreconcilable with the former. One has to recall, however, that our subdivision of the entire ensemble into sub-ensembles $B$ and $A$ represents only a simplifying model for the actually occurring reversible scatterings. In the real system the stochastic forces of the $B$-type become automatically forces of the $A$-type and vice versa within the characteristic time $\tau$ so that the change of the velocity $\Delta {\un v}$ in either sub-ensemble is $[\Delta {\un v}_A+\Delta {\un v}_B]/2$ within a time span $\Delta t\gg \tau$ which, however, must be small compared to time intervals within which the quantities of interest change sizeably. The situation is similar to that encountered in diffusion theory where we have
$$
\frac{\partial \rho}{\partial t}=\nu\,\Delta \rho\,.
$$ 
This equation is obtained from the equation of continuity for ${\un v}_B={\un u}$ and ${\un u}=-\nu\,\nabla \rho/\rho$ with the latter equation based on similar considerations as the derivation of Eq.(\ref{eqn:Navier_Stokes}) invoking Einstein's law (\ref{eqn:EinsteinLaw}) which implies $\Delta t\gg \tau$. The above equation of diffusion hence describes changes that are actually defined only on a coarse grain time scale and its validity is confined to changes that are sufficiently slow on that time scale. As we have already discussed in Section \ref{kapitel3}, this is also the assumption underlying our derivation of Newton's modified second law (\ref{eqn:Bohm_equation2}). \\
We temporarily rewrite the two equations (\ref{eqn:Stoch_equation1}) and (\ref{eqn:Stoch_equation2}) for an - in that sense -  ``appropriately long, but sufficiently short time interval'' $\Delta t$ in the form
\begin{eqnarray*}
\label{eqn:DifferenzenglgB_A} \Delta{\un v}_{B/A}({\un r},t+\Delta t)={\un R}_{B/A}({\un r},t)\,\Delta t
\end{eqnarray*}
where
\begin{eqnarray}
\label{eqn:DefinitionR_BA} 
{\un R}_{B/A}({\un r},t)=\frac{1}{m_0}\left(-\nabla[V({\un r})+V_{QP}({\un r},t)]\pm \vec{\Omega}({\un r},t)\right)\,.
\end{eqnarray}
and
\begin{eqnarray*}
\label{eqn:Omega} \vec{\Omega}=\frac{\partial{\un u}}{\partial t}+({\un
v}\cdot\nabla)\,{\un u}-({\un u}\cdot\nabla)\, {\un v}+\nu\,\Delta\,{\un v}\,.
\end{eqnarray*}
Here we have already used $V_{QP}$ instead of $V_{stoch}$, but still denoted the prefactor of $\Delta {\un v}$ by $\nu$ to demonstrate that $\vec{\Omega}$ (and consequently ${\un u}$) changes sign when $\nu$ changes sign. It should be noticed that according to Eq.(\ref{eqn:Quantum_potential}) $V_{QP}$ has the property \\ $V_{QP}(\rho_B({\un r},t))=V_{QP}(\rho_A({\un r},t))=V_{QP}(\rho({\un r},t))$ since at the time $t$ under consideration we have $\rho_A({\un r},t)=\rho_B({\un r},t)=\frac{1}{2}\,\rho({\un r},t)$.\\
After the elapse of a time $\Delta t$ within which each of the $N/2$ particles in the two  sub-systems has changed its affiliation (B from A or vice versa) we have
\begin{eqnarray*}
\label{eqn:Geschwin_nach_Delta_t1} {\un v}_{B/A}({\un r},t+\Delta t)={\un
v}_{B/A}({\un r},t)+\Delta{\un v}({\un r},t+\Delta t)\,.
\end{eqnarray*}
where
\begin{eqnarray}
\label{eqn:DefinitionDelta_v}
\Delta{\un v}({\un r},t+\Delta t)=\overline{\Delta{\un R}_{B/A}}({\un r},t+\Delta t)=\nonumber \\
\frac{1}{m_0}\left(-\nabla[V({\un
r})+V_{PQ}({\un r},t)]\right)\,\Delta t
\end{eqnarray}
with $\overline{\Delta{\un R}_{B/A}}$ denoting a time average.
\\[0.2cm]
$\gg$ {\large We consider Eqs.(\ref{eqn:DefinitionR_BA}) and (\ref{eqn:DefinitionDelta_v}) as implicitly defining ``motion under reversible scattering''}$\ll$.
\\[0.2cm]
If we now form the average over the entire ensemble we get
\begin{eqnarray*}
\label{eqn:Geschw_nach_Delta_t2}
{\un v}({\un r},t+\Delta t)=\frac{1}{2}{\un v}_B({\un r},t+\Delta t)+\frac{1}{2}{\un v}_A({\un r},t+
\Delta t)=\qquad \qquad \nonumber\\
{\un v}({\un r},t)+\frac{1}{m_0}\left(-\nabla[V({\un r})+V_{PQ}({\un r},t)]\right)\,\Delta t\,.\qquad
\end{eqnarray*}
Thus we have
$$
{\un v}_{B}({\un r},t+\Delta
t)={\un v}_{A}({\un r},t+\Delta t)={\un v}({\un r,}t+\Delta t)\,.
$$
We want to demonstrate that the densities behave analogously. For this reason we resort to the equation of continuity (\ref{eqn:equation_continuity}) which holds for each sub-ensemble
\begin{eqnarray}
\label{eqn:Equation_continuity_A/B}
\frac{\partial \rho_{B/A}}{\partial t}+\nabla \cdot (\rho_{B/A}\,{\un v}_{A/B})=0\,.
\end{eqnarray} 
It describes the conservation of the number of particles in each of the two subsystems. We conclude from this equation that $\dot{\rho}_{B}({\un r},t)=\dot{\rho}_{A}({\un r},t)$, if $\rho_{B}({\un r},t)=\rho_{A}({\un r},t)$ and ${\un v}_{B}({\un r},t)={\un v}_{A}({\un r},t)$.
If one differentiates Eq.(\ref{eqn:Equation_continuity_A/B}) with respect to time and uses $\dot{{\un v}}_{B}({\un r},t)=\dot{{\un v}}_{A}({\un r},t)=\dot{{\un v}}({\un r},t)$ as a result of the preceding considerations, we may conclude 
$\ddot{\rho}_{B}({\un r},t)=\ddot{\rho}_{A}({\un r},t)$. One can carry this conclusion further to any order of time derivative. Thus, the Taylor-expansions of $\rho_{B}({\un r},t+\Delta t)$ and $\rho_{A}({\un r},t+\Delta t)$ agree for any length of the time interval 
$\Delta t$ if $\rho_{B}, \rho_{A}$ and ${\un v}_{B}, {\un v}_{A}$ agree at time $t$. 
\\[0.4cm]
\section{Conservative diffusion}\label{kapitel9}
We want to prove the validity of Eq.(\ref{eqn:Expect_value_vacuum_force}) which constitutes a necessary condition for the preservation of classical motional behavior on the average. To see this more clearly, we first consider one particle (the $i$-th) in the cube $\Delta^3r$ around ${\un r}$ acted upon by the external force ${\un F}({\un r})$ and the stochastic force ${\un F}_{s\,i}(t)$. According to Newton's second law we have
$$
\frac{d}{d t}\,m_0\,{\un v}_i(t)={\un F}({\un r}_i)+{\un F}_{s\,i}(t)\,.
$$
If we sum this equation over the $n({\un r},t)$ particles contained in $\Delta^3r$, divide by $N$ and form ensemble averages similar to Eqs.(\ref{eqn:def_density}) and (\ref{eqn:average_v}) we obtain
\begin{eqnarray}
\label{eqn:AverageNewtonEquation}
\frac{\partial}{\partial t}\,m_0\,\frac{1}{N}\,\underbrace{\sum_{i=1}^{n({\un r},t)}{\un v}_i(t)}_{=n({\un r},t)\,{\un v}({\un r},t)}=\frac{1}{N}\,\underbrace{\sum_{i=1}^{n({\un r},t)}{\un F}({\un r}_i)}_{=n({\un r},t)\,{\un F}({\un r})}\nonumber\\
+\frac{1}{N}\,\underbrace{\sum_{i=1}^{n({\un r},t)}{\un F}_{s\,i}(t)}_{=n({\un r},t)\,{\un F}_s({\un r},t)}\,.
\end{eqnarray}
Here the summation runs over all particles in the cell irrespective of whether they belong to the first or second sub-ensemble.\\
The idea of ``conservative diffusion'' implies that the $N=\sum_{{\un r}}^{N_{{\un r}}}\,n({\un r},t)$ particles of the entire ensemble do not feel a stochastic force on the average although ${\un F}_s({\un r},t)$ does locally not vanish in general. Thus, ${\un F}_s({\un r},t)$ is required to have the property
\begin{eqnarray}
\label{eqn:StochasticForceZero}
\sum_{{\un r}}^{N_{\un r}}\frac{n({\un r},t)}{N}\,{\un F}_s({\un r},t)=\int_{\cal{V}}\rho({\un r},t)\,\underbrace{{\un F}_{QP}({\un r},t)}_{\equiv{\un F}_s({\un r},t)}\,d^3r=0\; \forall\, t\,,
\end{eqnarray}
as a result of which Eq.(\ref{eqn:AverageNewtonEquation}) yields after summation over all elementary cells
\begin{eqnarray*}
\frac{\partial}{\partial t}\,\sum_{{\un r}}^{N_{{\un r}}} m_0\,\frac{n({\un r},t)}{N}\,{\un v}({\un r},t)=\frac{d}
{d t}\,\underbrace{\int_{{\cal{V}}} \rho({\un r},t)\,m_0\,{\un v}({\un r},t)\,d^3r}_{\equiv\langle {\un p}(t)\rangle}=\\
\underbrace{\sum_{{\un r}}^{N_{{\un r}}} \frac{n({\un r},t)}{N}\,{\un F}({\un r})}_{=\int \rho({\un r},t)\,{\un F}({\un r})\,d^3r=\langle {\un F}\rangle}\,.
\end{eqnarray*}
We thus obtain as a consequence of the required property of ${\un F}_s({\un r},t)$
\begin{eqnarray}
\label{eqn:Ehrenfest_1}
\frac{d}{d t}\,\langle {\un p}(t)\rangle=\langle {\un F}\rangle
\end{eqnarray}
which is Ehrenfest's first theorem.\\
In case of a force-free particle for which $\langle {\un F}\rangle=0$, Eq.(\ref{eqn:Ehrenfest_1}) yields
$$
\langle {\un p}(t)\rangle=const.
$$
which demonstrates that a free particle exposed to Brownian/anti-Brownian stochastic forces does not change its momentum on the average, as opposed to a particle that moves in a classical ``Brownian'' environment.
\\[0.2cm]
We now want to show that the expectation value of ``Newton's modified second law'' that we have derived in the form of  Eq.(\ref{eqn:Bohm_equation2}), attains, in fact, exactly the form of Eq.(\ref{eqn:Ehrenfest_1}). To this end it is convenient to recast  Eq.(\ref{eqn:Quantum_potential}) as
\begin{eqnarray*}
V_{QP}=\frac{\hbar^2}{4\,m_0}\, \left[\frac{1}{2}
\left(\frac{\nabla \rho }{\rho}\right)^2-\frac{\nabla^2\rho }{\rho}\right]=\\
m_0\,\left[-\frac{{\un u}^2({\un r},t)}{2}+\frac{\hbar}{2m_0}\,\nabla\cdot {\un u}({\un r},t)\right]
\end{eqnarray*}
where we have used Eq.(\ref{eqn:osmotic_velocity}) defining ${\un u}({\un r},t)$. 
Hence 
\begin{eqnarray}
\label{eqn:Proof_vacuum_force_zero}
\int\rho({\un r},t)\,{\un F}_{QM}({\un r},t)\,d^3r= 
\end{eqnarray}
$$
 =m_0\,\int\left[\frac{1}{2}\rho({\un r},t)\nabla{\un u}^2
({\un r},t)-\frac{\hbar}{2m_0}\,\rho({\un r},t)\Delta {\un u}({\un r},t)\right]\,d^3r\,. 
$$
We rewrite the integral over the second term on the right-hand side using Gauss' theorem
$$
\int_{\cal{V}}\rho\,\underbrace{\nabla \cdot(\nabla {\un u})}_{=\Delta {\un u}}\,d^3r=\underbrace{\int_{\cal{V}}\nabla\cdot(\rho\,\nabla{\un u})\,d^3r}_{=\int_{\cal{F}}\rho\,\nabla {\un u}
\cdot d^{2}{\un r}}-\int_{\cal{V}}\nabla\rho\cdot \nabla{\un u}\,d^3r\,.
$$
We assume that $\rho({\un r},t)$ differs sizeably from zero only within a volume that lies completely within the finite space and drops sufficiently fast to zero toward infinity so that the surface integral vanishes. Using again Eq.(\ref{eqn:osmotic_velocity}) we hence arrive at
$$
-\int_{\cal{V}}\nabla \rho \cdot \nabla {\un u}\,d^3r=\frac{2m_0}{\hbar}\,\int_{\cal{V}}\rho \underbrace{({\un u}\cdot \nabla){\un u}}_{=\frac{1}{2}\,\nabla{\un u}^2} d^3r
$$
which shows that, in fact, the right-hand side of Eq.(\ref{eqn:Proof_vacuum_force_zero}) equals zero. Thus the expectation value of the right-hand side of ``Newton's modified second law'', Eq.(\ref{eqn:Bohm_equation2}), becomes equal to $\langle {\un F} \rangle$. However, we have on the left-hand side $\langle \frac{d}{dt}\,m_0\,{\un v}\rangle$ instead of $\frac{d}{d t}\,\langle m_0\,{\un v}\rangle$. Nevertheless, the two expressions are equal as follows from multiplying $\frac{d}{dt}\,m_0\,{\un v}$ by $\rho({\un r},t)$ and observing that ${\un v}({\un r},t)$ is curl-free. Because of the latter we have
$$
\frac{d}{dt}\,{\un v}=\frac{\partial}{\partial t}{\un v}+\frac{1}{2}\nabla {\un v}^2
$$
which can be recast as
$$
m_0\,\rho\,\frac{d}{dt}\,{\un v}=m_0\,\rho\,\frac{\partial}{\partial t}{\un v}+\left[m_0\,{\un v}\,\frac{\partial\rho}{\partial t}-m_0\,{\un v}\,\frac{\partial\rho}{\partial t}\right]+\frac{m_0}{2}\,\rho\,\nabla {\un v}^2
$$ 
where we have added zero in the form of the bracketed expression. The real-space integral over this equation may be written after reordering 
\begin{eqnarray}
\label{eqn:ExpectationAcceleration}
\underbrace{\frac{\partial}{\partial t}\,\int_{\cal{V}}\rho({\un r},t)\,m_0\,{\un v}({\un r},t)\,d^3r}_{=\frac{d}
{d t}\,\int_{\cal{V}}\rho({\un r},t)\,m_0\,{\un v}({\un r},t)\,d^3r}=\langle \frac{d}{dt}\,m_0\,{\un v}\rangle\nonumber\\
 +\frac{m_0}{2}\,\int_{\cal{V}}\,\left[2{\un v}\,\frac{\partial\rho}{\partial t}-\rho\,\nabla{\un v}^2\right]\,d^3r\,.
\end{eqnarray} 
The integral on the right-hand side vanishes because of the equation of continuity
\begin{eqnarray}
\label{eqn:EquationContinuity}
\frac{\partial\rho}{\partial t}+\nabla\cdot(\rho\,{\un v})=0\,.
\end{eqnarray}
This follows from multiplying this equation by ${\un v}$ and performing a real-space integration. We then have
\begin{eqnarray}
\label{eqn:Nullstrom} \int_{\cal{V}}{\un v}\,\frac{\partial \rho}{\partial
t}\,d^3\,r=-\sum_{\nu=1}^{3} {\un e}_{\nu}\int_{\cal{V}}v_{\nu}\,\nabla\cdot
(\rho\,{\un v})\,d^3\,r
\end{eqnarray}
$$
=-\sum_{\nu=1}^{3}{\un e}_{\nu}\underbrace{\int_{\cal{V}}\nabla\cdot
(v_{\nu}\,\rho\, {\un v})\,d^3\,r}_{=\int_{\cal{A}}v_{\nu}\,\rho\,{\un v}\cdot
d^2\,{\un r}}+\underbrace{\sum_{\nu=1}^{3}{\un e}_{\nu}\int_{\cal{V}}\rho
\,{\un v}\cdot \nabla v_{\nu}\,d^3\,r}_{=\int_{\cal{V}}\rho\,({\un v}\cdot
\nabla)\,{\un v}\,d^3\,r} 
$$
with ${\un e}_{\nu}$ denoting unit vectors.
The surface integral as been obtained by invoking Gauss' theorem. It vanishes since we may assume $\rho\,|{\un v}|$ to vanish sufficiently toward infinity. Again exploiting the property of ${\un v}$ being curl-free the second integral on the right-hand side can be written
$$
\int_{\cal{V}}\rho\,({\un v}\cdot \nabla)\,{\un
v}\,d^3\,r=\frac{1}{2}\,\int_{\cal{V}}\rho\,\nabla{\un v}^2\,d^3\,r\,.
$$ 
It follows then from Eq.(\ref{eqn:Nullstrom}) that the integral on the right-hand side of Eq.(\ref{eqn:ExpectationAcceleration}) is, in fact, equal to zero. Thus we have shown that the expectation value of the ``vacuum force'' ${\un F}_{QM}({\un r},t)$ vanishes
$$
\int\rho({\un r},t)\,{\un F}_{QM}({\un r},t)\,d^3r=0
$$
which plays also a central role in information theory (s.~e.~g.~Garbaczewski \cite{Garbaczewski}).
\\[0.4cm]

\section{The time-dependent Schr\"odinger equation in the presence of an electromagnetic field}\label{kapitel10}

In going through the various steps that led from Eq.(\ref{eqn:Bohm_equation2}) (``Newton's modified second law'') to the time-dependent Schr\"odinger equation (\ref{eqn:Schroedingerglg3}) one recognizes that we implied nowhere that ${\un F}$ has to be time-independent. Hence one is justified in allowing ${\un F}$ in Eq.(\ref{eqn:Bohm_equation2}) to be time-dependent and attain the particular form
\begin{eqnarray}
\label{eqn:Electromagnetic_force}
{\un F}({\un r},t)=-\nabla V_{cons}({\un r})+e\,\widehat{{\un E}}({\un r},t)\qquad\qquad\nonumber\\
+e\,{\un v}({\un r},t)\times{\un B}
({\un r},t)
\end{eqnarray}
if the particle under study possesses the charge $e$ and is acted upon by an electric field $\widehat{{\un E}}({\un r},t)$ and a magnetic field ${\un B}({\un r},t)$. The quantity $V_{cons}({\un r})$  denotes the potential of an additional conservative field (e.$\,$g. the gravitational field) which we include to ensure full generality, and ${\un v}({\un r},t)$ is the ensemble average defined by Eq.(\ref{eqn:average_v}). From ${\un B}=\nabla \times {\un A}$ and Faraday's law of induction we have $\nabla \times (\widehat{{\un E}}+\dot{{\un A}})=0$, and hence $\widehat{{\un E}}+\dot{{\un A}}$ may be expressed as a gradient of a scalar function which we denote by  $-\frac{1}{e}\,V_{el}({\un r},t)$. Thus
\begin{eqnarray}
\label{eqn:E_Feld_Potential} e\,\widehat{{\un E}}({\un r},t)=-e\,\dot{\un A}({\un r},t)-\nabla V_{el}({\un r},t)\,.
\end{eqnarray}
If the magnetic field is switched on, it induces a voltage $V_R$ along any circular path $C$
$$
V_R=\oint_{C}\widehat{{\un E}}_{ind.}({\un r}',t)\cdot d{\un
r}'=-\frac{\partial}{\partial\,t}\int_{\cal{A}}{\un B}({\un r}',t)\cdot d^{2}{\un r}'
$$ 
where $C$ is the rim of the surface $\cal{A}$. On multiplying this equation by $e$ and observing that $e\,\widehat{{\un E}}_{ind.}$ represents an additional force that changes the momentum of the particle, we obtain
\begin{eqnarray*}
\oint_{C}\dot{{\un p}}({\un r}',t')\cdot d{\un r}'=\oint_{C}e\,\widehat{{\un
E}}_{ind.}({\un r}',t')\cdot d{\un r}'=\qquad \qquad\\
 -\frac{\partial}{\partial\,t'}\,\oint_{C}e\,{\un A}({\un r}',t')\cdot d{\un
r}'\,.
\end{eqnarray*}
Integrating this equation from $t_0$ to $t$ and assuming ${\un A}({\un r}',t_0)\equiv 0$ we obtain
\begin{eqnarray*}
-\oint_{C}{\un p}({\un r}',t_0)\cdot d{\un r}'+\oint_{C}{\un p}({\un
r}',t)\cdot d{\un r}'=\qquad \qquad\\
-\oint_{C}e\,{\un A}({\un r}',t)\cdot d{\un r}'\,.
\end{eqnarray*}
where
$$
\oint_{C}{\un p}({\un r}',t_0)\cdot d{\un r}'=m_0\,\oint_{C}{\un v}({\un r}',t_0)\cdot d{\un r}'=0\,,
$$
which follows from Eq.(\ref{eqn:curl_zero}). Thus
$$
\oint_{C}\left[{\un v}({\un r}',t)+\frac {e}{m_0}\,{\un
A}({\un r}',t)\right]\cdot d{\un r}'=0\quad\forall\; t
$$
which means that the curl of the integrand vanishes:
\begin{eqnarray}
\label{eqn:vanishing_curl}
\nabla\times\left[{\un v}({\un r},t)+\frac{e}{m_0}\,{\un A} ({\un
r},t)\right]\equiv 0\,.
\end{eqnarray}
Consequently, it can be expressed as a gradient of a scalar function which we denote by $(\hbar/m_0)\,\varphi({\un r},t)$. Hence we arrive at
\begin{eqnarray}
\label{eqn:modifGeschwPotential} {\un v}({\un r},t)+\frac{e}{m_0}\,{\un A}({\un
r},t)=\frac{\hbar}{m_0}\,\nabla\varphi({\un r},t)\,.
\end{eqnarray}
which now stands in place of Eq.(\ref{eqn:Gradphi}).\\
We note here only in passing that we have because of $\psi({\un r})=|\psi({\un r})|e^{i\varphi({\un r})}$
$$
\frac{1}{2i}\,[\psi^*\nabla \psi-\psi\nabla \psi^*]=|\psi({\un r})|^2\,\nabla \varphi\,.
$$
Using Eq.(\ref{eqn:modifGeschwPotential}) one can recast this as
$$
\frac{\hbar}{2m_0}\,[\psi^*\nabla \psi-\psi\nabla \psi^*]=\rho\,{\un v}+\frac{e}{m_0}\,|\psi|^2{\un A}
$$
or equivalently
$$
\rho\,{\un v}=\frac{1}{2m_0}\,[\psi^*\widehat{{\un P}} \psi+c.c.]
$$
where $\widehat{{\un P}}$ is short-hand for $\widehat{{\un p}}-e\,{\un A}$.
After real-space integration and an integration by parts one arrives at
\begin{eqnarray}
\label{eqn:Expect_v_general}
\langle {\un v}\rangle=\frac{1}{m_0}\,\int\psi^*({\un r},t)\widehat{{\un P}}\,\psi({\un r},t)\,.
\end{eqnarray}
Because of Eq.(\ref{eqn:vanishing_curl}) the expression $({\un v}\cdot \nabla)\,{\un v}$ which appears in
\begin{eqnarray}
\label{eqn:NewtonModified(3)}
m_0\,\frac{d}{d t}\,{\un v}({\un r},t)=m_0\left[\frac{\partial}{\partial t}\,{\un v}+({\un v}\cdot \nabla)\,
{\un v}\right]=\qquad \qquad\nonumber \\
{\un F}({\un r},t)+{\un F}_{QP}({\un r},t)
\end{eqnarray}
cannot be replaced by $\small{\frac{1}{2}}\nabla {\un v}^2$ any more. Because of the generally valid relation
\begin{eqnarray*}
({\un a}\cdot \nabla)\,{\un a}=\nabla\,\frac{{\un
a}^2}{2}-{\un a}\times (\nabla \times {\un a})
\end{eqnarray*}
and because of Eq.(\ref{eqn:vanishing_curl}) we now have
\begin{eqnarray*}
({\un v}\cdot \nabla)\,{\un v}=\frac{1}{2}\nabla {\un v}^2-{\un v}\times
(\nabla\times{\un v})=\qquad\qquad\\
\frac{1}{2}\nabla {\un v}^2+\frac{e}{m_0}{\un v}\times
(\nabla\times{\un A})\,.
\end{eqnarray*}
Using $\nabla \times {\un A}={\un B}$ we may recast this as
\begin{eqnarray*}
({\un v}\cdot \nabla)\,{\un v}=\frac{1}{2}\nabla {\un v}^2+\frac{e}{m_0}{\un
v}\times {\un B}\,.
\end{eqnarray*}
Inserting this result together with Eq.(\ref{eqn:Electromagnetic_force}) and ${\un F}_{QP}=-\nabla V_{QP}$ into Eq.(\ref{eqn:NewtonModified(3)}) we notice that the Lorentz-force $e\,{\un v}({\un r},t)\times{\un B}
({\un r},t)$ drops out in favor of ${\un A}({\un r},t)$, and we get
\begin{eqnarray}
\label{eqn:kanonischeBeschleunigung} \frac{\partial}{\partial\,t}({\un
v}+\frac{e}{m_0}\,{\un A})=-\frac{1}{m_0}\nabla V-\frac{1}{2}\nabla{\un v}^2\nonumber \\
+\frac{1}{2}\,\nabla{\un u}^2-\frac{\hbar}{2\,m_0}\Delta {\un u}
\end{eqnarray}
where we have introduced
\begin{eqnarray}
\label{eqn:Definitio_V} 
V({\un r},t)=V_{cons}({\un r})+V_{el}({\un r},t)\,.
\end{eqnarray}
We now multiply Eq.(\ref{eqn:kanonischeBeschleunigung}) by the imaginary unit $i$ and subtract  Eq.(\ref{eqn:Zeitableitung_u2}) which gives in complete analogy to Eq.(\ref{eqn:komplexeBeschleunigung}) 
\begin{eqnarray}
\label{eqn:komplexeBeschleunigung1} \frac{\partial}{\partial\,t}\left[-{\un
u}+i({\un v}+\frac{e}{m_0}\,{\un A})\right]=
\frac{i}{2}\,\nabla (-{\un u}+i\,{\un v})^2 \nonumber \\
+\frac{i\,\hbar}{2\,m_0}\nabla\left[\nabla \cdot(-{\un u}+i\,{\un
v})\right]-\frac{i}{m_0}\,\nabla V.
\end{eqnarray}
We mention here only in passing that Eq.(\ref{eqn:Zeitableitung_u2}) is equivalent to Fick's law and is hence not affected by the presence of an electromagnetic field as long as Einstein's law (\ref{eqn:EinsteinLaw}) remains unchanged which is obvious from his derivation. (S. also Fritsche and Haugk \cite{FritscheHaugk}.)\\
As in the case without electromagnetic field we absorb the two independent scalar informations $\rho({\un r},t)$ and $\varphi({\un r},t)$ into one complex-valued function
\begin{eqnarray}
\label{eqn:DefinitionPsi1} \psi({\un r},t)=\pm \sqrt{\rho({\un
r},t)}\,e^{i\varphi({\un r},t)}\,.
\end{eqnarray}
As ${\un v}$ is no longer equal to $\small{\frac{\hbar}{m_0}}\nabla \varphi$ we have now in place of Eq.(\ref{eqn:komplexeGeschwindigkeit}) 
\begin{eqnarray}
\label{eqn:komplexeGeschwindigkeit1}
\frac{\hbar}{m_0}\,\nabla(\ln\psi/\sqrt{\rho_0})=-{\un u}+i({\un
v}+\frac{e}{m_0}\,{\un A})\,.
\end{eqnarray}
The left-hand side of Eq.(\ref{eqn:komplexeBeschleunigung1}) is obviously the time-derivative hereof. It will be useful to notice that
\begin{eqnarray}
\label{eqn:ZeitablkomplexeGeschwindigkeit1}
\frac{\partial}{\partial\,t}\,\frac{\hbar}{m_0}\,\nabla(\ln\psi/\sqrt{\rho_0})=\nabla
(\frac{1}{m_{0}\,\psi}\,\hbar \frac{\partial}{\partial\,t}\psi)\,.
\end{eqnarray}
We may also use Eq.(\ref{eqn:komplexeGeschwindigkeit1}) to recast the first expression on the right-hand side of Eq.(\ref{eqn:komplexeBeschleunigung1})
\begin{eqnarray*}
\frac{i}{2}\,\nabla (-{\un u}+i\,{\un v})^2=\frac{i}{2}\frac{\hbar^2}{m_{0}^2}
\nabla[\nabla \ln(\psi/\sqrt{\rho_0})]^2\\
+\frac{\hbar}{m_0}\,\frac{e}{m_0}\nabla[{\un A}\cdot \nabla
\ln(\psi/\sqrt{\rho_0})]-\frac{i}{2}\nabla(\frac{e}{m_0}{\un A})^2\,.
\end{eqnarray*}
If one observes that
\begin{eqnarray}
\label{eqn:nabla_psi}
\nabla[\nabla \ln(\psi/\sqrt{\rho_0})]=\frac{1}{\psi}\,\Delta
\psi-[\nabla\ln(\psi/\sqrt{\rho_0})]^2\,,
\end{eqnarray}
the second expression on the right-hand side of Eq.(\ref{eqn:komplexeBeschleunigung1}) can be written
\begin{eqnarray*}
i\frac{\hbar}{2\,m_0}\nabla\left[\nabla \cdot(-{\un u}+i\,{\un v})\right]=-\frac{i}{2}\,\frac{\hbar^2}{m_{0}^2}\,\nabla[\nabla \ln(\psi/\sqrt{\rho_0})]^2\\
+i\nabla\left[\frac{1}{\psi}(\frac{1}{2}\,\frac{\hbar^2}{m_{0}^2}\,\Delta\psi-
\frac{\hbar}{2\,m_0}\,\frac{e}{m_0}\,\underbrace{\psi\,\nabla \cdot {\un
A}}_{=(\nabla \cdot {\un A})\psi- ({\un A}\cdot\nabla)\psi})\right]\,.
\end{eqnarray*}
Hence we obtain
\begin{eqnarray*}
\frac{i}{2}\,\nabla (-{\un u}+i\,{\un v})^2+\frac{i\,\hbar}{2\,m_0}\nabla
\left[\nabla(-{\un u}+i\,{\un v})\right]= \nonumber \\
-i\,\nabla\left[\frac{1}{m_{0}\,\psi}\cal{G}\,\psi\right]
\end{eqnarray*}
where
$$
-\frac{\hbar^2}{2\,m_{0}}\,\Delta\psi+i\,
\frac{\hbar}{2}\frac{e}{m_0}\,\nabla \cdot {\un
A}+i\,2\,\frac{\hbar}{2}\,\frac{e}{m_0}{\un
A}\cdot\nabla+\frac{1}{2\,m_0}\,(e\,{\un A})^2\,.
$$
The right-hand side of his equation may be compactified by using the momentum operator (\ref{eqn:Momentum_operator}) as a convenient short-hand notation
\begin{eqnarray*}
\frac{i}{2}\,\nabla (-{\un u}+i\,{\un
v})^2+\frac{i\,\hbar}{2\,m_0}\nabla\left[\nabla(-{\un u}+i\,{\un v})\right]=\\
-i\,\nabla\left[\frac{1}{m_{0}\,\psi}\frac{(\widehat{{\un p}}-e\,{\un
A})^2}{2\,m_0}\,\psi\right]\,.
\end{eqnarray*}
Inserting this result into Eq.(\ref{eqn:komplexeBeschleunigung1}) which derives from Eq.(\ref{eqn:NewtonModified(3)}) (``Newton's modi\-fied second law'') and Eq.(\ref{eqn:diffusion_currentdensity}) ($\equiv$ Fick's law) and exploiting the Eqs.(\ref{eqn:ZeitablkomplexeGeschwindigkeit1}) and (\ref{eqn:nabla_psi}) we arrive at
\begin{eqnarray}
\label{eqn:zeitabhSchroedingerglg5} i\hbar\,\frac{\partial\,\psi({\un
r},t)}{\partial\,t}=\widehat{H}({\un r},t)\,\psi({\un r},t)
\end{eqnarray}
where
\begin{eqnarray*}
\label{eqn:HamiltOp_mitFeld} \widehat{H}({\un r},t)=\frac{\widehat{{\un
P}}\,^2}{2\,m_0}+V({\un r},t)\quad \mbox{and}\quad \widehat{{\un
P}}=\widehat{{\un p}}-e\,{\un A}({\un r},t)\,.
\end{eqnarray*}

\section{A model for non-Markovian diffusion 
illustrating the origin of non-locality}\label{kapitel11}

It is instructive to consider a model illustrating ``conservative diffusion''. The latter is a consequence of forming the arithmetic mean of Eqs.(\ref{eqn:Stoch_equation1}) and (\ref{eqn:Stoch_equation2}) which leads to Eq.(\ref{eqn:Bohm_equation1}). If one were to follow the motion of an individual particle, just one member out of the total ensemble, one would directly see the effect of stochastic forces changing back and forth from ``Brownian'' to  ``anti-Brownian'' with the latter causing a motion enhancement after the former have effected a slow down of the particle motion.
Figure \ref{Trajectories} shows three situation of the (free) particle which moves within a two-dimensional frame where a two-slit diaphragm has been inserted on the left-hand side. 
\begin{figure}[ht]    
\hspace*{0.8cm} \epsfig{file=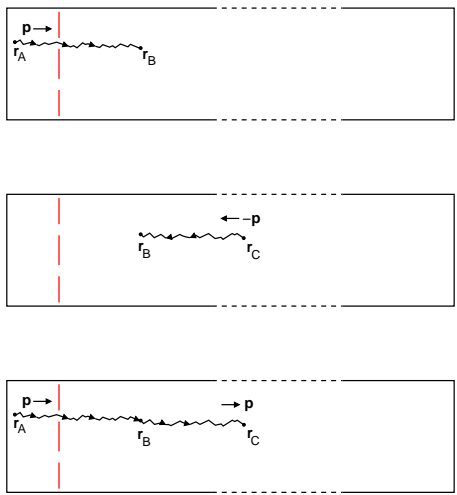,width=6cm}\vspace*{0.4cm}
\caption[Trajectories] {\label{Trajectories} Trajectory of the test particle undergoing reversible scatterings}
\end{figure}
The ``walls'' of the frame are assumed elastically reflecting. The  stochastic forces acting on the particle are simulated by a two-dimensional gas of $N$ identical point masses ($N \gg 1$) that interact via Lennard-Jones pair-potentials with each other and with the particle under study as well. The latter will henceforth be referred to as ``test particle''. It is this situation which the original derivation of Einstein's law (\ref{eqn:EinsteinLaw}) refers to where the motion of the test particle is described by a Langevin equation into which the embedding of the particle enters through a stochastic force. (The practical calculations have been performed with slightly modified Lennard-Jones potentials that were truncated at twice the average particle distance.) In our model the particle motion of the embedding gas results from a molecular dynamics simulation which one starts by first keeping the test particle fixed at the point ${\un r}_A$ and letting the $N$ gas particles start from some corner of the frame with equal absolute values of their momenta.
Thereby one defines a certain value of their total kinetic energy $E_{kin}^{gas}$.  After a short simulation time the gas particles are uniformly distributed within the frame and their distribution in the momentum space has become Maxwellian. The latter is associated with a certain temperature such that the thermodynamical expectation value of the kinetic energy equals $E_{kin}^{gas}$. It is this temperature which finally shows up in Einstein's law (\ref{eqn:EinsteinLaw}). After the embedding gas has ``thermalized'' one imparts a certain momentum ${\un p}$ on the test particle  and continues the molecular dynamics simulation with the test particle now included. As indicated in the upper panel of the figure, it performs an irregular (Brownian) motion and loses momentum to the embedding gas whose particles are not shown in the figure. We have chosen the starting point ${\un r}_A$ such that the particle moves through the upper slit of the diaphragm and reaches the point ${\un r}_B$ after a simulation time $\Delta t$ of the order of $\tau$ which is the time constant of a freely moving particle in a gaseous medium with friction. We now look for a point ${\un r}_C$ further to the right in the forward direction of the test particle (s. panel in the middle of the figure). At this point we impart a momentum $-{\un p}$ on the particle (after thermalization of the embedding gas), i.e. just the reverse of the momentum at ${\un r}_A$.\\ 
The point ${\un r}_C$ is chosen such that the trajectory ends - again after an identical simulation time of $\Delta t$ seconds - at point ${\un r}_B$. At this point the test particle has lost its original momentum $-{\un p}$ almost completely. If one now turns the velocities of \textbf{all} particles around by 180$^0$ and starts the simulation again with the time running forward as before, the test particle continues its motion from point ${\un r}_B$ and moves exactly along the trajectory it had formerly followed in the opposite direction coming from ${\un r}_C$. When it has reached ${\un r}_C$ again, it has regained the previously lost momentum, but this time with the sign reversed. Hence, in moving from ${\un r }_A$ to ${\un r}_C$ the particle undergoes scattering processes  that are in alternating succession Brownian and anti-Brownian within a time interval of the order $\tau$. Thereby the average momentum of the particle is conserved. This is illustrated in the third panel (bottom).
A striking feature of the momentum reconstruction by the above scattering processes is the occurrence of non-locality. This can be demonstrated by repeating the procedure that led to the trajectory portion from ${\un r}_C$ to ${\un r}_B$ with a crucial modification: If one closes the lower slit of the diaphragm and starts then with \textbf{the same position/velocity configuration} of all particles as before, the trajectory of the test particle evolves now differently and does no longer join the previously generated trajectory portion at ${\un r}_B$. This is what the molecular dynamics simulation clearly yields. On the other hand, this is to be expected anyway because every momentum transfer from the test particle to the gas spreads with sound velocity throughout the entire structure and probes the change that has been introduced. The stochastic forces acting on the test particle are modified by such a change when these sound waves are reflected back on the particle. If one wants the modified trajectory to join the first trajectory portion at ${\un r}_B$ again, one has to choose a different starting point ${\un r}'_C$. Once the test particle has arrived at ${\un r}_B$, one inverts all the velocities as before, and the particle will now recover the momentum ${\un p}$ on its modified trajectory toward ${\un r}'_C$. Note: this change in the course of the particle motion results just from closing the \textbf{lower} slit although the particle definitely traverses the \textbf{upper} slit. One is tempted to surmise that this mechanism of probing the environment ``in real life'' as the particle  exchanges temporarily momentum with the {\bf vacuum}, occurs at light velocity. The latter would impose a limit on the distance beyond which a previously passed potential structure can no longer affect the evolution of the particle's trajectory at its current position.\\
If one were dealing with Brownian scattering only, the succession of scattering events could be classed as ``Markovian''. (Shorthand definition: given the presence, future and past are independent.) However, the overall character of the combined Brownian/anti-Brownian scattering processes is obviously non-Markovian. It is true that the particle has almost completely lost its memory of its original momentum when it arrives at ${\un r}_B$, but its future time evolution while moving toward ${\un r}_C$ reconstructs, so to speak, past scattering events. The particle's momentum ${\un p}(0)$ when it is  at ${\un r}_A$, and its momentum ${\un p}(t)$ at ${\un r}_C$ are {\bf strongly correlated}.\\
This does not apply to the positions ${\un r}_A(0)$ and ${\un r}_C(t)$: if one repeats the experiment and lets the particle start at ${\un r}_A$ with the same momentum ${\un p}(0)$ as before, but with the thermalization process of the embedding gas started some time interval earlier, the particle's trajectory will now be different and lead to a point different from ${\un r}_C$ though it regains its original momentum after (approximately) the same traveling time.\\
Obviously, the non-Markovian (reversible) character of particle motion which results from such a combination of scattering processes can only show up on a coarse grain time scale which is the crucial assumption underlying our derivation of Newton's modified second laws Eqs.(\ref{eqn:Bohm_equation2}) and (\ref{eqn:NewtonModified(3)}) together with (\ref{eqn:Electromagnetic_force}). \\

\section{Operators and commutators}\label{kapitel11.1}

An important advantage of our approach may be seen in the derivability of Hermitian operators which in standard quantum mechanics can merely be obtained from educated guessing employing Jordan's replacement rules. In Section \ref{kapitel6} we have already derived the momentum operator
$$
\widehat{{\un p}}=-i\,\hbar\nabla
$$
exploiting our expression (\ref{eqn:Gradphi}) for ${\un v}({\un r},t)$ and  ${\un j}({\un r},t)=\rho({\un r},t)\,{\un v}({\un r},t)$. The same arguments used in deriving $\widehat{{\un p}}$ apply to the angular momentum operator $\widehat{{\un L}}$ which occurs on forming the expectation value of the angular momentum of a particle with respect to a center located at ${\un r}=0$. This expectation value $\langle {\un L} \rangle$ is primarily 
defined as a real-space integral over the angular momentum density ${\un r}\times m_0\,{\un j}$:
\begin{eqnarray}
\label{eqn:angular_momentum_1}
\langle {\un L}(t) \rangle=\int {\un r}\times m_0\,{\un j}({\un r},t)\,d^3r 
\end{eqnarray}
If one here inserts ${\un j}$ from Eq.(\ref{eqn:Definition2_j}), integrates by parts and requires $\psi({\un r},t)$ to vanish sufficiently toward infinity, the result may be written
\begin{eqnarray}
\label{eqn:angular_momentum_2}
\langle {\un L}(t) \rangle=\int \psi^{*}({\un r},t)\,({\un r}\times\widehat{{\un p}})\,\psi({\un r},t)\,d^3r
\end{eqnarray}
which justifies terming $\widehat{{\un L}}\equiv {\un r}\times\widehat{{\un p}}$ $\;$``angular momentum operator''.\\
The kinetic energy of an individual particle, labeled by the index j, is defined as the work performed on that particle by the external force ${\un F}$ in accelerating it from zero velocity at time $t=0$ to its velocity ${\un v}_{cj}(t)$ at time $t$, which yields
$$
E^{j}_{kin}=\frac{m_0}{2}\,{\un v}^{2}_{cj}(t) \,.
$$
Forming the ensemble average according to Eq.(\ref{eqn:average_v}) one obtains $E_{kin}=\frac{m_0}{2}\,{\un v}^2_c({\un r},t)$. Thus, the density of the kinetic energy is given by
\begin{eqnarray*}
\label{eqn:kinetischeEnergiedichte} \epsilon_{kin}({\un r},
t)=\frac{m_0}{2}\,\rho({\un r}, t)\left[{\un v}_{c}({\un r}, t)\right]^2\,.
\end{eqnarray*}
In the two subsystems ``B'' and ``A'' we are considering the velocity ${\un v}$ is always the same, whereas the convective velocity 
${\un v}_c$ is different, and therefore we distinguish 
${\un v}^{B}_c$ from ${\un v}^{A}_c$ and form the ensemble average over the two subensembles:
\begin{eqnarray}
\label{eqn:kinetischeEnergiedichte0} \epsilon_{kin}({\un r},
t)=m_0\,{\textstyle\frac{\rho({\un r},t)}{2}\,\frac{1}{2}}\left(\left[{\un v}_{c}^{B}({\un
r}, t)\right]^2+\left[{\un v}_{c}^{A}({\un r}, t)\right]^2\right)
\end{eqnarray}
According to Eq.(\ref{eqn:convectivevelocity}) which still refers to the ``B''-system, we have
$$
{\un v}_{c}^{B}={\un v}-{\un u} \quad \mbox{and therefore}\quad {\un v}_{c}^{A}={\un v}+{\un u}\,.
$$
Consequently Eq.(\ref{eqn:kinetischeEnergiedichte0}) may be cast 
\begin{eqnarray}
\label{eqn:kinetischeEnergiedichte} \epsilon_{kin.}({\un r},
t)=m_0\,\frac{\rho({\un r},t)}{2}\,\left(\left[{\un v}({\un
r},t)\right]^2+\left[{\un u}({\un r},t)\right]^2\right)\,.
\end{eqnarray}
From Eq.(\ref{eqn:komplexeGeschwindigkeit}) we have
\begin{eqnarray*}
-{\un u}+i\,{\un v}=\frac{\hbar}{m_0}\,\frac{1}{\psi}\,\nabla\psi\,.
\end{eqnarray*}
The modulus square of this equation times $m_0\,\rho/2$  is equal to the right-hand side of Eq.(\ref{eqn:kinetischeEnergiedichte}), that is
\begin{eqnarray*}
\label{eqn:kinetischeEnergiedichte_mit_psi} \epsilon_{kin.}({\un
r},t)=\frac{\hbar^2}{2\,m_0}\,|\nabla \psi({\un r},t)|^2\,.
\end{eqnarray*}
Taking the real-space integral of this expression one obtains the kinetic energy
\begin{eqnarray}
\label{eqn:kinetischeEnergie0}
E_{kin}\equiv \langle T(t)\rangle=\int_{\cal{V}}\frac{\hbar^2}{2\,m_0}\,\nabla \psi^{*}({\un
r},t)\cdot \nabla \psi({\un r},t)\,d^3r
\end{eqnarray}
which by employing Green's theorem may be given the familiar form
\begin{eqnarray*}
\int_{\cal{V}}\frac{\hbar^2}{2\,m_0}\,\nabla\psi^{*}({\un r},t)\cdot \nabla \psi({\un r},t)\,d^3r=\\
\int_{\cal{V}}\psi^{*}
({\un r},t)\,\left[-\frac{\hbar^2\,\nabla^2}{2\,m_0}\right]\,\psi({\un r},t)\,d^3r\,,
\end{eqnarray*}
and hence
\begin{eqnarray*}
 \langle T(t)\rangle=\int_{\cal{V}}\psi^{*}({\un r},t)\,\frac{\widehat{{\un p}}^2}{2\,m_0}\,\psi({\un r},t)\,d^3r\,,
\end{eqnarray*}
which justifies terming $\widehat{{\un p}}^2/2\,m_0$ $\,$ ``kinetic energy operator''.\\
In practical calculations one often benefits from the fact that $E_{kin}$ may alternatively be cast as in Eq.(\ref{eqn:kinetischeEnergie0}) where the integrand is real-valued and may immediately be interpreted as ``kinetic energy density''.
\\[0.2cm]
The statistical operator is a particular example of derivability from a simple concept. We confine ourselves here to the case of a quantum mechanical system of a bound particle in contact with a heat bath of temperature $T$. In a stationary state the latter constantly exchanges energy with the system, in the simplest case photons. Hence, the wave function of that system cannot be one of its eigenstates any more, but rather represents a solution to the time-dependent Schr\"odinger equation and can be expanded in terms of eigenfunctions $\psi_n({\un r})$
\begin{eqnarray}
\label{eqn:timedependent_psi}
\psi({\un r},t)=\sum_{n}\,c_n(t)\,\psi_n({\un r})\,e^{-\frac{i}{\hbar}\,E_{n}\,t}
\end{eqnarray}
where $E_n$ denotes eigenvalues of the unperturbed one-particle Hamiltonian $\widehat{H}$.\\
To make the external system classifiable as a heat bath, the time-averaged coupling energy of the two systems must be negligibly small compared to the difference $E_{n'}-E_n$ of any two eigenvalues. The particle's thermodynamical expectation value of its energy (indicated by double brackets) is given by
\begin{eqnarray}
\label{eqn:U_1}
\langle\langle \widehat{H}\rangle\rangle\equiv U=\qquad \qquad\qquad \nonumber\\
\frac{1}{\tau}\,\int_{t}^{t+\tau}\left[\int \psi^{*}({\un r},t')\,\widehat{H}\,
\psi({\un r},t')\,d^3r\right]\,dt'
\end{eqnarray}
where $\tau$ (not to be confused with the slow-down time in Section \ref{kapitel3}) has to be chosen sufficiently large such that $U$ does not depend on $t$ any more. Quantities that derive from $U$ like the specific heat, are only defined as time-averages of this kind.\\
Inserting Eq.(\ref{eqn:timedependent_psi}) into (\ref{eqn:U_1}) we obtain
\begin{eqnarray}
\label{eqn:U_2}
U=\sum_{n}E_n\,\{\frac{1}{\tau}\,\int_{t}^{t+\tau}|c_n(t')|^2\,dt'\}
\end{eqnarray}
where the expression in curly brackets may be interpreted as the relative frequency of the system of being in the $n$-th eigenstate.\\
Straight-forward thermodynamics yields for a system that possesses energy levels $E_n$ 
\begin{eqnarray}
\label{eqn:U_3}
U=\sum_{n}E_n\,\frac{1}{\sigma}\,e^{-\beta\,E_n}\:, \quad \beta=\frac{1}{k_B\,T}\,,
\end{eqnarray}
where
$$
\sigma=\sum_{n}e^{-\beta\,E_n}\,.
$$
Thus, we have from Eq.(\ref{eqn:U_2})
$$
\frac{1}{\tau}\,\int_{t}^{t+\tau}|c_n(t')|^2\,dt'=\frac{1}{\sigma}\,e^{-\beta\,E_n}\,.
$$
If one defines a statistical operator
$$
\widehat{\rho}=\frac{1}{\sigma}\,e^{-\beta\,\widehat{H}}
$$
Eq.(\ref{eqn:U_3}) can alternatively be cast as
$$
U=\sum_{n}\langle\psi_n|\widehat{\rho}\,\widehat{H}|\psi_n\rangle \equiv \mbox{Tr}(\widehat{\rho}\,\widehat{H})\,.
$$
\\[0.2cm]
{\bf Commutation rules} for the operators apply when the potential $V({\un r})$ in the time-independent Schr\"odinger equation (\ref{eqn:DglgPhi}) possesses a certain symmetry. If $V({\un r})$ is spherically symmetric, for example, one verifies simply by performing partial differentiations that
$$
\left(\widehat{H}\widehat{{\un L}}^2-\widehat{{\un L}}^2\widehat{H}\right)\,\psi_n({\un r})\equiv [\widehat{H},\widehat{{\un L}}^2]\,\psi_n({\un r})=0
$$
and similarly
$$
[\widehat{H},\widehat{{\un L}}_z]\,\psi_n({\un r})=0
$$
if $\psi_n({\un r})$ is an eigenfunction of $\widehat{H}$.
\\[0.2cm]
If one is dealing with some operator $\widehat{A}$ which represents just some analytical expression in ${\un r}$ and $\widehat{{\un p}}$, the time dependence of its expectation value $\langle \widehat{A} \rangle$ can be determined by employing the time-dependent Schr\"odinger equation which gives
\begin{eqnarray*}
\frac{d}{dt}\int \psi^{*}({\un r},t)\,\widehat{A}\,\psi({\un r},t)\,d^3r=\qquad \qquad\qquad  \\
\int\psi^{*}({\un r},t)\frac{i}{\hbar}[\widehat{H},\widehat{A}]\,
\psi({\un r},t)\,d^3r\,,
\end{eqnarray*}
in short-hand notation
$$
\frac{d}{d t}\,\widehat{A}=\frac{i}{\hbar}[\widehat{H},\widehat{A}]\,.
$$
Commutation rules of the above kind, again in short-hand notation
$$
[\widehat{H},\widehat{{\un L}}^2]=0\;;\;\;[\widehat{H},\widehat{{\un L}}_z]=0\,,
$$
similarly
$$
[\widehat{H},\widehat{{\un p}}]=0\;\;\mbox{if}\;\; 
V({\un r})=\mbox{const.}\,,
$$
but also
$$
[\widehat{p}_j,x_k]=\frac{\hbar}{i}\, \delta_{j\,k}\quad\mbox{where}\quad j=1,2,3\,;\;k=1,2,3
$$
constitute fundamental elements of standard quantum mechanics and are discussed as pivotal in the context of measurement. From our point of view they are just byproducts of the Schr\"odinger equation and do not contain any more physics than has already gone into the derivation of the Schr\"odinger equation. In practice it is impossible to find quantum systems where eigenvalues of $\widehat{H}$ and $\widehat{L}_z$, for example, can be measured simultaneously  although there is a widespread belief to the contrary. It is not even possible, for example, to measure the eigenvalues of $\widehat{H}$ for a hydrogen atom which - in clamped proton approximation - represents the archetypal one-particle system and the starting point of quantum mechanics. The lines one observes in its discrete optical spectrum refer to eigenvalue differences and possess - different from true eigenvalues - a natural line width which goes to zero only in the hypothetical case of zero radiation coupling, that is when the lines cannot be observed any more. \\
There is a remark by Wigner \cite{Wigner1} which reveals exactly that lack of stringency and consistency in the foundation of orthodox quantum mechanics: {\it `` All these are concrete and clearly demonstrated limitations on the measurability of operators. They should not obscure the other, perhaps even more fundamental weakness of the standard theory, that it postulates the measurability of operators but does not give directions as to how the measurement should be carried out.''}

\section{Collaps of the wave function and the node problem}\label{kapitel11.2} 

A vital point of the Copenhagen interpretation consists in the notion that the wave function of a stationary one-particle state collapses on performing a measurement on the position of the particle, for example. Within our approach a phenomenon of this kind cannot occur. First of all, in our view ``measurement'' is not a process of something foreign intruding the realm of quantum mechanics but is rather a part of it. If one calculates, for example, the time-independent wave function $\psi({\un r})$ for a stationary situation where electrons in a diffraction chamber leave a tunneling cathode, sufficiently far behind each other, run through a two-slit diaphragm and finally hit a fluorescent screen, $|\psi({\un r})|^2$ will display the familiar diffraction pattern behind the diaphragm and in particular on the screen. But clearly, the structure of this pattern reflects the distribution of the entire ensemble of electrons that leave the cathode, and a particular electron, that hits the screen some place, is only one member out of this ensemble. Hence its capture on the screen does not destroy the properties of the ensemble. The electron capture by an atom of the screen constitutes a process that has only marginally to do with the diffraction state in that the latter determines the probability of the electron being at that particular atom. Otherwise the capture process is governed by the time dependent Schr\"odinger equation and the perturbation caused by the electromagnetic field of the outgoing photon. All this is completely independent of the possible presence of an ``observer'' who might see that photon. 
\\[0.2cm]
Despite deceptive similarities the situation becomes conceptually different when one replaces the tunneling tip in the otherwise unchanged diffraction chamber by a light source that emits, again in sufficiently large time intervals, photons of the same wave length as the previously considered electrons. Since the space-time structure of the wave (in principle $\propto \cos[{\un k}\cdot{\un r}-\omega\,t])$ with which each photon is associated, is not defined as the property of an ensemble of mechanical objects but rather by classical electrodynamics (Maxwell's equations), it will, in fact, disappear on the disappearance of the photon in question. There are a couple of properties by which photons differ crucially from massive particles: They move always at light velocity along straight lines in vacuo and the associated waves are vector-valued functions. By contrast, the de Broglie waves of massive particles are in general complex-valued functions, and the average velocity of the particles is given by the gradient of the functions' phase. Their interaction with other particles and parts of an experimental setup is described by the Schr\"odinger equation and the potentials therein. On the other hand, the interaction of photons with polarizers, mirrors, quaterwave plates, filters etc. is governed by classical electrodynamics. Malus' law, for example, constitutes a law of classical optics. Because of these rather fundamental differences any analysis of photon correlation experiments, for example, should critically be scrutinized whether a transfer to analogous experiments with massive particles is truly justified. In Section \ref{kapitel20}  we shall draw on the familiar example of the Stern-Gerlach experiment to demonstrate that the selection mechanism for up-spin and down-spin particles in the Stern-Gerlach magnet has nothing to do with the mechanism separating horizontally and vertically polarized photons in a polarizing beam spitter. 
\\[0.2cm]
According to Mielnik and Tengstrand \cite{Mielnik} excited stationary states appear to pose a serious problem in that $\psi({\un r})$ possesses nodal surfaces at which the normal derivative $\frac{\partial}{\partial n} \rho({\un r})$ vanishes but the normal component of the osmotic velocity
$$
{\un u}_n({\un r})=-\frac{\hbar}{2m_0}\,\frac{\frac{\partial}{\partial n} \rho({\un r})}{\rho({\un r})}\,{\un e}_n
$$
becomes formally infinite. Moreover, at surfaces across which $\rho({\un r})$ attains a maximum, $\frac{\partial}{\partial n} \rho({\un r})$ vanishes as well, but ${\un u}_n({\un r})$ becomes now zero. If $\psi({\un r})$ is real-valued then ${\un v}({\un r})$ vanishes everywhere, and therefore  we have on such surfaces with maximum probability density
$$
{\un v}={\un u}_n=0\,.
$$
In the 2s-state of a hydrogen electron, for example, one has a spherical surface of this kind. Hence, it seems that this sphere separates two regions of space that are mutually inaccessible for the electron. But the above velocities are only ensemble averages or - in the spirit of the definition (\ref{eqn:time-average_v}) - averages of non-vanishing velocities ${\un v}(t_i), {\un u}_n(t_i)$ of different directions over a sufficiently long time $T$.
\\[0.2cm]
As for $|{\un u}_n({\un r})|$ going to infinity as one crosses a nodal surface of $\psi({\un r})$, one has to keep in mind that stationary excited states (excited eigenstates) are highly fictional and do actually not exist in nature. Because of $\Delta E\,\Delta t\approx \hbar$ and $\Delta E=0$ for an eigenstate it would take an infinite time to prepare them. Hence, truly existing excited states do not possess nodal surfaces where $\psi({\un r})$ vanishes exactly. But even if one would allow them to exist, the kinetic energy density $\frac{m_0}{2}\,\rho({\un r})\,{\un u}^2
({\un r})=\frac{\hbar^2}{2m_0}|\nabla \psi({\un r})|^2$ remains finite and hence ensures a physically meaningful behavior even for this idealized situation.

\section{The Feynman path integral} \label{kapitel16}

As our concept builds on the existence of particle trajectories one might surmise that there should be some affinity to Feynman's path integral method \cite{Feynman} which also relates to possible paths a particle might take. We shall outline that there is neither any formal kinship nor does Feynman name any cause for the possible occurrence of non-classical trajectories. In so doing we limit ourselves, as Feynman in his article, to the one-dimensional case of a particle that moves non-relativistically in a potential $V(x)$. Feynman's considerations are based on two hypotheses that may be summarized by stating that the wave function $\psi(x,t+\Delta t)$ of the particle at some point $x$ and time $t+\Delta t$ is connected with the wave function $\psi(x-\sigma,t)$ at a previous point $x-\sigma$ and earlier time $t$ by an integral equation similar to the Smoluchowski equation (\ref{eqn:Smoluchowski}) of the ensuing section, viz.
\begin{eqnarray}
\label{eqn:Feynman_Smoluchowski}
\psi(x,t+\Delta t)=\int \psi(x-\sigma,t)\,F(x,x-\sigma,t,\Delta t)\,d\sigma
\end{eqnarray}
where $F(x,x-\sigma,t,\Delta t)$ is the function that brings in classical mechanics. It is defined as
$$
F(x,x-\sigma,t,\Delta t)=\frac{1}{A}\;e^{\frac{i}{\hbar}\,S(x,\,x-\sigma,\,t,\,\Delta t)}
$$
where 
$$
 A=\left(\frac{2\pi\hbar\,i\,\Delta t}{m_0}\right)^{\frac{1}{2}}\,.
$$
Here $S(x,x-\sigma,t,\Delta t)$ denotes Hamilton's first principle function for a particle moving classically in a potential $V(x)$ along a trajectory from a point $x-\sigma$ to $x$ within an infinitesimally small time span $\Delta t$. Hence
\begin{eqnarray*}
\label{eqn:S}
S(x,x-\sigma,t,\Delta t)=\qquad\qquad\qquad \qquad\qquad\qquad\nonumber \\
\mbox{Min.}\,\int_t^{t+\Delta t}\left[\frac{m_0}{2}\,\dot{\sigma}^2-V(x-\sigma(t'))\right]\,dt'
\end{eqnarray*}
where
$$
L(\dot{\sigma}(t),\sigma(t))=\frac{m_0}{2}\,\dot{\sigma}^2-V(x-\sigma(t))
$$
denotes the Lagrangean.\\
As $\Delta t$ is infinitesimally small $S(x,x-\sigma,t,\Delta t)$ may be approximated
$$
S=\Delta t\,\left[\frac{m_0}{2}\,\left(\frac{\sigma}{\Delta t}\right)^2-V(x)\right]\,.
$$
Hence one has
$$
F=\frac{1}{A}\,\left[e^{\frac{i\,m_0}{2\hbar\,\Delta t}\,\sigma^2}\cdot e^{-\frac{i\,V(x)\,\Delta t}{\hbar}}\right]\,.
$$
The first exponential oscillates rapidly as a function of $\sigma$ because of the prefactor $1/\Delta t$ in the exponent, whereas, by comparison,  $\psi(x-\sigma,t)$ may be assumed slowly varying as a function of $\sigma$. The value of the integral in Eq.(\ref{eqn:Feynman_Smoluchowski}) depends therefore only on a small interval of $\sigma$ around the point $x$. Within this interval $\psi(x-\sigma,t)$ may be expanded as
\begin{eqnarray*}
\label{eqn:psi_expansion}
\psi(x-\sigma,t)=\psi(x,t)-\frac{d\psi}{dx}\,\sigma+\frac{1}{2}\,\frac{d^2\psi}{dx^2}\,\sigma^2\,.
\end{eqnarray*}
If one inserts this into Eq.(\ref{eqn:Feynman_Smoluchowski}), observes 
$$
\frac{1}{A}\,\int_{-\infty}^{\infty}e^{\frac{i\,m_0}{2\hbar\,\Delta t}\,\sigma^2}\,d\sigma=1\,;
$$
(note that this equation defines $A$!), further
$$
\frac{1}{A}\,\int_{-\infty}^{\infty}e^{\frac{i\,m_0}{2\hbar\,\Delta t}\,\sigma^2}\,\sigma\,d\sigma=0\;,
$$
$$ 
\frac{1}{A}\,\int_{-\infty}^{\infty}e^{\frac{i\,m_0}{2\hbar\,\Delta t}\,\sigma^2}\,\sigma^2\,d\sigma=\frac{i\hbar}{m_0}\,\Delta t
$$
and uses
$$
e^{-\frac{i\,V(x)\,\Delta t}{\hbar}}\approx 1-\frac{i\,V(x)\,\Delta t}{\hbar}\,,
$$
one obtains
\begin{eqnarray*}
\psi(x,t+\Delta t)=\psi(x,t)\left(1-\frac{i}{\hbar}\,V(x)\,\Delta t\right)+\qquad \qquad\\
\frac{1}{2}\,\frac{d^2\psi}{dx^2}\cdot\,\frac{i\hbar}{m_0}\,\Delta t\left(1-\frac{i\,V(x)\,\Delta t}{\hbar}\right)\,.
\end{eqnarray*}
Multiplying this equation by $\frac{i\hbar}{\Delta t}$ and letting $\Delta t$ tend to zero one arrives at the time dependent Schr\"odinger equation
$$
i\hbar\frac{\partial}{\partial t}\,\psi(x,t)=\left[-\frac{\hbar^2}{2\,m_0}\,\frac{\partial^2}{ \partial x^2}+V(x)\right]\psi(x,t)\,.
$$
Though the Schr\"odinger equation is obviously recovered following this line of argument, it remains unclear why Hamilton's classical first principle function should appear in the exponent of $F(x,x-\sigma,t,\Delta t)$. Feynman's considerations lean closely on arguments of measurement typical of the Copenhagen school of thought, cast into an axiomatic framework notably by v.~Neumann \cite{v_Neumann}.  But $\psi(x,t)$ may, for example, describe the motion of a harmonic oscillator in the absence of any measurement. In fact, if one were to perform a measurement on the harmonic oscillator the Schr\"odinger equation would contain a perturbative extra term that would give rise to a different wave function. Clearly, as already stated in Section\ref{kapitel1} the probabilistic character of $\psi(x,t)$ does not originate from indeterminacies caused by the process of measurement. The complex-valuedness of the wave function in the form $\psi(x,t)=|\psi(x,t)|\,e^{i\varphi(x,t)}$ comes about by incorporating two autonomous real-valued informations: the probability density $|\psi(x,t)|^2$ of the particle being at $x$ and time $t$ and the ensemble average $v(x,t)=\frac{\hbar}{m_0}\frac{d}{dx}\,\varphi(x,t)$ of its velocity. For that reason our derivation of the Schr\"odinger equation requires two Smoluchowski equations for the real-valued functions $\rho({\un r},t)$ and ${\un v}({\un r},t)$ instead of Feynman's single Eq.(\ref{eqn:Feynman_Smoluchowski}). We believe, therefore, that in our derivation the connection to classical mechanics becomes definitely more transparent and convincing.
\section{The time-dependent N-particle Schr\"odinger equation}\label{kapitel12}

So far we have merely been concerned with a single particle whose stochastic behavior was described by regarding it as a member of $N$ identically prepared, but statistically independent one-particle systems under the supposition that $N$ be sufficiently large. To avoid confusion we shall henceforth rename that number by $\cal{N}$. Instead of a single particle we now consider $N$ particles that interact via pair-forces. Each of these particles is individually a member of $\cal{N}$ statistically independent one-particle systems where the $N-1$ remaining particles appear at fixed positions ${\un r}_2,{\un r}_2, \ldots {\un r}_N$  if the particle under consideration, picked at will, just happens to be ``number 1''.   
The considerations of Sections \ref{kapitel6} and \ref{kapitel10} carry over to this $N$-particle system. To see that one simply has to replace the 3-dimensional real-space of the single particle discussed as yet by a $3N$-dimensional space where the $N$ particles appear as one point again. Instead of the probability density $\rho({\un r},t)$ one is now dealing with
\begin{eqnarray}
\label{eqn:N_density}
\rho({\un r}_1,{\un r}_2, \ldots {\un r}_N,t)=\rho({\un r}^N,t)\,;\nonumber\\
\mbox{where}\quad\int \rho({\un r}^N,t)\,d^3r_1\,d^3r_2\, \ldots d^3r_N =1
\end{eqnarray}
and
\begin{eqnarray*}
{\un r}^N=({\un r}_1, {\un r}_2, \dots {\un
r}_N)=\sum_{j=1}^{N}\sum_{k=1}^{3}x_{j\,k}\,{\un e}_{j\,k} 
\end{eqnarray*}
with $j=1,2,\dots N$ numbering the particles and $x_{j\,k}$ denoting Cartesian coordinates which are associated with orthogonal unit vectors 
${\un e}_{j\,k}$. The quantities $\nabla^{N}$, ${\un u}^{N}$ and ${\un
v}^{N}$ are defined analogously.\\
Instead of $\varphi({\un r},t)$ we now have $\phi({\un r}^{N},t)$. Thus 
\begin{eqnarray}
\label{eqn:3NTot_velocity}
{\un v}^{N}({\un r}^{N},t)=\frac{\hbar}{m_0}\,\nabla^{N}\,\phi({\un
r}^{N},t)\,.
\end{eqnarray}
Correspondingly, the 3$N$-dimensional osmotic velocity has the form
\begin{eqnarray}
\label{eqn:3NOsm_velocity}
{\un u}^{N}({\un r}^{N},t)=-\frac{\hbar}{2\,m_0}\,\nabla^{N}\,
\ln[\rho({\un r}^{N},t)/\rho_0]\,,
\end{eqnarray}
and hence we have similar to the single-particle case
\begin{eqnarray}
\label{eqn:time_derivative_u1}
\frac{\partial {\un u}^N}{\partial t}=-\frac{\hbar}{2\,m_0}\,\nabla^N\frac{\partial}{\partial t}\,[\ln\rho/\rho_0]=\nonumber \\-\frac{\hbar}{2\,m_0}\,\nabla^N\,\left[\frac{1}{\rho}\,\frac{\partial \rho}{\partial t}\right]\,.
\end{eqnarray}
Invoking the equation of continuity 
$$
\frac{\partial \rho}{\partial t}+\underbrace{\nabla^N \cdot (\rho\,{\un v}^N)}_{=\rho\,\nabla^N \cdot {\un v}^N+{\un v}^N \cdot\nabla^N \rho}=0
$$
and using the definition (\ref{eqn:3NOsm_velocity}), Eq.(\ref{eqn:time_derivative_u1}) can be cast as
\begin{eqnarray}
\label{eqn:time_derivative_u2}
\frac{\partial {\un u}^N}{\partial t}=-\frac{\hbar}{2\,m_0}\,\nabla^N\,[(\nabla^N \cdot {\un v}^N)-({\un u}^N \cdot {\un v}^N)]\,.
\end{eqnarray}
In the following we first confine ourselves to time-independent conservative forces which - in the spirit of our notation - may be written
$$
{\un F}_{ext.}^{N}({\un r}^{N})=\sum_{j=1}^{N}\sum_{k=1}^{3}F_{k}^{ext.}({\un
r}_j)\,{\un e}_{j\,k}
$$
where 
$$
F_{k}^{ext.}({\un r}_j)=-\frac{\partial}{\partial x_{j\,k}}\,V_{ext.}({\un
r}_j)
$$
with $V_{ext.}({\un r})$ denoting an external potential. Hence ${\un
F}_{ext.}^{N}$ may alternatively be written
$$
{\un F}_{ext.}^{N}({\un r}^{N})=-\nabla^{N}\widehat{V}_{ext.}({\un r}_1, {\un
r}_2, \dots {\un r}_N)
$$
where
$$
\widehat{V}_{ext.}({\un r}_1, {\un r}_2, \dots {\un
r}_N)=\sum_{j=1}^{N}V_{ext.}({\un r}_j)\,.
$$
The force exerted on the $j$-th particle due to pair-interaction with the $N-1$ remaining particles is given by
$$
F_{j\,k}^{inter}({\un
r}_j)=-\frac{\partial}{\partial\,x_{j\,k}}\sum_{\stackrel{i=1}{i\not=j}}^{N}V(|{\un
r}_j-{\un r}_i|)\,,
$$
where $V(|{\un r}_j-{\un r}_i|)$ denotes the interaction potential. The generalized total force in the $3N$-dimensional space may therefore be cast as
$$
{\un F}^{N}({\un r}^{N})=-\nabla^{N}\widehat{V}({\un r}_1, {\un r}_2, \dots
{\un r_N})\,,
$$
where $\widehat{V}({\un r}_1, {\un r}_2, \dots {\un r_N})$ is defined by
\begin{eqnarray}
\label{eqn:manyparticle_potential}
\widehat{V}({\un r}_1, {\un r}_2, \dots {\un r_N})=\sum_{j=1}^{N}V_{ext.}({\un
r}_j)+\qquad \qquad \qquad \nonumber \\
\frac{1}{2}\sum_{j=1}^{N}\sum_{\stackrel{i=1}{i\not=j}}^{N}V(|{\un
r}_j-{\un r}_i|)\,.
\end{eqnarray} 
Newton's modified second law (9.2,I) hence attains the form
\begin{eqnarray}
\label{eqn:3NNewton_modified}
\frac{\partial\,{\un v}^{N}}{\partial\,t}=-\nabla^{N}\left[\frac{1}{m_0}\,\widehat{V}+\frac{1}{2}\,({\un v}^{N})^2-\frac{1}{2}\,({\un u}^{N})^2+\right.\nonumber \\
\left.\frac{\hbar}{2\,m_0}\,\nabla^{N}\cdot \,{\un u}^{N}\right]\,.
\end{eqnarray}
As in the one-particle case the two scalar functions $\rho^{N}({\un r}^{N},t)$ and $\phi({\un r}^{N},t)$ can be absorbed into a complex-valued function $\Psi({\un r}_1, {\un r}_2, \dots {\un r_N}, t)$ defined by
\begin{eqnarray*}
\Psi({\un r}_1, {\un r}_2, \dots {\un r_N}, t)=\pm \sqrt{\rho({\un r}_1, {\un
r}_2, \dots {\un r_N}, t)}\times \qquad \qquad\\
\exp\,[i\,\phi({\un r}_1, {\un r}_2, \dots {\un
r_N}, t)]\,.
\end{eqnarray*}
This is equivalent to
\begin{eqnarray*}
\label{eqn:komplexeGeschwindigkeit2}
-{\un u}^N({\un r}^{N},t)+i\,{\un v}^N({\un r}^{N},t)=\frac{\hbar}{m_0}\,\nabla^N(\ln\Psi
({\un r}^{N},t)/\sqrt{\rho_0})\\
=\frac{\hbar}{m_0}\,\frac{\nabla^N\Psi}{\Psi}
\end{eqnarray*}
which is the analogue to Eq.(9.9,I), and we obtain accordingly
\begin{eqnarray*}
\frac{\partial}{\partial t}(-{\un u}^N+i\,{\un v}^N)=\nabla^N\left(\frac{\hbar}{m_0}\,\frac{1}{\Psi}\,
\frac{\partial \Psi}{\partial t}\right)\,.
\end{eqnarray*}
If we here insert Eqs.(\ref{eqn:time_derivative_u2}) and(\ref{eqn:3NNewton_modified}) for $\frac{\partial\,{\un v}^{N}}{\partial\,t}$ and $\frac{\partial\,{\un u}^{N}}{\partial\,t}$ and proceed exactly as in the single-particle case we arrive at the $N$-particle Schr\"odinger equation 
\begin{eqnarray}
\label{eqn:zeitabhSchroedingerN}
\underbrace{\left[\widehat{H}_0+\frac{1}{2}\sum_{\stackrel{i,j}{i\not= j}}V(|{\un r}_j-{\un
r_i}|)\right]}_{=\widehat {H}}\,\Psi({\un r}_1, {\un r}_2, \dots {\un r_N},
t)= \nonumber \\
i\,\hbar\,\frac{\partial}{\partial\,t}\Psi({\un r}_1, {\un r}_2, \dots {\un r_N}, t)\,.
\end{eqnarray}
Here $\widehat{H}_0$ denotes the ``free Hamiltonian''
\begin{eqnarray}
\label{eqn:HamiltonN}
\widehat{H}_0=\sum_{j=1}^{N}\widehat{H}_j \quad \mbox{where} \quad \widehat{H}_j=\left[\frac{\widehat{{\un
p}}_j^{2}}{2\,m_0}+V_{ext.}({\un r}_j)\right].
\end{eqnarray}
Because of Eq.(\ref{eqn:3NTot_velocity}) the phase of the wave function may still depend on time when $\rho$ and ${\un v}^N$ are time-independent:
$$
\phi({\un r}^N,t)=\phi_0({\un r}^N)+f(t)\,.
$$
Thus we have in this case
\begin{eqnarray*}
\Psi({\un r}^N,t)=\Psi_0({\un r}^N)\,e^{-if(t)}\;;\\
\Psi_0({\un r}^N)=\pm \sqrt{\rho({\un r}^N)}\,\exp\,[i\,\phi_0({\un r}^N)] 
\end{eqnarray*}
which on insertion into Eq.(\ref{eqn:zeitabhSchroedingerN}) yields
\begin{eqnarray*}
\widehat{H}\,\Psi_0({\un r}^N)=\hbar\,\dot{f}\,\Psi_0({\un r}^N)\quad \hookrightarrow \quad \hbar\,\dot{f}=const.=E\\
\hookrightarrow f(t)=\frac{E}{\hbar}\,t\,,
\end{eqnarray*}
whereby Eq.(\ref{eqn:zeitabhSchroedingerN}) becomes the time-independent Schr\"odinger equation
\begin{eqnarray}
\label{eqn:zeitunabhSchroedingerN}
\widehat{H}\,\Psi_0({\un r}^N)=E\,\Psi_0({\un r}^N)\,.
\end{eqnarray}

\section{States of identical particles and entanglement} \label{kapitel14}

If the particles are non-interacting, one would na\"ively expect their motions to be completely uncorrelated which means
\begin{eqnarray}
\label{eqn:DensityProduct}
\rho({\un r}_1,{\un r}_2, \ldots {\un r}_N,t)=\prod_{j=1}^{N} \rho_j({\un r_j},t)\,,
\end{eqnarray}
and
\begin{eqnarray}
\label{eqn:PhaseSum}
\phi({\un r}_1,{\un r}_2, \ldots {\un r}_N,t)=\sum_{j=1}^{N} \varphi_j({\un r_j},t)\,.
\end{eqnarray}
In that case Eq.(\ref{eqn:3NTot_velocity}) attains the form
\begin{eqnarray*}
({\un v}_1({\un r}_1,t),{\un v}_2({\un r}_2,t), \ldots {\un v}_N({\un r}_N,t))=\qquad \qquad\qquad \qquad\qquad\qquad\qquad \\
\frac{\hbar}{m_0}\,\left(\nabla_{1}\varphi_1({\un r}_1,t),\nabla_{2}\varphi_2({\un r}_2,t),\ldots \nabla_{N}\varphi_N
({\un r}_N,t)\right)\,.\qquad\qquad\qquad \qquad \qquad  
\end{eqnarray*}
Likewise, Eq.(\ref{eqn:3NOsm_velocity}) becomes
\begin{eqnarray*}
({\un u}_1({\un r}_1,t),{\un u}_2({\un r}_2,t), \ldots {\un u}_N({\un r}_N,t))=\qquad\qquad\qquad\\
-\frac{\hbar}{2\,m_0}\,\left(\nabla_{1}\ln [\rho_1({\un r}_1,t)/\rho_{01}],\right.
\end{eqnarray*}
$\left.\nabla_{2}\ln [\rho_2({\un r}_2,t)/\rho_{02}],\ldots \nabla_{N}
[\ln \rho_N({\un r}_N,t)/\rho_{0N}]\right)\,.$
\\[0.2cm]
Newton's modified second law (\ref{eqn:3NNewton_modified}) decomposes accordingly into $N$ analogous equations for single particles, as has to be expected. Each of these equations can be subjected to a Madelung transform which yields time-dependent one-particle Schr\"odinger equations solved by one-particle wave functions $\psi_j({\un r}_j,t)$. 
If one multiplies
\begin{eqnarray}
\label{eqn:SingleParticleEq}
\widehat{H}_j({\un r}_j)\,\psi_j({\un r}_j,t)=i\hbar\,\frac{\partial}{\partial\,t}\,\psi_j({\un r}_j,t)
\end{eqnarray}
by $\prod_{\stackrel{i=1}{i\not =j}}^{N}\psi_i({\un r}_i,t)$ one obtains
$$
\widehat{H}_j({\un r}_j)\,\prod_{i=1}^{N}\,\psi_i({\un r}_i,t)=\prod_{\stackrel{i=1}{i\not =j}}^{N}\psi_i
({\un r}_i,t)\,i\hbar\,\frac{\partial}{\partial\,t}\,\psi_j({\un r}_i,t)
$$
which on forming the sum $\sum_{j=1}^{N}$ yields, in fact,
\begin{eqnarray}
\label{eqn:Productwave function}
\widehat{H}_0\,\Psi({\un r}_1,{\un r}_2, \ldots {\un r}_N,t)=i\hbar\,\frac{\partial}{\partial\,t}\,\Psi({\un r}_1,{\un r}_2, \ldots {\un r}_N,t)\nonumber \\
\quad \mbox{where} \quad \Psi({\un r}_1,{\un r}_2, \ldots {\un r}_N,t)=\prod_{j=1}^{N}\,\psi_j({\un r}_j,t)\,. \qquad \quad \quad \,\,\,
\end{eqnarray}
Hence, the above time-dependent $N$-particle Schr\"odinger equation is solved by the product of individually time-dependent  wave functions $\psi_j({\un r}_j,t)$. 
\\[0.2cm]
Obviously, the density (\ref{eqn:DensityProduct}) that results from this wave function is {\bf not} invariant against interchange of any two particles if they are in different states, say $\psi_{k_n}({\un r}_k,t)$ and $\psi_{l_m}({\un r}_l,t)$ where $k_n \not= l_m$. \\
It is not exactly physical wisdom but rather firm belief that even non-interacting massive particles, though non-existing in nature, do not perform an uncorrelated motion and can, therefore, not be described by the wave function (\ref{eqn:Productwave function}).  This belief is based on the idea that the particles cannot be tracked individually as they move (contrary to classical particles) because the uncertainty relation ``forbids'' the existence of trajectories. Our approach to the many-particle problem is characterized by the plausible assumption that each particle can be identified any time by an affix if it has been assigned to a certain number at some chosen instant since each particle follows an individual trajectory. The quantity ${\un v}_1({\un r}_1,{\un r}_2,\ldots {\un r}_N,t)$, for example, represents the average over all particle velocities at ${\un r}_1$ and time $t$ of the ensemble associated with particle ``number 1''. In forming this average the positions ${\un r}_2,{\un r}_3,\ldots {\un r}_N$ of the $N-1$ remaining particles are kept fixed, that is, the average results from the entire set of ``number 1''-trajectories that occur in the ``number 1''-ensemble while the particles  ``number 2, 3 ...$N$'' are at fixed positions. Clearly, if one or more of those particles are kept at different positions and if all particles interact, ${\un v}_1$ will in general be different at ${\un r}_1$ and time $t$. Hence in our view there is no extra quantum phenomenon of indiscernibility. As in classical mechanics it is entirely sufficient to characterize identical particles merely by their property of having the same mass and charge.    
\\[0.2cm]
If one insists, however, on ``quantum indiscernibility'' also for non-interacting particles, that is, on the invariance of $\rho({\un r}_1,{\un r}_2, \ldots {\un r}_N)$ against interchange of any two particles, one has to replace (\ref{eqn:Productwave function}) with a renormalized linear combination of all $N!$ products that differ in the interchange  of two particles
\begin{eqnarray}
\label{eqn:LinCombProductwave function}
\Psi({\un r}_1,{\un r}_2, \ldots {\un r}_N)=\qquad\qquad\qquad\qquad\qquad\nonumber\\
\frac{1}{\sqrt{N!}}\sum_{P=1}^{N!} (\pm 1)^{P}\hat{P}(k,l)\prod_{j=1}^{N}\,\psi_{n_j}({\un r}_j)\,. 
\end{eqnarray}
where $\hat{P}(k,l)$ is the permutation operator exchanging the particle referring to $j=k$ with that for $j=l$, and $P$ numbers the permutations. 
\\[0.2cm]
{\bf If the particles interact} and are bound in an external potential or move in a parallelepiped where $\Psi({\un r}_1,{\un r}_2, \ldots {\un r}_N)$ is subjected to periodic boundary conditions, each particle is constantly scattered, and hence the probability of some particle, say ``number k'', being within an elementary volume $\Delta^3r$ around ${\un r}$ is given by:
\begin{eqnarray*}
P({\un r})=\int|\Psi({\un r}_1,\ldots {\un r}_{k-1},{\un r},{\un r}_{k+1}\,\ldots {\un r}_N)|^2\,d^3r_1\ldots\\ d^3r_{k-1}d^3r_{k+1}\ldots d^3r_N\,.
\end{eqnarray*}
{\bf Indiscernibility means} that $P({\un r})$ is the same for any particle one picks, that is, each particle appears at ${\un r}$ with the same probability. Hence we have
$$
\rho({\un r})=N\,P({\un r})
$$ 
with $\rho({\un r})\,\Delta^3r$ denoting the probability of {\bf any} of the $N$ electrons being in $\,\Delta^3r$.\\
The function $\rho({\un r}_1,{\un r}_2, \ldots {\un r}_N)$ is now naturally invariant against interchange of any two particles. \\
An important property of particles is their spin which will be discussed farther below in this article. In the present context it may be sufficient to introduce 
$$
{\un x}=({\un r},\sigma)
$$
as a generalized particle coordinate where $\sigma=\pm1$ denotes its discrete spin coordinate and refers to parallel or anti-parallel orientation with respect to a global axis. The wave function (\ref{eqn:LinCombProductwave function}) for non-interacting particles then takes the form
\begin{eqnarray}
\label{eqn:LinCombProductwave function2}
\Psi({\un x}_1,{\un x}_2, \ldots {\un x}_N)=\qquad\qquad\qquad\qquad\qquad\nonumber\\
\frac{1}{\sqrt{N!}}\sum_{P=1}^{N!} (\pm 1)^{P}\hat{P}(k,l)\prod_{j=1}^{N}\,\psi_{n_j}({\un x}_j)\,. 
\end{eqnarray}
The alternative in the sign under the sum is related to the two fundamentally different species of particles: The plus sign in $(\pm 1)^{P}$ characterizes bosons, the minus sign fermions. Hence the latter are associated with a wave function that changes sign on interchanging any two particles. This property persists when $\Psi({\un x}_1,{\un x}_2, \ldots {\un x}_N)$  describes $N$ interacting fermions. Antisymmetry of the wave function gives rise to a peculiar behavior of the so-called pair-density
\begin{eqnarray*}
\rho_2({\un x},{\un x}')\stackrel{\mbox{{\tiny Def}}}{=}\qquad\qquad\qquad\qquad\qquad\\
N(N-1)\,\int|\Psi({\un x},{\un x}',{\un x}_3,\ldots{\un x}_N)|^2\,d^4x_3\ldots d^4x_N
\end{eqnarray*}
where 
$$\int \ldots d^4x=\sum_{\sigma} \ldots d^3r\,.
$$
Obviously
\begin{eqnarray*}
\Psi({\un x}_1,\ldots {\un x}_{\nu},{\un x}_{\nu+1},\ldots{\un x}_N)\equiv\Psi({\un x}_1,\ldots {\un x}_{\nu+1},{\un x}_{\nu},\ldots{\un x}_N) \\
\mbox{if} \quad {\un x}_{\nu}={\un x}_{\nu+1}\,.
\end{eqnarray*}
On the other hand, $\Psi$ is required to change sign on interchanging two particles, and hence the above equation can only hold if $\Psi$ equals zero if the coordinates of any two particles are equal. Thus
$$
\rho_2({\un x},{\un x}')=0\quad \mbox{if} \quad {\un x}'={\un x}\,.
$$
This indicates the occurrence of the so-called Fermi-hole which is absent in bose-particle systems.\\
The form of the wave function (\ref{eqn:LinCombProductwave function2}) may be cast as a determinant, named after J.~C.~Slater. In so-called EPRB-experiments (EPRB=Einstein, Podolsky, Rosen \cite{Einstein1}, Bohm \cite{Bohm}) which were originally devised to test possible correlations between two macroscopically distant fermions in a singlet state, the associated wave function is just a $2\times 2$ determinant. The respective two one-particle states are in this context commonly referred to as ``entangled states''.   
\\[0.2cm]
The requirement of antisymmetry, which is equivalent to the Pauli exclusion principle, is a strong subsidiary condition in solving the Schr\"odinger equation (\ref{eqn:zeitunabhSchroedingerN}). Wave functions associated with fermions constitute only a small subset of the set of functions that satisfy the Schr\"odinger equation (\ref{eqn:zeitunabhSchroedingerN}). 
\\[0.2cm]
It should clearly be stated that the antisymmetry of the wave function is definitely not a consequence of our stochastic approach, but rather has to be required as an additional property, as in standard quantum mechanics.
\\[0.2cm]
The derivation of the time-dependent Schr\"odinger equation (\ref{eqn:zeitabhSchroedingerN}) can again be extended to the case where the particles move in an electromagnetic field. The external potential becomes time-dependent then ($V_{ext.}({\un r})\rightarrow V({\un r},t)$) and $\widehat{{\un p}}_j$ has to replaced with $\widehat{{\un P}}_j({\un r},t)=\widehat{{\un p}}_j-e\,{\un A}({\un r}_j,t)$. 

\section{A borderline case of entanglement} \label{kapitel15}

We consider a hydrogen molecule whose nuclei are located at ${\un R}_A$ and ${\un R}_B$, respectively. The Hamiltonian of the two electrons is given by
\begin{eqnarray}
\label{eqn:Hamilton_Operator}
\hat{H} =\sum_{k=1}^{2}
\left[\frac{(-i\hbar\,\nabla_{k}-e{\un A}({\un r}_k,t))^2}{2\,m_0}+V(\un{r}_k)\right]\nonumber\\
+\frac{e^2}{4\pi\,\epsilon_0}\sum_{k,l\neq k}\frac{1}{|\un{r}_{k}-\un{r}_{l}|}
\end{eqnarray}
where
\begin{eqnarray*}
\label{eqn:Potential}
V({\un r})=-\frac{e^2}{4\pi\,\epsilon_0|{\un r}-{\un R}_A|}-\frac{e^2}{4\pi\,\epsilon_0|{\un r}-{\un R}_B|}\,, 
\end{eqnarray*}
and $\epsilon_0$ denotes the electric constant.\\
We first assume that there is no external  field (${\un A}({\un r},t)\equiv 0$) and that the 2-electron wave function has for large proton-proton separation, that is when $R_{AB}=|{\un R}_A-{\un R}_B|\gg$Bohr radius, still  the entangled form of a {\bf singlet state} dictated by the Pauli principle
\begin{eqnarray}
\label{eqn:Verschraenkung}
\Psi({\un r}_1,{\un r}_2)=\qquad\qquad\qquad\nonumber\\
\frac{1}{\sqrt 2}\,[\underline{\psi}
({\un r}_{1}\,,\uparrow)\otimes\underline{\psi}({\un r}_2\,,\downarrow)-\underline{\psi}
({\un r}_{2}\,,\downarrow)\otimes\underline{\psi}({\un r}_1\,,\uparrow)]
\end{eqnarray}
where
\begin{eqnarray}
\label{eqn:Orbitalform}
\underline{\psi}({\un r},\sigma)=[a_{\sigma}(R_{AB})\,\varphi_A({\un r})+b_{\sigma}(R_{AB})\,\varphi_B({\un r})]\,\underline{\chi}(\sigma)\;
\end{eqnarray}
and 
$$
\sigma=\uparrow(\downarrow)\,;\quad a_{\sigma}^2+b_{\sigma}^2=1\,;\quad\varphi_{A/B}({\un r})=\varphi({\un r}-{\un R}_{A/B})
$$
with the property 
\begin{eqnarray*}
\int |\varphi({\un r}-{\un R}_{A/B})|^2\,d^3r=1\,. 
\end{eqnarray*}
Here the integrand denotes the electronic 1s-orbital of a single hydrogen atom, and $\rho_{A/B}({\un r},t)=|\varphi({\un r}-{\un R}_{A/B})|^2$ is the associated probability density. Furthermore, the unit spinors $\underline{\chi}(\sigma)$ have the property
$$
\underline{\chi}^{\dagger}(\sigma')\,\underline{\chi}(\sigma)=\delta_{\sigma'\sigma}\,.
$$
Under the supposition that $R_{AB}=|{\un R}_A-{\un R}_B|$ is sufficiently large, say 10$\,$cm or even larger, the expectation value $\langle \widehat{H}\rangle$ attains a minimum for either 
$$
(a_{\uparrow}\rightarrow 1\,, a_{\downarrow}\rightarrow 0) \hookrightarrow (b_{\uparrow}\rightarrow 0\,, b_{\downarrow}\rightarrow 1)\quad \mbox{``case l''}
$$
or
$$
(a_{\uparrow}\rightarrow 0\,, a_{\downarrow}\rightarrow 1) \hookrightarrow (b_{\uparrow}\rightarrow 1\,, b_{\downarrow}\rightarrow 0)\quad \mbox{``case r''}\,.
$$
For both cases $\langle \widehat{H}\rangle$ yields the correct value, viz. -2$\,$Ryd, as has to be expected for two hydrogen atoms, each of which possesses the energy -1$\,$Ryd. Since  the spin structure does not reflect the symmetry of the potential, one forms a symmetry-adapted linear combination
$$
\Psi_i({\un r}_1,{\un r}_2)=\frac{1}{\sqrt{2}}\,[\Psi_l({\un r}_1,{\un r}_2)+\Psi_r({\un r}_1,{\un r}_2)]\,,
$$
where
$$
\Psi_l({\un r}_1,{\un r}_2)=\frac{1}{\sqrt{2}} \left|\begin{array}{ll}
\varphi_A({\un r}_1)\,\underline{\chi}(\uparrow) & \varphi_A({\un r}_2)\,\underline{\chi}(\uparrow)\\
\varphi_B({\un r}_1)\,\underline{\chi}(\downarrow) & \varphi_B({\un r}_2)\,\underline{\chi}(\downarrow)
\end{array} \right|
$$
and
$$
\quad  \Psi_r({\un r}_1,{\un r}_2)=\frac{1}{\sqrt{2}} \left|\begin{array}{ll}
\varphi_B({\un r}_1)\,\underline{\chi}(\uparrow) & \varphi_B({\un r}_2)\,\underline{\chi}(\uparrow)\\
\varphi_A({\un r}_1)\,\underline{\chi}(\downarrow) & \varphi_A({\un r}_2)\,\underline{\chi}(\downarrow)
\end{array} \right|\,.
$$
The two 2-electron functions are associated with the same energy which hence applies to $\Psi_i({\un r}_1,{\un r}_2)$ as well. As a consequence of the symmetry of $\Psi_i({\un r}_1,{\un r}_2)$ in ${\un r}_1$ and ${\un r}_2$ we have
\begin{eqnarray*}
\rho({\un r}_1)=\int|\Psi({\un r}_1,{\un r}_2)|^2\,d^3r_2= \rho_A({\un r}_1)+\rho_B({\un r}_1) 
\end{eqnarray*}
and
$$
\rho({\un r}_2)=\int|\Psi({\un r}_1,{\un r}_2)|^2\,d^3r_1= \rho_A({\un r}_2)+\rho_B({\un r}_2)\,.
$$
Moreover
\begin{eqnarray}
\label{eqn:DensityIntegrals}
\int \rho_{A/B}({\un r}_{1/2})\,d^3r_{1/2}=\frac{1}{2}\quad\mbox{and} \nonumber\\
\int\rho_{A/B}({\un r}_1)\,d^3r_1+\int\rho_{A/B}({\un r}_2)\,d^3r_2=1\,.
\end{eqnarray}
That means that each electron appears in each of the atoms (A and B) with the same probability. This has rather implausible consequences if one exposes, for example, {\bf one} of the atoms (say A) to a Laser puls of frequency $\omega$. Now ${\un A}({\un r},t)$ is no longer zero. The associate perturbation operator has the form
\begin{eqnarray*}
\label{eqn:Stoeroperator}
 V_{perturb}({\un r}_1,{\un r}_2,t)=\qquad \qquad\qquad \qquad\nonumber \\
\left\{\begin{array}{r@{\quad \;}l}
\frac{ie\hbar}{m_0}\,\sum_{k=1}^{2}{\un A}({\un r}_k,t)\cdot \nabla_k  & \mbox{if}\; {\un r}_1, {\un r_2}\; \mbox{in or near atom A} \\
0 \qquad \qquad & \qquad \qquad\mbox{else}
\end{array} \right. 
\end{eqnarray*}
which promotes the 2-electron system to an excited state
$$
\Psi_f({\un r}_1,{\un r}_2)=\frac{1}{\sqrt{2}}\,[\Psi^{(f)}_l({\un r}_1,{\un r}_2)+\Psi^{(f)}_r({\un r}_1,{\un r}_2)]
$$
where
$$
\Psi^{(f)}_l({\un r}_1,{\un r}_2)=\frac{1}{\sqrt{2}} \left|\begin{array}{rr}
\varphi^{(f)}_A({\un r}_1)\,\underline{\chi}(\uparrow) & \varphi^{(f)}_A({\un r}_2)\,\underline{\chi}(\uparrow)\\
\varphi_B({\un r}_1)\,\underline{\chi}(\downarrow) & \varphi_B({\un r}_2)\,\underline{\chi}(\downarrow)
\end{array} \right|
$$
and
$$
\Psi^{(f)}_r({\un r}_1,{\un r}_2)=\frac{1}{\sqrt{2}} \left|\begin{array}{rr}
\varphi^{(f)}_B({\un r}_1)\,\underline{\chi}(\uparrow) & \varphi^{(f)}_B({\un r}_2)\,\underline{\chi}(\uparrow)\\
\varphi_A({\un r}_1)\,\underline{\chi}(\downarrow) & \varphi_A({\un r}_2)\,\underline{\chi}(\downarrow)
\end{array} \right|\,.
$$
Here $\varphi^{(f)}_{A/B}({\un r})$ describes an outgoing wave which has in principle the asymptotic form
$$
\varphi^{(f)}_{A/B}({\un r})\cong \frac{1}{r_{A/B}}\,e^{ik\,r_{A/B}}\,Y_{10}(\widehat{r}_{A/B})\;;\quad r_{AB}\equiv |{\un r}-{\un R}_{A/B}|
$$
with $Y_{10}(\widehat{r}_{A/B})$ denoting the spherical harmonic for $l=1,m=0$, and $k$ is given by $\hbar^2\,k^2/2\,m_0=-1\,\mbox{Ryd}+\hbar\,\omega$. We have assumed linearly polarized Laser light with the quantization axis of $Y_{10}(\widehat{r}_{A/B})$ coinciding with the axis of polarization. Moreover we have disregarded the residual charge left with each atom as part of the electronic charge is emitted.\\
Although only the illuminated volume of atom A can contribute to the transition matrix element 
\begin{eqnarray*}
M_{fi}= \qquad \qquad \qquad \qquad\qquad \qquad\\
\int_{atom\,A}\int \Psi^{*}_f({\un r}_1,{\un r}_2)\,V_{perturb}({\un r}_1,{\un r}_2)\,\Psi_i({\un r}_1,{\un r}_2)\,d^3r_1\,d^3r_2  
\end{eqnarray*}
the final state $\Psi_f({\un r}_1,{\un r}_2)$ yields a current density 
\begin{eqnarray*}
{\un j}({\un r})=\frac{\hbar}{i\,m_0}\,\int[\Psi^{*}_f({\un r},{\un r}_2)\nabla \Psi_f({\un r},{\un r}_2)-c.c.]\,d^3r_2=\\
\frac{\hbar}{i\,m_0}\,\int[\Psi^{*}_f({\un r_1},{\un r})\nabla \Psi_f({\un r_1},{\un r})-c.c.]\,d^3r_1=\\
{\un j}_A({\un r}-{\un R}_A)+
{\un j}_B({\un r}-{\un R}_B)
\end{eqnarray*}
where ${\un j}_{A/B}$ is associated with $\varphi^{(f)}_{A/B}$, and hence ${\un j}({\un r})$ contains also a photo emission current coming from the non-illuminated atom B at a distance of 10$\,$cm away from A. Similar considerations apply if one excites the molecule to a bound state which would spontaneously decay back then to the ground-state by emitting fluorescent light. If one repeats the excitation sufficiently often one would obtain as many fluorescence photons coming from the illuminated atom as from the non-illuminated one. There is no experimental evidence that anything like that could ever happen. \\
We are hence led to conclude that the concept of entanglement (i.$\,$e. the Pauli exclusion principle when dealing with fermions) does not apply anymore if the atoms are separated by a macroscopic distance. The reason may be tracked down to the definition (\ref{eqn:time_average_rho}) of the probability density $\rho({\un r})$ as the relative residence time that a particle spends in an elementary volume $\Delta^3r$ around ${\un r}$, provided it is bound in a potential and thus  occurs repeatedly in that volume. On pulling the two atoms of a H$_2$-molecule gradually apart one arrives at a situation where one of the two electrons remains captured near the nucleus of atom A for a while, and accordingly the second electron stays captured near the nucleus of atom B for the same time. The Coulomb repulsion between the two electrons effects a correlated separation of the two electrons into the two regions.\footnote{The possibility that both electrons accumulate in one of the atoms can safely be excluded. In such a case the other atom would be left ionized requiring an energy $\Delta E$ of about 1Ryd. Within a time $\Delta t$ that excess energy must disappear again where $\Delta t$ results from $\Delta E\,\Delta t\approx\hbar$. This yields $\Delta t\approx 5\cdot 10^{-17}\,s$, thus excluding the possibility for one of the electrons to go back to the ionized atom 10$\,$cm away at a speed well below light velocity.} If the inter-nuclear distance becomes large compared to the linear dimensions of the atoms, the time spans for tunneling of the ``A-electron'' (marked by the index ``1'') into the B-region and vice versa become enormously long compared to the time required to traverse the associated atom. The time $T$ for the photo-excitation process will therefore be many orders of magnitude shorter than the tunneling time. Given this situation, the definition (\ref{eqn:time_average_rho}) yields
$$
\rho({\un r}_{1/2})=\left\{\begin{array}{r@{\quad \quad}l}
\,\rho_{A/B}({\un r}_{1/2}) & \mbox{for}\; {\un r}_{1/2} \;\mbox{around nucleus A/B} \\
0 \qquad& \qquad \qquad \qquad\mbox{else}
\end{array} \right. 
$$
where - different from Eq.(\ref{eqn:DensityIntegrals}) - the densities $\rho_{A/B}({\un r}_{1/2})$ now integrate to unity. The two electrons do not appear entangled any more, and only the A-atom will now emit an electron under the exposure of light.

\section{Decomposing an experimental setup into the quantum system under study and a remainder. Schrödinger's cat} \label{kapitel15.1}

One of the puzzling credos of the Copenhagen interpretation of quantum mechanics consists in the conviction that an experimental setup for performing measurements on microscopic particles, has to be subdivided ``somehow'' into the particles under study and a remainder that functions as a classical system. This decomposition is known under the name ``Heisenberg-cut''. Yet from an unbiased point of view it appears to be self-evident that an experimental setup as a whole represents a many-particle system each part of which is subjected to the same laws of quantum mechanics as the particular portion that constitutes the object under study, an electron in a diffraction chamber, for example. We shall use this example to demonstrate the consistency of this standpoint, but we limit ourselves to considering a system that merely consists of just one specific apparatus plus a particle undergoing diffraction in it. The generalization to the inclusion of the entire environment is obvious from the ensuing considerations.\\
We assume that the system is made up of $N$ particles, a subset consisting of atomic nuclei which we number by a label $\alpha$, and $N_e$ electrons, one of which representing the single particle of interest, the ``test particle''. To keep the notation simple, we limit ourselves to considering only electrostatic particle interactions of the kind described by the many-body potential (\ref{eqn:manyparticle_potential}). If the test particle has left the cathode of the setup it is kept by electrodes, diaphragms and lenses at a macroscopic distance away from all kinds of surfaces it might strike and where it might get captured. Thus, the associated one-particle wave function $\psi_e({\un r},t)$ which describes the electron on its way through the apparatus to the screen or detector, has de facto zero overlap with the wave function of the $N-1$ remaining particles of the apparatus. Still, in standard setups it is intended that the particle hits a secluded portion of material on its way to the monitoring device, a diffracting single crystalline foil of metal, for example. But in the majority of cases the contact time is so short compared to the electronic excitation times of the material that the test electron cannot mingle with the other electrons. Further below we shall briefly discuss prominent exceptions.\\
Similar to the case of the H$_2$-molecule with macroscopically distant nuclei, one is justified then in assuming a factorization of the total wave function
\begin{eqnarray}
\label{eqn:factorization}
\Psi_N({\un r},{\un r}_2,\ldots {\un r}_N,t)=\qquad \qquad \qquad \qquad \qquad \nonumber\\
\psi_e({\un r},t)\,\Psi_{N-1}({\un r}_2,{\un r}_3,\ldots {\un r}_N,t)
\end{eqnarray}
where the spin coordinates have been suppressed for simplicity. If we insert this into the $N$-particle Schr\"odinger equation (\ref{eqn:zeitabhSchroedingerN}) we obtain
\begin{eqnarray}
\label{eqn:factorization_2}
\Psi_{N-1}\,i\hbar\,\dot{\psi_e}+\psi_e\,i\hbar\,\dot{\Psi}_{N-1}=\qquad \qquad \nonumber \\
\psi_e\,\widehat{H}_0^{(N-1)}\,\Psi_{N-1}+\Psi_{N-1}\,\widehat{H}_0^{e}\,\psi_e+V_{tot}\psi_e\,\Psi_{N-1}
\end{eqnarray}
where $V_{tot}$ denotes the total (Coulombic) interaction potential between all particles
\begin{eqnarray}
\label{eqn:Potentialdecomp}
V_{tot}=\frac{1}{2}\,\sum_{\stackrel{i,j}{i\not= j}}V(|{\un r}_j-{\un r_i}|)=\qquad \nonumber \\
V_{apparatus}({\un r}_2 \ldots 
{\un r}_N)+\sum_{j=2}^{N}\frac{e^2\,Z_j}{4\pi\,\epsilon_0\,|{\un r}_j-{\un r}|}
\end{eqnarray}
and from Eq.(\ref{eqn:HamiltonN})
\begin{eqnarray*}
\widehat{H}_0=\sum_{j=1}^{N}\left[\frac{\widehat{{\un
p}}_j^{2}}{2\,m_0}+V_{ext.}({\un r}_j)\right]=\qquad\qquad\\
\widehat{H}_{0}^{N-1}+\underbrace{\left[\frac{\widehat{{\un
p}}^2}{2\,m_0}+V_{ext.}({\un r})\right]}_{=\widehat{H}^{e}_{0}}\,.
\end{eqnarray*}
In Eq.(\ref{eqn:Potentialdecomp}) $|Z_j|$ stands for the number of elementary charges, i.$\,$e.
$$
Z_j=\left\{\begin{array}{r@{\quad \quad}l}
-Z_{\alpha}  & \mbox{if}\, j \,\,\mbox{runs over the}\,\alpha^{th}\, \mbox{nucleus}\\
\qquad 1 \,\,& \mbox{if}\,j\,\,\mbox{refers to an electron}\,. 
\end{array} \right. 
$$
If one multiplies Eq.(\ref{eqn:factorization_2}) by $\Psi_{N-1}^{*}$ and performs an integration with respect to ${\un r}_2,\ldots {\un r}_N$ one obtains
\begin{eqnarray*}
i\hbar\,\dot{\psi}_e+\psi_e\,\int\Psi_{N-1}^{*}\left[i\hbar\,\frac{\partial}
{\partial t}\Psi_{N-1}-\right.\qquad\\ \left.\left(\widehat{H}_0^{(N-1)}\,+V_{apparatus}\right)\,\Psi_{N-1}\right]\,d^3r_2\ldots d^3r_N=\\
\widehat{H}_0^{e}\,\psi_e+\widehat{V}_{e}\,\psi_e \qquad \qquad 
\end{eqnarray*}
where $\widehat{V}_e({\un r},t)$ represents a one-electron potential defined as
\begin{eqnarray}
\label{eqn:Effective_potential}
\widehat{V}_e({\un r},t)=\qquad \qquad \qquad\qquad \nonumber \\  \int\Psi_{N-1}^{*}\sum_{j=2}^{N}\frac{e^2\,Z_j}{4\pi\,\epsilon_0\,|{\un r}_j-{\un r}|}\,\Psi_{N-1}\,d^3r_2\ldots d^3r_N\,.
\end{eqnarray}
Since the bracketed expression under the integral in the above equation vanishes, we arrive at
$$
i\hbar\,\frac{\partial}{\partial t}\psi_e({\un r})=[\widehat{H}_0^{e}+\widehat{V}_e({\un r},t)]\,\psi_e({\un r})\,.
$$
Thus, the wave function of the electron under study obeys, in fact, a one-particle Schr\"odinger equation.\\
There are certain cases in which the contact time of the test particle is not short enough, and hence there is a non-vanishing probability that the particle mingles with those of the target. To get a rough picture of this situation, we describe the wave function instead of (\ref{eqn:factorization}) by
\\[0.2cm]
$\Psi_N({\un r},{\un r}_2,\ldots {\un r}_N,t)=
c_0(t)\,\psi_e({\un r},t)\,\Psi_{N-1}({\un r}_2,{\un r}_3,\ldots {\un r}_N,t)$
\\[-0.4cm]
\begin{eqnarray}
\label{eqn:mingled_wavefunction}
+c_1(t)\,\Psi^{capt}_N({\un r},{\un r}_2,\ldots {\un r}_N,t)\qquad \qquad 
\end{eqnarray}
where $c_0(t)$ and $c_1(t)$ are real-valued functions with the property $|c_0(t)|^2+|c_1(t)|^2=1$, in particular 
$$
c_0(t)=e^{-\frac{t}{2\tau}}
$$
and hence
\begin{eqnarray}
\label{eqn:decay_form}
\quad |c_0(t)|^2=e^{-\frac{t}{\tau}}\,; \quad |c_1(t)|^2=1-e^{-\frac{t}{\tau}}\,.
\end{eqnarray}
Here $\tau$ refers to a characteristic interaction time with the target, and $|c_1(t)|^2$  is the probability with which the test electron is captured by the target. Thereby it loses its identity as the ``test electron''. The latter effect is expressed by the property of $\Psi^{capt}_N$ being antisymmetric with respect to interchange of {\bf any} two particles out of the set of $N$ electrons.\\
Inserting $\Psi_N$ from Eq.(\ref{eqn:mingled_wavefunction}) into the Schr\"odinger equation (\ref{eqn:zeitabhSchroedingerN}) we obtain
\\[0.2cm]
$c_0(t)\left[\Psi_{N-1}\,i\hbar\,\dot{\psi_e}+\psi_e\,i\hbar\,\dot{\Psi}_{N-1}\right]+$
\\[0.2cm]   $i\hbar\,\dot{c}_0(t)\,\psi_e\Psi_{N-1}+c_1(t)\,i\hbar\,\dot{\Psi}^{capt}_{N}+i\hbar\,\dot{c}_1(t)\,\Psi^{capt}_{N}= $
\\[0.2cm]
$c_0(t)\left[\psi_e\,\widehat{H}_0^{(N-1)}\,\Psi_{N-1}+\Psi_{N-1}\,\widehat{H}_0^{e}\,\psi_e+V_{tot}\psi_e\,\Psi_{N-1}\right]$
\\[-0.4cm]
\begin{eqnarray}
\label{eqn:decay_form_Schrodinger_eq}
 +c_1(t)\,\widehat{H}\,\Psi^{capt}_N\,.\qquad\qquad \qquad\qquad \qquad\qquad 
\end{eqnarray}
The functions $\Psi_{N-1}$ and $\Psi^{capt}_N$ satisfy the associated time-dependent Schr\"odinger equations
\begin{eqnarray}
\label{eqn:Schroedinger_restequation}
i\hbar\,\dot{\Psi}_{N-1}=\left[\widehat{H}_0^{(N-1)}+V_{apparatus}\right]\,\Psi_{N-1}
\end{eqnarray}
and
$$
i\hbar\,\dot{\Psi}^{capt}_{N}=\widehat{H}\,\Psi^{capt}_{N}\,.
$$
If we insert this into Eq.(\ref{eqn:decay_form_Schrodinger_eq}), multiply the result in front by $\Psi^{*}_{N-1}$ and perform an integration over ${\un r}_2, {\un r}_3, \ldots {\un r}_N$, we obtain
\\[0.2cm]
$c_0(t)\,\left[i\hbar\,\frac{\partial}{\partial t}-\widehat{H}^{e}_0-V'_e({\un r},t)\right]\psi_e({\un r},t)+\qquad \qquad\qquad\qquad \qquad\qquad\qquad$
\\[-0.6cm]
\begin{eqnarray}
\label{eqn:effective_Schrodinger_eq}
i\hbar\,\dot{c}_1(t)\,\int \Psi^{*}_{N-1}({\un r}_2,\ldots {\un r}_N,t)\,\qquad\qquad\qquad\nonumber\\
\times\Psi^{capt}_{N}({\un r},{\un r}_2\,\ldots {\un r}_N,t)\,d^3r_2\,d^3r_3\ldots d^3r_N=0\,.
\end{eqnarray}
Here we have used $i\hbar\,\frac{\partial}{\partial t}\,c_0(t)=-i\,\frac{\hbar}{2\tau}\,c_0(t)$  and set 
$$
V'_e({\un r},t)=V_e({\un r},t)+i\,\tilde{V}_e \quad \mbox{where}\quad \frac{\tilde{V}_e}{\hbar}\equiv \frac{1}{2\tau}\,.
$$
The imaginary part of $V'_e({\un r},t)$ is commonly referred to as ``optical potential''. \\
According to our classification of the electron under study as either ``distinguishable'' or ``non-distinguishable'' the associated total  probability density $\rho({\un r},t)$ splits (almost quantitatively) into the ``either- and or-probability''
\begin{eqnarray*}
\rho({\un r},t)=\underbrace{|c_0(t)|^2\,|\psi_e({\un r},t)|^2}_{=\rho_0({\un r},t)}+\qquad\qquad \\
\underbrace{|c_1(t)|^2\,\int|\Psi^{capt}_N({\un r}, {\un r}_2, \ldots 
{\un r}_N,t)|^2\,d^{3}r_2, d^{3}r_3\ldots d^{3}r_N}_{\equiv \rho_{1}({\un r},t)}
\end{eqnarray*}
which means
\begin{eqnarray*}
S_e({\un r},t)=\int\Psi^{*}_{N-1}({\un r}_2,{\un r}_3, \ldots{\un r}_N) \qquad \qquad\\
\times\Psi^{capt}_{N}({\un r},{\un r}_2,{\un r}_3,\ldots{\un r}_N)\,d^{3}r_2, d^{3}r_3\ldots d^{3}r_N\approx0\, \forall\, {\un r}, t\,.
\end{eqnarray*}
It follows then from Eq.(\ref{eqn:effective_Schrodinger_eq}) that $\psi_e({\un r},t)$ solves the modified Schr\"odinger equation
\begin{eqnarray}
\label{eqn:Schroedinger_testparticle}
\left[i\hbar\,\frac{\partial}{\partial t}-\widehat{H}^{e}_0-V'_e({\un r},t)\right]\psi_e({\un r},t)=0\,,
\end{eqnarray} 
which describes situations one encounters, for example, in experiments on low energy electron diffraction (LEED) at surfaces of solids.\\
The mod squared of the actually not completely vanishing overlap
$$
S(t)=\int S_e({\un r},t)\,\psi^{*}_e({\un r},t)\,d^3r
$$
determines the transition probability $1/\tau$. Eq.(\ref{eqn:mingled_wavefunction}) and the resulting Eqs.(\ref{eqn:Schroedinger_restequation}) and (\ref{eqn:Schroedinger_testparticle}) are pivotal in describing generic quantum mechanical processes, a subset of which plays the role of ``measurements''\footnote{We side here emphatically with John Bell \cite {Bell1} who pleads in his article{\it ``Against Measurement''} for more common sense in describing what is actually happening: the time evolution of a particular {\bf experiment}.}. For example, when one is dealing with a setup where an electron traverses the legendary double slit diaphragm, defined by the potential (\ref{eqn:Effective_potential}), the function $\Psi^{capt}_N({\un r},{\un r}_2,\ldots {\un r}_N,t)$ describes the situation when the electron has been captured by the detector which is a part of the ``apparatus''. In spirit this in keeping with a statement by Hartle and Gell-Mann \cite{Hartle}: {\it ``In a theory of the whole thing there can be no fundamental division into observer and observed''} Our approach reflects even more directly the standpoint taken by v.~Kampen \cite{Kampen}: {\it ``The measuring act is fully described by the Schroedinger equation for object and apparatus together...''} 
\\[0.2cm]
It is the archetypal combination of a particular setup and ``pointer readings'' of a detector that enables the experimentalist to determine certain properties of the one-particle quantum system by extracting the sought-for information from the solution to the corresponding Schrödinger (or Pauli) equation that yields ${\un j}({\un r})=\rho({\un r},t)\,{\un v}({\un r},t)$ at ${\un r}_{detector}$. Eigenvalues of hermitian operators can only be obtained via this detour, and for fundamental, mostly experimental, reasons, only with limited accuracy.
\\[0.2cm]
The paradoxical situation which one runs into if one endows the ``observer'' (or ``measurer'') with an unrealistic meaning, is illustrated by Schrödinger's cat example \cite{Schroedinger1}: An alpha-particle emitted from some radioactive material triggers a device that kills a cat in a closed box by releasing a poisonous gas. Of course, the moment of radioactive decay does in no way depend on the particular setup. Our description of this process would be based on Eq.(\ref{eqn:mingled_wavefunction}) where $\Psi_N({\un r},{\un r}_2,\ldots {\un r}_N,t)$ on the left-hand side now represents the wave function $\Psi_{gas+cat}(t)$ of the system cat plus gas, $\psi_e({\un r},t)$ has to be replaced by an $N$-particle wave function $\psi_{poison}$ referring to the only weakly ``cat-overlapping'' molecules of the poisonous gas set free by the device, and $\Psi_{N-1}({\un r}_2,{\un r}_3,\ldots {\un r}_N,t)$ has to be identified with the many-particle wave function $\Psi_{cat}$ of the live cat. After the elapse of a time $\approx \tau$ the system's wave function $\Psi_{gas+cat}(t)$ has attained the form $\Psi_{capture}(t)$ where the poisonous molecules are now a part of the cat. It solves the time-dependent Schrödinger equation of the united system. The  time-evolution of $\Psi_{capture}(t)$ describes all the atomic (chemical) processes that eventually lead to the cat's death. It is this time-dependent process that is familiar from ab initio calculations on chemical reactions. The latter are completely ``self-controlled''. There is definitely no ``observer-induced'' influence. From this point of view it appears to be rather absurd that orthodox quantum mechanics interprets Eq.(\ref{eqn:mingled_wavefunction}) with the explained new meaning of the wave functions as a superposition of a ``live'' and a ``dead''-state of the cat, and only on opening the lid of the box by an observer, $\Psi_{gas+cat}$ collapses onto the wave function of a live or dead cat.

\section{The origin of particle spin}\label{kapitel16}

In 1925 Uhlenbeck and Goudsmit \cite{Uhlenbeck} suggested in a widely recognized paper that Pauli's idea \cite{Pauli} of a fourth quantum number in the description of electronic states of atoms might be associated with the rotation of an electron about its own axis thus giving rise to an extra angular moment. From the analysis of atomic spectra it was clear that the magnetic moment generated by such a rotation of the electron as a charged sphere had to exactly equal the Bohr magneton
$$
\mu_B=\frac{e\hbar}{2m_0}\,.
$$
There was also experimental evidence that the associated mechanical spin moment $\vec{S}$ - different from the atomic orbital momentum - would not obey the classical law of magneto-mechanical parallelism according to which $\mu_B$ should differ from $\vec{S}$ by a factor $\frac{e}{2m_0}$. In actual fact this factor had been found to be $\frac{e}{m_0}$ instead so that
$$
|\vec{S}|=\frac{\hbar}{2}\,.
$$
and hence
\begin{eqnarray}
\label{eqn:Definition_g}
\mu_B=g\,\frac{e}{2m_0}\,\frac{\hbar}{2}\,.
\end{eqnarray}
We ignore here and in the following the minute departure of $g$ from 2 due to quantum electrodynamical corrections.\\
The radius of the rotating electron sphere was identified with the classical electron radius $2.8\cdot 10^{-13}\,$cm. Lorentz immediately demonstrated to Uhlenbeck and Goudsmit that the electron mass would actually be larger than that of a proton if the magnetic moment of a Bohr magneton would be confined to that sphere. Moreover, the speed at the equator of the rotating sphere would by far exceed the velocity of light. Although these objections definitely disqualified the rotating sphere as a model of electron spin, it is still used, tacitly implied or appears concealed as ``intrinsic property'' in the analysis of most of the present-day experiments involving spin-orientation or spin flips.\\
If an ``eigen-rotation'' cannot explain the occurrence of a mechanical spin moment associated with a gyratory electronic motion, what else can be responsible for it? The following considerations are based on the idea that {\bf ``spin'' is not a property of the particle} but is rather a property of its quantum mechanical state.\\
Our description of particle motion as modified by stochastic vacuum forces makes it particularly suggestive to correlate - similar to the explanation of zero-point motion of oscillators - particle spin with the quivering motion that results from those forces and vanishes as $\hbar$ tends to zero. (This applies, of course, to all the other quantum mechanical ground-state properties as well.) To illustrate this we consider the simplest case of a hydrogen electron exposed to a magnetic field ${\un B}=B_Z\,{\un e}_z$ in its ground state $\psi_0({\un r})$. 
\begin{figure}[ht]    
 \hspace*{0.0cm} \epsfig{file=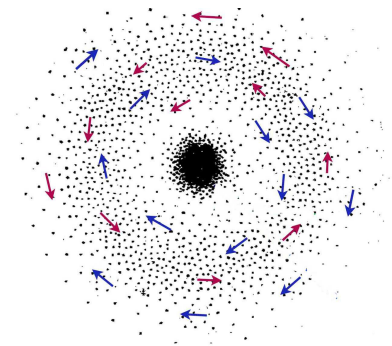,width=5cm}\vspace*{0.4cm}
\caption[SpinPfeile] {\label{SpinPfeile} Spin effective components of the quivering motion}
\end{figure}
In Fig.\ref{SpinPfeile} we show a schematic distribution of positions that the electron has successively taken at times $t_i$ and equal time intervals $t_{i+1}-t_i=\Delta t$  where $\Delta t$ is very small compared to $T$. This time span has been introduced in Section \ref{kapitel2} in connection with defining the probability density $\rho({\un r})$. The $z-$axis is thought to run through the atomic center perpendicular to the plotting plane. At each of the points the electron possesses a velocity which we decompose into a radial and a $z-$component, and in a component perpendicular to the $z-$axis. Only the latter are indicated by arrows. For symmetry reasons there will be as many positive as negative radial and $z-$components in the elementary volume around each point. They average out. We subdivide the set of arrows into two subsets associated with left-hand and right-hand circular motion, respectively. One might surmise that in the presence of the magnetic field one of the sets becomes empty and the other set now gives rise to a net circular current so as to minimize the total energy in that electronic state. As a defining property, the energy gain is proportional to $B_z$ and vanishes as $\hbar\rightarrow 0$. As we know from Section \ref{kapitel10} the magnetic field causes in the state $\psi_0({\un r})$ a current density ${\un j}({\un r})=\frac{e}{m_0}|\psi_0({\un r})|^2\,{\un A}({\un r})$, but the energy gain from this goes as $B_{z}^2$ because of $\Delta E=\int{\un j}\cdot{\un A}\,d^3r$ and ${\un B}=\nabla\times {\un A}$. The omission of the empty subset of arrows does not at all change the distribution of points defining $\psi_0({\un r})$. The energy gain is provided by the vacuum fluctuations in complete analogy to the zero-point energy of oscillators. \\
We thus arrive at the conclusion that the circular current which occurs on allowing the quivering motion of the electron to become asymmetric does not change the probability density which is characteristic of real-valued solutions to the Schr\"odinger equation. But it definitely yields a physical effect that has so far been outside our formal framework. In discussing certain properties of solutions to the Dirac equation Schr\"odinger \cite{Schroedinger} was led to a similar interpretation of particle spin and coined the irregular particle motion causing it ``Zitterbewegung''.\\
The additional spin-dependent interaction with a magnetic field occurs also in complex-valued states $\psi({\un r})=|\psi({\un r})|e^{i\varphi({\un r})}$. In that case there is an additional set of arrows superimposed on those shown in Fig.1. That set consists of arrows depicting ${\un v}({\un r}) =\frac{\hbar}{m_0}\,\nabla \varphi({\un r})$ at the various points distributed according to $|\psi({\un r})|^2$. Clearly, a linear superposition of those arrows is only possible as long as the velocities are within the non-relativistic regime. Otherwise the superposition is affected by spin-orbit coupling as a result of which $\varphi_{\uparrow}({\un r})$  and $\varphi_{\downarrow}({\un r})$ now become different. This point will be taken up again in Section \ref{kapitel20.0}.

\section{Generalizing one-particle quantum mechanics by including particle spin} \label{kapitel16}

Several suggestions have already been made to incorporate particle spin into a theory that is akin to the ideas of the present article (s. e.$\,$g. Dankel \cite{Dankel}, Dohrn et al. \cite{Dohrn}, Nelson \cite{Nelson}). We believe, however, that our approach offers - in the spirit of a statement by v.$\,$Weizs\"acker \cite{Weizsaecker}$^{,}$\footnote{``... What we are dissatisfied with is basically not that the old perceptions  
have failed but that they could not be substituted by something immediately comprehensible.''} -  ``something immediately comprehensible''.
\\[0.2cm]
The points associated with the two subset of arrows in Fig.1 define probability densities $\rho_{\uparrow}({\un r})$ and $\rho_{\downarrow}({\un r})$ with $\uparrow$ and $\downarrow$ referring to the respective direction of the spin moment. Both densities integrate to unity
\begin{eqnarray}
\label{eqn:spin_norm}
\int\rho_{\uparrow(\downarrow)}({\un r})\,d^3r=1\,.
\end{eqnarray}
To keep the formalism flexible at the outset we consider a situation where the total probability density is not yet a pure ``up'' or ``down'' density
\begin{eqnarray}
\label{eqn:spin_density_sum}
\rho({\un r})=|a|^2\,\rho_{\uparrow}({\un r})+|b|^2\,\rho_{\downarrow}({\un r})
\end{eqnarray}
with $a$ and $b$ denoting coefficients whose modulus squares sum up to unity
\begin{eqnarray}
\label{eqn:spin_coefficients}
|a|^2+|b|^2=1\,.
\end{eqnarray}
It is obviously not possible to partition the wave function analogously: $\psi({\un r})=a\,\psi_{\uparrow}({\un r}) +b\,\psi_{\downarrow}({\un r})$ because $\psi^*({\un r})\psi({\un r})$ would contain cross-terms. However, if one introduces a two-component spinor of the form 
\begin{eqnarray}
\label{eqn:spinor_definition}
\ul{\psi}({\un
r})=\left({a\,\psi_\uparrow({\un r}) \atop b\,\psi_\downarrow({\un r})}
\right)=a\,\psi_\uparrow({\un r}) \left( {1 \atop 0} \right)
+b\,\psi_\downarrow({\un r}) \left( {0 \atop 1} \right)
\end{eqnarray}
and its adjoint $\ul{\psi}^{\dagger}({\un
r})=\left(a^{*}\psi_{\uparrow}^{*}({\un r}),\,b^{*}\psi_{\downarrow}^{*} ({\un
r})\right)$ where
\begin{eqnarray}
\label{eqn:spinorfunction_norm}
\int |\psi_{\uparrow(\downarrow)}({\un r})|^2\,d^3r=1\,,
\end{eqnarray}
one obtains as intended
\begin{eqnarray*}
\ul{\psi}^{\dagger}({\un r})\ul{\psi}({\un r})=|a|^2\,|\psi_{\uparrow}({\un r})|^2+|b|^2\,|\psi_{\downarrow}({\un r})|^2=\rho({\un r})
\end{eqnarray*}
and
\begin{eqnarray}
\label{eqn:spinor_norm}
\int \ul{\psi}^{\dagger}({\un r})\ul{\psi}({\un r})\,d^3r=1\,.
\end{eqnarray}
We consider the Bohr magneton as known from the experiment. Thus, the energy densities of the interaction with the magnetic field for ``up``- and ``down''-spin may be cast as
$$
-\mu_{B}B_z\,|a|^2\,\psi_{\uparrow}^{*}({\un r})\,\psi_{\uparrow}({\un r})\quad
\mbox{and}\quad +\mu_{B}B_z\,|b|^2\,\psi_{\downarrow}^{*}({\un r})\,\psi_{\downarrow}({\un r})\,.
$$ from which the total interaction density results as
\begin{eqnarray}
\label{eqn:interaction_energy}
 u_{magn.}({\un r})=-\ul{\psi}^{\dagger}({\un r})\,\mu_{B}\ul{\ul{B}}\;\ul{\psi}({\un r})
\end{eqnarray}
where we have introduced a matrix
\begin{eqnarray}
\label{eqn:B_matrix_definition}
\ul{\ul{B}}=\left( \begin{array}{c c}
B_z & 0 \\
0 & -B_z
\end{array} \right)\,.
\end{eqnarray}
Likewise, we may cast the non spin-dependent energy density of the electron as
$$
\ul{\psi}^{\dagger}({\un r})\,\widehat{H}\;\ul{\psi}({\un r})
$$
where 
\begin{eqnarray}
\label{eqn:Definition_H}
\widehat{H}=\widehat{H}_0+V({\un r})\quad \mbox{and} \quad \widehat{H}_0=\frac{(\widehat{{\un p}}-e{\un A}({\un r}))^2}{2\,m_0}\,,
\end{eqnarray}
and with $V({\un r})$ denoting some potential in which the electron moves. 

\section{The time-dependent non-relativistic Pauli equation} \label{kapitel17}

The basic two constituents of our approach, viz.~$|\psi({\un r},t)|^2$ and $\nabla \varphi({\un r},t)$ remain unaffected by our incorporation of  spin. Hence, it is completely in line with the conceptual idea of our approach to assume that the two theorems of Ehrenfest stay unaffected as well. That means, according to Ehrenfest's Second Theorem
\begin{eqnarray}
\label{eqn:Ehrenfest_2}
\langle {\un v} \rangle=\frac{d}{d t}\,\langle{\un r}\,\rangle=\int \left[\dot{\ul{\psi}}^{\dagger}({\un r},t)\,{\un r}\,\ul{\psi}({\un r},t)\right.\qquad\qquad\nonumber\\
\left.+\ul{\psi}^{\dagger}({\un r},t)\,{\un r}\,
\dot{\ul{\psi}}({\un r},t)\right]\,d^3r\,,
\end{eqnarray}
and we have alternatively from Eq.(\ref{eqn:Expect_v_general}) 
\begin{eqnarray}
\label{eqn:Expect_v_general_2}
\langle {\un v}\rangle=\frac{1}{m_0}\,\int\ul{\psi}^{\dagger}({\un r},t)\widehat{{\un P}}\,\ul{\psi}({\un r},t)\,,
\end{eqnarray}
which holds without modification also for the spinors we have introduced.
Exploiting the relation
$$
[\widehat{H}_0\,{\un r}-{\un r}\,\widehat{H}_0]\,\ul{\psi}({\un r},t)=-i\,\frac{\hbar}{m_0}\,\widehat{{\un P}}\, \ul{\psi}({\un r},t)\,,
$$
which follows from simply applying the chain rule, we may combine Eqs.(\ref{eqn:Ehrenfest_2}) and (\ref{eqn:Expect_v_general_2}) to obtain
\begin{eqnarray}
\label{eqn:Zero_velocity}
\int\left(\left[\widehat{H}_0+i\hbar\,\frac{\partial}{\partial t}\right]\ul{\psi}^{\dagger}({\un r},t)\right)\,{\un r}\,\ul{\psi}({\un r},t)\,d^3r-\nonumber \\
\int\ul{\psi}^{\dagger}({\un r},t)\,{\un r}\,\left[\widehat{H}_{0}-i\,\hbar\frac{\partial}{\partial t}\right]\,\ul{\psi}({\un r},t)\,d^3r=0\,.
\end{eqnarray}
This equation holds for any $t$ if $\ul{\psi}({\un r},t)$ satisfies
\\[0.2cm]
$\left[\widehat{H}_0-i\,\hbar\frac{\partial}{\partial t}\right]\,\ul{\psi}({\un r},t)=-\ul{\ul{D}}\;\ul{\psi}({\un r}) -F({\un r},t)\,\ul{\psi}({\un r},t)\,,$
\\[-0.4cm]
\begin{eqnarray}
\label{eqn:Pauli_equation_with_F}
\qquad
\end{eqnarray}
and correspondingly
\begin{eqnarray*}
\left[\widehat{H}_0+i\,\hbar\frac{\partial}{\partial t}\right]\,\ul{\psi}^{\dagger}({\un r},t)=-\ul{\psi}^{\dagger}\,({\un r})\,\ul{\ul{D}} -\ul{\psi}^{\dagger}({\un r})\,F({\un r},t)\,,
\end{eqnarray*}
where $F({\un r},t)$ is some integrable real-valued function and $\ul{\ul{D}}$ denotes some unitary 2$\times$2-matrix that will be specified later to meet requirements of Ehrenfest's first theorem.   \\
The expectation value of the force exercised on an electron which moves in a potential $V({\un r})$ and simultaneously - through its magnetic moment - feels a force in a spatially varying magnetic field ${\un B}(z,t)=B_z(z,t)\,{\un e}_z$ may be cast as
\\[0.2cm]
$\langle{\un F}\rangle=-\int\ul{\psi}^{\dagger}({\un r},t)\,\left[\nabla \{V({\un r})+\mu_B\,\ul{\ul{B}}\}\right]\,\ul{\psi}({\un r},t)\,d^3r$
\\[-0.4cm]
\begin{eqnarray}
\label{eqn:ExpForce}
\qquad \qquad \qquad \qquad  -\langle e\dot{{\un A}}({\un r},t)\rangle\,.
\end{eqnarray}
The appearance of the induction-derived force $-\langle e\dot{{\un A}}({\un r},t)\rangle$ is a consequence of Eq.(\ref{eqn:E_Feld_Potential}).\\
We perform an integration by parts on the first integral and obtain
\\[0.2cm]
$\langle{\un F}\rangle=\int\left[\nabla\ul{\psi}^{\dagger}({\un r},t)\right]\{V({\un r})+\mu_B\,\ul{\ul{B}}\}\,
\ul{\psi}({\un r},t)\,d^3r+$
\\[0.2cm]
\hspace*{1.0cm}  $\int\ul{\psi}^{\dagger}({\un r},t)\,\{V({\un r})+\mu_B\,\ul{\ul{B}}\}\,\nabla\ul{\psi}({\un r},t)\,d^3r$
\\[-0.4cm]
\begin{eqnarray}
\label{Expect_value_force}
\qquad \qquad \qquad \qquad -\langle e\dot{{\un A}}({\un r},t)\rangle\,.
\end{eqnarray}
From Eq.(\ref{eqn:Expect_v_general}) we have
\begin{eqnarray*}
\langle \dot{{\un p}}\rangle=\frac{d}{d t}\int \ul{\psi}^{\dagger}({\un r},t)\,[-i\hbar\nabla -e\,{\un A}({\un r},t)]\,\ul{\psi}
({\un r},t)\,d^3r
\end{eqnarray*}
which we rewrite
\begin{eqnarray*}
\langle \dot{{\un p}}\rangle=\int \left(-i\hbar\frac{\partial}{\partial t}\,\ul{\psi}^{\dagger}\right)\nabla\ul{\psi}\,d^3r+\underbrace{\int\ul{\psi}^{\dagger}\,\nabla\left(-i\hbar\frac{\partial  }{\partial t}\,\ul{\psi}\right)\,d^3r}_{=-\int \nabla \ul{\psi}^{\dagger}\left(-i\hbar \frac{\partial}{\partial t}\ul{\psi}\right)\,d^3r}\\
-\int \ul{\psi}^{\dagger}\,\ul{\psi}\,e\,\dot{{\un A}}\,d^3r\,.
\end{eqnarray*}
On forming $\langle{\un F}\rangle-\langle\dot{\un p}\rangle=0$ (Ehrenfest's First Theorem) we obtain
\\[0.2cm]
\hspace*{0.5cm}$\int\left(\nabla \ul{\psi}^{\dagger}\right)\,\left\{V+\mu_B\,\ul{\ul {B}}-i\hbar\frac{\partial}{\partial t}\right\}\,\ul{\psi}\,d^3r+$
\\[0.2cm]
\hspace*{0.5cm}$\int\left[\left(+i\hbar\frac{\partial}{\partial t}\ul{\psi}^{\dagger}\right)\nabla \ul{\psi}+\ul{\psi}^{\dagger}\{V+\mu_B\,\ul{\ul{B}}\}\,\nabla\ul {\psi}\right]\,d^3r=0\,.$
\\[-0.4cm]
\begin{eqnarray}
\label{eqn:Zero_force}
\qquad
\end{eqnarray}
If we here eliminate the time-derivatives using Eqs.(\ref{eqn:Pauli_equation_with_F}) and identify $\ul{\ul{D}}$ with $\mu_B\,\ul{\ul{B}}$  this equation takes the form:
\\[0.2cm]
\hspace*{1.0cm}$-\int\left[(\nabla\ul{\psi}^{\dagger})\,\widehat{H}_0\,\ul{\psi}+(\widehat{H}_0\,\ul{\psi}^{\dagger})\nabla\ul{\psi}\right]\,d^3r+$ 
\\[0.2cm]
\hspace*{1.0cm}$\int\left[(\nabla\ul{\psi}^{\dagger}({\un r},t))\,\ul{\psi}({\un r},t)+\ul{\psi}^{\dagger}({\un r},t)\,\nabla\ul{\psi}({\un r},t)\right]$
\\[0.2cm]
\hspace*{4.0cm}$\times\{V({\un r})-F({\un r},t)\}\,d^3r=0\,.$
\\[0.2cm]
Since the first integral vanishes we arrive at 
\begin{eqnarray}
\label{eqn:Conclusion_time_dependence}
\int\nabla\rho({\un r},t)\{V({\un r})-F({\un r},t)\}\,d^3r=0 \quad \forall\, t\,.
\end{eqnarray}
We first consider the possibility that the expression in curly brackets does not vanish, but the integral does.
As we have emphasized in defining probability densities $\rho({\un r},t)$ and average velocities ${\un v}({\un r},t)$ through Eqs.(\ref{eqn:time_average_rho}) and (\ref{eqn:time-average_v}), non-stationary states require a certain sample-time $T$ of the particle under study to allow its time-derived probability density $\rho({\un r},t)$ to become quasi-stationary. Hence, if we introduce at $t=t_0$ a small perturbational potential $V({\un r})\rightarrow V({\un r})+\delta v({\un r},t)$ where $t_0\leq t\ll T$, the probability density $\rho({\un r},t)$, and thus its gradient remain practically unaffected, but the bracketed expression is now definitely different. We are hence led to conclude that Eq.(\ref{eqn:Conclusion_time_dependence}) can only be satisfied if $F({\un r},t)\equiv V({\un r})$ holds for any time. That means - because of Eq.(\ref{eqn:Pauli_equation_with_F}) - that the spinor function $\ul{\psi}({\un r},t)$ solves 
\begin{eqnarray}
\label{eqn:Pauli_equation_1}
\left[\widehat{H}_0+V({\un r})+\mu_B\,\ul{\ul{B}} \right]\,\ul{\psi}({\un r},t)\,=i\,\hbar\frac{\partial}{\partial t}\,\ul{\psi}({\un r},t)
\end{eqnarray}
with $\widehat{H}_0$ as defined in Eq.(\ref{eqn:Definition_H}). This constitutes the time-dependent non-relativistic Pauli equation.

\section{The Cayley-Klein parameters and Pauli spin matrices} \label{kapitel18}

We want to adapt Eq.(\ref{eqn:Pauli_equation_1}) to a situation where the direction of the magnetic field no longer coincides with the $z$-axis of the coordinate system. This can be achieved by exploiting a surprising alternative to the standard form of rotating the coordinate system by applying orthogonal 3$\times$3 matrices. The idea goes back to Felix Klein (S. Goldstein \cite{Goldstein}) and is related to earlier work of Cayley. He considers the rotation of the coordinate system ($x,y,z \rightarrow x',y',z'$) to be performed in three steps described by the Euler angles $\phi,\theta$ and $\psi$ shown in Fig.\ref{euler}.
\begin{figure}[ht]    
\hspace*{-1.0cm}\epsfig{file=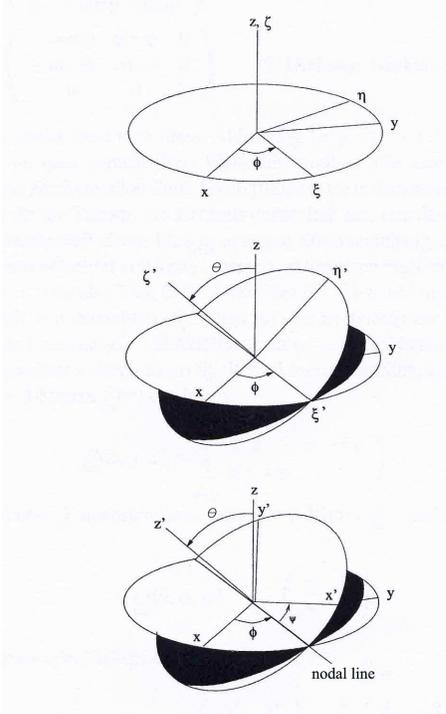,width=6cm}\vspace*{0.4cm}
\caption[Euler angles] {\label{euler} Euler angles}
\end{figure}
Instead of representing the position vector ${\un r}$ by a column matrix he uses a 2$\times$2-matrix $\ul{\ul{P}}({\un r})$ of the form
\begin{eqnarray}
\label{eqn:Klein_Ortsvektor}
\ul{\ul{P}}(x,y,z):=\left( \begin{array}{c c} z & x-i\,y \\
x+i\,y & -z \end{array} \right)\,. \label{eqn:Klein_x_y_z}
\end{eqnarray}
In place of the standard 3$\times$3-rotation matrix one now has a 2$\times$2-unimodular matrix 
\begin{eqnarray}
\label{eqn:coordinate_rotation}
 \ul{\ul{Q}}(\theta,\phi,\psi)&=&
\left( \begin{array}{c c} \alpha & \beta \\
\gamma & \delta \end{array} \right)
\end{eqnarray} 
whose elements - the so-called Cayley-Klein parameters - are connected to the Euler angles through
\begin{eqnarray}
\label{eqn:Matrixelemente}
\alpha &=& \; e^{\frac{i}{2}\;(\psi+\phi)}\,
\cos \frac{\theta}{2}
\nonumber \\
\beta &=& i e^{\frac{i}{2}\;(\psi-\phi)}\, \sin \frac{\theta}{2}
\nonumber \\
\gamma &=& i e^{-\frac{i}{2}(\psi-\phi)}\, \sin \frac{\theta}{2}
\nonumber \\
\delta &=& \;e^{-\frac{i}{2}(\psi+\phi)}\, \cos \frac{\theta}{2}\,.
\end{eqnarray} 
After the three steps of the rotation have been performed the original position vector ${\un r}=(x,y,z)$ is now associated with the new coordinates $x',y',z'$ that may be obtained from the transform
\begin{eqnarray}
\label{eqn:KleinTransformation}
\ul{\ul{Q}} \, \ul{\ul{P}} \, \ul{\ul{Q}}^{+} =
\ul{\ul{P}}'(x',y',z')=\left( \begin{array}{c c} z' & x'-i\,y' \\
x'+i\,y' & -z' \end{array} \right)
\end{eqnarray}
where $\ul{\ul{Q}}^{+}$ denotes the adjoint of $\ul{\ul{Q}}$, and we have
\begin{eqnarray}
\label{eqn:unitarydefinition}
\ul{\ul{Q}}^{+}\,\ul{\ul{Q}}=\ul{\ul{Q}}\,\ul{\ul{Q}}^{+}=\ul{\ul {1}}
\end{eqnarray}
The matrix $\ul{\ul{B}}$ had been defined in Eq.(\ref{eqn:B_matrix_definition}) as
\begin{eqnarray}
\label{eqn:matrixB_z_definition}
\ul{\ul{B}}= \left( \begin{array}{c c}
B_z & \;0 \\
0 & -B_z
\end{array} \right)=B_z\,\left( \begin{array}{c c}
1 & \;\;0 \\
0 & -1
\end{array} \right)\,.
\end{eqnarray} 
In Klein's representation the point ${\un r}=(0,0,z)$ attains the analogous form
$$
\ul{\ul{P}}({\un r})= z\,\left( \begin{array}{c c}
1 & \;\;0 \\
0 & -1
\end{array} \right)\,.
$$
Hence, $\ul{\ul{B}}$ has to be required to transform under coordinate rotation as $\ul{\ul{P}}$:
\begin{eqnarray}
\label{eqn:B_transform} 
\ul{\ul{B}}'=\ul{\ul{Q}} \,
\ul{\ul{B}} \, \ul{\ul{Q}}^+\,. 
\end{eqnarray}
For a general orientation of the coordinate system with respect to the magnetic field $\ul{\ul{B}}$ has the form analogous to $\ul{\ul{P}}$ in Eq.(\ref{eqn:Klein_Ortsvektor}), viz.
\begin{eqnarray}
\label{eqn:B_matrix_general}
\ul{\ul{B}}=\left( \begin{array}{c c} B_z & B_x-i\, B_y \\
B_x+i\, B_y & -B_z \end{array} \right)\,. 
\end{eqnarray}
This matrix can be decomposed 
\begin{eqnarray}
\label{eqn:ZerlegungMatrixB} 
\ul{\ul{B}}=B_x \,
\ul{\ul{\sigma}}_{\,x} + B_y \, \ul{\ul{\sigma}}_{\,y} + B_z \,
\ul{\ul{\sigma}}_{\,z}\,,
\end{eqnarray}
where the three matrices on the right-hand side are just the Pauli spin matrices
\\[0.2cm]
$\ul{\ul{\sigma}}_{\,x} = \left(
\begin{array}{c c} 0 & 1 \\ 1 & 0  \end{array} \right) \quad
\ul{\ul{\sigma}}_{\,y} = \left( \begin{array}{c c} 0 & -i \\ i & \quad 0
\end{array} \right) \quad \ul{\ul{\sigma}}_{\,z} = \left(
\begin{array}{c c} 1 & \; \;0 \\ 0 & -1  \end{array} \right)\,.$
\\[-0.4cm]
\begin{eqnarray}
\label{eqn:Pauli_Matrizen}
\qquad
\end{eqnarray} 
They are commonly lumped together in the form of a vector
\begin{eqnarray}
\label{eqn:Paulivektor}
\vec{\sigma}=\ul{\ul{\sigma}}_{\,x}\,{\un e}_x+ \ul{\ul{\sigma}}_{\,y}\,{\un
e}_y+\ul{\ul{\sigma}}_{\,z}\,{\un e}_z\,. 
\end{eqnarray} 
The matrix $\ul{\ul{B}}$ in Eq.(\ref{eqn:ZerlegungMatrixB}) may therefore be cast as
\begin{eqnarray}
\label{eqn:B_general_representation}
\ul{\ul{B}}=\vec{\sigma}\cdot {\un B}\,.
\end{eqnarray}
The Pauli equation (\ref{eqn:Pauli_equation_1}) then attains the familiar form
\begin{eqnarray}
\label{eqn:Pauli-equation_2}
\left[\widehat{H}_0+V({\un r})+\mu_{B}\,\vec{\sigma}\cdot
{\un B}\right]\ul{\psi}({\un r},t)=i\hbar\,\frac{\partial}{\partial t}\,\ul{\psi}({\un r},t)\,.
\end{eqnarray}
Actually, the spinor in this equation should be marked by a prime because it has changed under the transform as well. We have dropped the prime for simplicity. Since the density of the magnetic interaction energy is, of course, invariant under rotation of the coordinate system
\begin{eqnarray*}
\label{eqn:energy_density_transform}
u_{magn.}({\un r})=u'_{magn.}({\un r}')\,,
\end{eqnarray*}
it can be shown then that the new $\psi'$ is connected to the original $\psi$ through
\begin{eqnarray}
\label{eqn:SpinorTransformation} \ul{\psi}'&=&\ul{\ul{Q}} \, \ul{\psi}
\end{eqnarray}
and correspondingly
$$
\ul{\psi}'^{\dagger}\,=\,\left(\ul{\ul{Q}} \, \ul{\psi}\right)^{\dagger} =\ul{\psi}^{\dagger} \,
\ul{\ul{Q}}^+\,.
$$ 
This becomes obvious from forming
$$
\ul{\psi}'^{\dagger}\, \ul{\ul{B}}' \, \ul{\psi}'\; (=: u'_{mag}({\un r}',t)) =
\ul{\psi}^{\dagger}\, \ul{\ul{Q}}^+ \, \ul{\ul{B}}' \, \ul{\ul{Q}} \, \ul{\psi}\,.
$$
If we insert Eq.(\ref{eqn:B_transform}) on the right-hand side we obtain
$$
\ul{\psi}^{\dagger}\, \ul{\ul{Q}}^+ \, \ul{\ul{B}}' \, \ul{\ul{Q}} \, \ul{\psi}=\ul{\psi}^{\dagger}\, \underbrace{\ul{\ul{Q}}^+ \,\ul{\ul{Q}}}_{=\ul{\ul{1}}}\, \ul{\ul{B}} \,\underbrace{\ul{\ul{Q}}^+ \, \ul{\ul{Q}}}_{=\ul{\ul{1}}} \,\ul{\psi}=u_{mag}({\un r},t)\,.
$$
Because of
$$
\ul{\psi}'^{\dagger}\,\ul{\psi}'=\ul{\psi}^{\dagger}\, \ul{\ul{Q}}^+ \, \ul{\ul{Q}} \, \ul{\psi}=\ul{\psi}^{\dagger}\,\ul{\psi}=\rho({\un r},t)
$$
the probability density is also invariant under rotation of the coordinate system which is consistent with our idea of a spin-defining motional decomposition at the beginning of our considerations. Moreover, if the state of the particle in the original coordinate system has the form
\begin{eqnarray}
\label{eqn:oriented_spinors}
\psi_{\uparrow}({\un r})\,\left( {1 \atop 0} \right)\quad \mbox{or}\quad
\psi_{\downarrow}({\un r})\,\left( {0 \atop 1} \right)\,,
\end{eqnarray}
it becomes after coordinate rotation
\begin{eqnarray*}
\ul{\psi}_{\uparrow}'({\un r})=\psi_{\uparrow}({\un r})\,\ul{\ul{Q}}({\un
r})\left( {1 \atop 0} \right)=\qquad\qquad\qquad\qquad\qquad\qquad\\
\qquad\qquad\qquad\psi_{\uparrow}({\un r})\left[\alpha({\un
r})\,\left( {1 \atop 0} \right)-\beta^{*}({\un r})\,\left( {0 \atop 1}
\right)\right]
\end{eqnarray*}
or
\begin{eqnarray*}
\ul{\psi}_{\downarrow}'({\un r})=\psi_{\downarrow}({\un r})\,\ul{\ul{Q}}({\un
r})\left( {0 \atop 1} \right)=\qquad\qquad\qquad\qquad\qquad\qquad\\
\qquad\qquad\qquad\psi_{\downarrow}({\un r})\left[\beta({\un
r})\,\left( {0 \atop 1} \right)+\alpha^{*}({\un r})\,\left( {1 \atop 0}
\right)\right]\,,
\end{eqnarray*}
where $\alpha({\un r})$ and $\beta({\un r})$ are the Cayley-Klein parameters describing the rotation which we have allowed here to be different at different positions ${\un r}$.\\
The spin orientation with respect to the direction of a magnetic field is already uniquely defined by the two angles $\theta$ and $\phi$. Hence one is at liberty to choose $\psi$ at will without loss of generality. It is convenient to set $\psi=-\pi/2$. We consider the projection of the unit vector ${\un e}'_z$ onto the original $x/y$-plane where it makes an angle $\varphi$ with the $x$-axis. This angle and the Euler-angle $\phi$ are interrelated
$$
\phi=\varphi+\frac{\pi}{2}\,.
$$
If one inserts this relation into Eqs.(\ref{eqn:coordinate_rotation}) and (\ref{eqn:Matrixelemente}), $\ul{\ul{Q}}$ takes the familiar form
\\[0.2cm]
$$\ul{\ul{Q}}({\un r})=
\left( \begin{array}{c c} \exp[\frac{i}{2}\varphi({\un
r})]\,\cos\frac{\theta({\un r})}{2} & \exp[-\frac{i}{2}\varphi({\un
r})]\,\sin\frac{\theta({\un r})}{2}$$
\\[0.1cm]
$$-\exp[\frac{i}{2}\varphi({\un r})]\,\sin\frac{\theta({\un r})}{2}&
\exp[-\frac{i}{2}\varphi({\un r})]\,\cos\frac{\theta({\un r})}{2}\end{array}
\right)\,.$$
\\[-0.8cm]
\begin{eqnarray}
\label{eqn:Spindrehmatrix}
\qquad
\end{eqnarray}
The above considerations on the magnetic interaction energy starting with the expression (\ref{eqn:interaction_energy}) carry over to the spin momentum
\begin{eqnarray}
\label{eqn:Spinerwartungswert0}
\langle S_z\rangle=\frac{\hbar}{2}\,\int\left[|a|^2\,|\psi_{\uparrow}({\un
r})|^2-|b|^2\,|\psi_{\downarrow}({\un r})|^2\right]\,d^3r\,.
\end{eqnarray}
The symbol $S_z$ refers to the effective spin moment in the $z$-direction with respect to which the functions $\psi_{\uparrow(\downarrow)}({\un r})$ have been defined. In complete analogy to (\ref{eqn:matrixB_z_definition}) this expression can be compactified by introducing
\begin{eqnarray}
\label{eqn:DefS_z}
 \ul{\ul{S}}_{\,z}=\frac{\hbar}{2}\left( \begin{array}{c c}
1 &\quad 0 \\
0 & -1 
\end{array} \right)
\end{eqnarray}
so that
\begin{eqnarray}
\label{eqn:Spinerwartungswert1} \langle S_z\rangle=\int\ul{\psi}^{+}({\un
r})\,\ul{\ul{S}}_{\,z}\,\ul{\psi}({\un r})\,d^3r\,.
\end{eqnarray}
In case that the functions $\psi_{\uparrow(\downarrow)}({\un r})$ refer to a $z'$-direction that belongs to a rotated coordinate sytem $x',y',z'$, we have in analogy to Eq.(\ref{eqn:B_transform}) 
\begin{eqnarray*}
\ul{\ul{S}}_{\,z'}=\ul{\ul{Q}}\,\ul{\ul{S}}_{\,z}\,\ul{\ul{Q}}^+\,.
\end{eqnarray*}
If we use the analogous relations pertaining to Eqs.(\ref{eqn:B_matrix_general}) up to (\ref{eqn:B_general_representation}) we may cast $\ul{\ul{S}}_{\,z'}$ as 
\begin{eqnarray}
\label{eqn:S_zInKomponenten}
\ul{\ul{S}}_{\,z'}=\frac{\hbar}{2}\,[\hat{\alpha}_x\,\ul{\ul{\sigma}}_{\,x}+
\hat{\alpha}_y\,\ul{\ul{\sigma}}_{\,y}+ \hat{\alpha}_z\,\ul{\ul{\sigma}}_{\,z}]
\end{eqnarray}
with $\hat{\alpha}_x, \hat{\alpha}_y, \hat{\alpha}_z$ denoting the component of the unit vector ${\un e}_{z'}$ in the $z'$-direction
\begin{eqnarray*}
\hat{\alpha}_x=\cos\varphi\,\sin\theta\\
\hat{\alpha}_y=\sin\varphi\,\sin\theta\\
\hat{\alpha}_z=\cos\theta\,.\quad \;\;\,
\end{eqnarray*}
It is convenient to introduce a vector $\vec{S}$ (commonly referred to as ``spin operator''), which is analogous to $\vec{\sigma}$, by setting
\begin{eqnarray}
\label{eqn:Def_S} \vec{S}=\frac{\hbar}{2}\,\vec{\sigma}\,.
\end{eqnarray}
Eq.(\ref{eqn:S_zInKomponenten}) may then be cast
\begin{eqnarray*}
\ul{\ul{S}}_{\,z'}={\un e}_{z'} \cdot\vec{S}\,,
\end{eqnarray*}
and hence we have
\begin{eqnarray*}
\langle S_{z'}\rangle=\int\ul{\psi}^{+}({\un r})\,\ul{\ul{S}}_{\;z'}\,\ul{\psi}({\un r})\,d^3r={\un e}_{z'}\cdot\langle \vec{S}\rangle
\end{eqnarray*}
where
\begin{eqnarray}
\label{eqn:Spinrichtung} 
\langle \vec{S}\rangle=\int\ul{\psi}^{+}({\un
r})\,\vec{S}\,\ul{\psi}({\un r})\,d^3r\,.
\end{eqnarray} 
If $\ul{\psi}({\un r},t)$ has the form (\ref{eqn:oriented_spinors}), Eq.(\ref{eqn:Spinrichtung}) yields $\langle \vec{S}\rangle=\pm \frac{\hbar}{2}\,{\un e}_z$. On the other hand, if $\ul{\psi}({\un r},t)$ possesses two non-vanishing components, there will always be a coordinate system that is rotated with respect to the present one, in which $\langle \vec{S}\rangle$ becomes $\pm \frac{\hbar}{2}\,{\un e}_{z'}$. One only has to turn the pertinent $z'$-axis in the plane spanned by the original direction of $\langle \vec{S}\rangle$ and the original $z$-axis until ${\un e}_{z'}$ is parallel or anti-parallel to $\langle \vec{S}\rangle$.

\section{Spin precession in a magnetic field} \label{kapitel19}

So far we have assumed the magnetic field and the spin direction to be collinear. As an example for a non-collinear situation we consider an electron that is bound within an atom where it is initially exposed to a magnetic field along some direction. We omit here discussing the details of its spin alignment due to some minute time dependent perturbations and simply assume that it has eventually attained a stationary spinor state in which its spin momentum points parallel or anti-parallel to the direction of the magnetic field. If one now changes non-adiabatically the direction (and in general inevitably also the magnitude) of the magnetic field, the spin momentum can - without an appropriate external torque - not adjust to the new field direction, and hence the previously existing collinearity no longer obtains. As we shall show by discussing the pertinent solution to the time-dependent Pauli equation (\ref{eqn:Pauli-equation_2}), the spin momentum now precesses about the new direction of the magnetic field in a completely classical way.\\
We identify the initial direction of the magnetic field with the $z'$-axis of a ``primed'' coordinate system in which the spin-aligned state of the electron has the form
\begin{eqnarray}
\label{eqn:Praezession1} 
\ul{\psi}'({\un r}')=\psi'_{0}({\un r}')\left({1\atop
0}\right)
\end{eqnarray}
where $\psi'_{0}({\un r}')$ is the energetically lowest lying solution to the Schr\"odinger equation of the one-particle system under study. We denote this solution by $\psi_0({\un r})$ in the unprimed coordinate system in which the new magnetic field lies along the $z$-direction and in which the spinor (\ref{eqn:Praezession1}) can be cast as
\begin{eqnarray}
\label{eqn:Praezession2} \ul{\psi}({\un r})=\ul{\ul{Q}}^+\,\ul{\psi}'({\un
r}')=\underbrace{\psi_0({\un r})\,e^{-i\frac{\varphi}{2}}\,\cos
\frac{\theta}{2}\,\left({1\atop 0}\right)}_{=\ul{\psi}_{0\uparrow}({\un
r})}+\qquad \qquad \nonumber \\
\underbrace{\psi_0({\un r})\,e^{i\frac{\varphi}{2}}\,\sin
\frac{\theta}{2}\,\left({0\atop 1}\right)}_{=\ul{\psi}_{0\downarrow}({\un
r})}\,,\qquad
\end{eqnarray}
where $\theta,\varphi,\psi(=0)$ are the Euler angles that refer to the interrelation $(x',y',z')\rightarrow(x,y,z))$. Hence we have
\begin{eqnarray}
\label{eqn:SummeSpinoren}
 \ul{\psi}({\un r})=\ul{\psi}_{0\uparrow}({\un r})+\ul{\psi}_{0\downarrow}({\un r})\,.
\end{eqnarray}
Note that the unit spinors in Eq.(\ref{eqn:Praezession2}) are now referenced to the new $z$-axis!\\ 
We now consider the Pauli equation (\ref{eqn:Pauli-equation_2}) for the time-independent case in the absence of a magnetic field in which case $\ul{\psi}_{0\uparrow(\downarrow)}({\un r})$ are independent degenerate solutions and $\psi_0({\un r})$ satisfies the associated Schr\"odinger equation
$$
\widehat{H}({\un r})\,\psi_0({\un r})=E_0\,\psi_0({\un r})\,.
$$
For $B_z\not=0$ the two spinors belong to different energies $E_{0\uparrow(\downarrow)}=E_0 \pm \mu_B\,B_z$ and 
their sum does not satisfy the time-independent Pauli equation any more. However
\begin{eqnarray}
\label{eqn:SummeZeitabhLoesungn1} \ul{\psi}({\un
r},t)=\ul{\psi}_{0\uparrow}({\un
r})\,e^{-\frac{i}{\hbar}E_{0\uparrow}\,t}+\ul{\psi}_{0\downarrow}({\un
r})\,e^{-\frac{i}{\hbar}E_{0\downarrow}\,t}
\end{eqnarray}
solves the time-dependent Pauli equation (\ref{eqn:Pauli_equation_1}) if we disregard effects of second and higher order in the magnetic field. We now insert the definitions of $\ul{\psi}_{0\uparrow(\downarrow)}({\un r})$ from above and obtain
\begin{eqnarray}
\label{eqn:SummeZeitabhLoesungn2} \ul{\psi}({\un r},t)=\psi_0({\un
r})\,\left[e^{-i\frac{(\varphi-\omega_{L}\,t)}{2}}\,\cos
\frac{\theta}{2}\,\left({1\atop
0}\right)+\right. \qquad \qquad \nonumber \\
\left. e^{i\frac{(\varphi-\omega_{L}\,t)}{2}}\,\sin
\frac{\theta}{2}\,\left({0\atop 1}\right)\right]e^{-\frac{i}{\hbar}E_{0}\,t}.\quad
\end{eqnarray}
Here we have made use of $E_{0\uparrow(\downarrow)}=E_0 \pm \mu_B\,B_z$ and introduced the frequency $\omega_L$ which is defined through
\begin{eqnarray}
\label{eqn:DefinitionOmega}
E_{0\uparrow}-E_{0\downarrow}=2\,\mu_BB_z=\hbar\omega_L\,.
\end{eqnarray}
In complete analogy to Eq.(\ref{eqn:Praezession2}) we form
$$
\ul{\psi}^{+}({\un r},t)=\ul{\psi}'^{+}({\un r}',t)\,\ul{\ul{Q}}
$$
and calculate the expectation value of $\vec{S}$
\\[0.2cm]
\hspace*{1.0cm}$\langle \vec{S}\rangle =\int\ul{\psi}^{+}({\un
r},t)\,\vec{S}\,\ul{\psi}({\un r},t)\,d^3r=$
\\[0.2cm]
\hspace*{1.0cm}$=\int\ul{\psi}'^{+}\,({\un
r}',t)\,\ul{\ul{Q}}\,\vec{S}\,\ul{\ul{Q}}^{+}\,\ul{\psi}'({\un r}',t)\,d^3r'$
\\[0.2cm]
\hspace*{1.0cm}$=\frac{\hbar}{2}\left[\cos(\varphi-\omega_L t)\,\sin\theta\,{\un
e}_x+\right.$
\\[0.2cm]
\hspace*{3.0cm}$\left.\sin(\varphi-\omega_L t)\,\sin\theta\,{\un e}_y+\cos\theta\,{\un
e}_z\right]\,.$
\\[-0.4cm]
\begin{eqnarray}
\label{eqn:precess_vectorS}
\qquad
\end{eqnarray}
Thus, the vector $\langle\vec{S}\rangle$ of the spin momentum moves on a circular cone with an apex angle of $2\,\theta$ about the direction of the magnetic field and its projection onto the $x/y$-plane rotates at an angular frequency $\omega_L$, the ``Larmor frequency'', about the $z$-axis. According to Eq.(\ref{eqn:DefinitionOmega}) this frequency is given by
\begin{eqnarray}
\label{eqn:LarmorFrequenz} \omega_L=\frac{\mu_B B_z}{\hbar/2}\,.
\end{eqnarray}
The spin precession is completely analogous to that of a classical spinning top which rotates about its symmetry axis at an angular frequency $\omega$ and is exposed to the gravitational field of the earth. The precession frequency $\omega_P$ is in this case given by
$$
\omega_P=\frac{F\,r_s}{L}\,,
$$
where $L$ denotes the absolute value of the angular momentum, $F$ is the absolute value of the gravitational force acting on the top's centroid, and $r_s$ is the distance of the centroid from the point of support. In case one has instead of a gravitational field a magnetic field and if the spinning top possesses a  magnetic moment $\mu_B$, one has $F\,r_s=\mu_B B_z$. Inserting this into the classical equation for $\omega_P$ and setting $L=\hbar/2$ one obtains exactly the expression (\ref{eqn:LarmorFrequenz}) for the Larmor frequency. If the magneto-mechanical parallelism would also hold for the spin momentum, that is if $g$ in Eq.(\ref{eqn:Definition_g}) were equal to one, the magnetic moment would be $\frac{1}{2}\,\mu_B$, and the precession frequency would be smaller by a factor 2 in striking disagreement with the experiment.\\
The completely classical behavior of a precessing spin moment in a magnetic field can also be made evident by the following consideration.\\
Using Eqs.(\ref{eqn:Definition_g}) and (\ref{eqn:precess_vectorS}) we may express the spin-derived magnetic moment $\vec{M}_{Spin}$ as
\\[0.2cm]
$\vec{M}_{Spin}=\mu_B\,\left[\cos(\varphi-\omega_L t)\,\sin\theta\,{\un
e}_x+\right.$
\\[0.2cm]
\hspace*{3.5cm}$\left.\sin(\varphi-\omega_L t)\,\sin\theta\,{\un e}_y+\cos\theta\,{\un
e}_z\right]\,.$
\\[0.2cm]
The time derivative of Eq.(\ref{eqn:precess_vectorS}) can be written 
\\[0.2cm]
$\frac{d}{dt}<\vec{S}>=\mu_B\,B_z\,\left[\sin(\varphi-\omega_L t)\,\sin\theta\,
{\un e}_x-\right.$
\\[0.2cm]
\hspace*{4.0cm}$\left.\cos(\varphi-\omega_L t)\,\sin\theta\,{\un e}_y\right]\,,$
\\[-0.4cm]
\begin{eqnarray}
\label{eqn:zeitlAenderungSpin}
\qquad
\end{eqnarray}
where we have used $\omega_L=2\mu\,B_z/\hbar$. We observe that ${\un B}=B_z\,{\un e}_z$ and
\begin{eqnarray*}
{\un e}_x\times{\un e}_z=-{\un e}_y\,;\quad {\un e}_y\times{\un e}_z={\un e}_x\,;\quad{\un e}_z\times{\un e}_z=0\,.
\end{eqnarray*}   
Hence, the right-hand side of Eq.(\ref{eqn:zeitlAenderungSpin}) can be cast as
\\[0.2cm]
$\mu_B\,B_z\,\left[\sin(\varphi-\omega_L t)\,\sin\theta\,{\un
e}_x-\cos(\varphi-\omega_L t)\,\sin\theta\,{\un e}_y\right]=$
\\[0.2cm]
\hspace*{6.5cm}$\vec{M}_{Spin}\times {\un B}\,.$
\\[0.2cm]
The result may be written
\begin{eqnarray}
\label{eqn:BewGlg_Kreisel} 
\frac{d}{dt}<\vec{S}>=\vec{M}_{Spin}\times {\un B}\,.
\end{eqnarray} 
This is identical with the classical equation of motion describing the temporal behavior of a spinning top that is acted upon by a torque $\vec{M}_{Spin}\times {\un B}$. It corresponds to Ehrenfest's First Theorem, and it is this equation (\ref{eqn:BewGlg_Kreisel}) which governs the phenomena encountered in electron and nuclear spin resonance. (S. e.~g.~Slichter \cite{Slichter}.) In applying magnetic resonance techniques one has to supplement Eq.(\ref{eqn:BewGlg_Kreisel}) by perturbational terms that cause a change of the precession cone. An equation of this kind was put forward by Bloch \cite{Bloch} in 1945. If the atom is not exposed to a time-dependent perturbation the spin keeps precessing on the cone without changing its apex angle even when the strength of the magnetic field adiabatically increases or decreases. A change of the absolute value of ${\un B}$ only changes the Larmor frequency $\omega_L$. 
\\[0.2cm]
As opposed to the impression that is commonly invited by even the most recent literature, Eq.(\ref{eqn:BewGlg_Kreisel}) constitutes a purely quantum mechanical result and is in no ways ``semi-classical'' or ``macroscopical''. The fact that from our derivation $|\langle \vec{S}\rangle_z|$ may attain any value, seems to contradict the principle of ``orientation quantization'' according to which $|\langle \vec{S}\rangle_z|$ may equal only integer multiples of $\hbar/2$. Clearly, if $\langle \vec{S}\rangle_z$ is not parallel or anti-parallel to ${\un B}$ but rather precesses about the direction of the latter, the electron emits magnetic dipole radiation until its spin is aligned. But this is a weak interaction, and therefore the state of non-alignment may well be regarded as meta-stable in certain experimental situations.
\\[0.2cm]
Spin precession in a magnetic field exhibits a peculiar feature that relates to the occurrence of the argument $\frac{\varphi}{2}$ in the exponential functions of Eq.(\ref{eqn:Praezession2}). To see that we assume $\ul{\psi}({\un r})$ to represent a wavepacket of a free particle that traverses a homogeneous magnetic field in an orthogonal direction. When the wavepacket enters the magnetic field the spin component perpendicular to the field may point in the x-direction which is also the direction of flight. We then have
$$
\varphi=0\quad \mbox{and hence} \quad e^{\pm i\,\frac{\varphi}{2}}=1\,.
$$
During the flight $\theta$ stays constant. When the wavepacket leaves the magnetic field after a full precession period we have
$$
\varphi=2\pi\quad \mbox{which means} \quad e^{\pm i\,\frac{\varphi}{2}}=-1\,.
$$
Hence $\ul{\psi}({\un r})$ has changed its sign, or one may just as well say, its phase has been shifted by $\pi$. However, as can be seen from Eq.(\ref{eqn:precess_vectorS}), $\langle\vec{S}\rangle$ points in the same direction as at the beginning of the precession. This phase shift is well detectable in double-beam experiments with spin-polarized neutrons (s. e.~g.~Rauch \cite{Rauch}, Werner et al. \cite{Werner}).

\section{A theory of the Stern-Gerlach experiment} \label{kapitel20}

{\it ``...Phenomena of this kind made physicists despair of finding any consistent space-time picture of what goes on the atomic and subatomic scale...many came to hold not only that it is difficult to find a coherent picture but that it is wrong to look for one...''}
\\[0.2cm]
\hspace*{4.0cm} John Bell \cite{Bell}
\\[0.2cm]
Quite a few attempts have already been made on a theory of the Stern-Gerlach (SG-) experiment \cite{Stern}. For a recent rather complete update of the pertinent literature see Home et al.\cite{Home}. But a coherent picture of the fundamental mechanism is still missing. Most physicists seem to favor the idea that the electronic state of the atom on entering the magnet constitutes a linear combination of spin states ``up`` and ``down'', and the modulus square of the associated coefficients defines the probability of the atom for being either pulled up or down, that is parallel or anti-parallel to the magnetic field gradient. On detection of the atom in the ``up''- or ``down''-beam the atomic wave function collapses onto the respective component of the linear combination. From our point of view this is unjustifiably associating the process of detection with some mystical influence of ``observing'', based on pure claim: the atomic beam would behave differently if it would not be detected.  By contrast, we believe that the outcome of the experiment is completely determined by the time-dependent Pauli equation and is hence a result of a ``quantum mechanics without observer''.\\
Our approach implies a linear combination of spin states as well, that is we describe the electronic 1s-state of the atom that we shall consider below by
$$
{\ul \psi}_{atom}({\un r},t)=\psi_{1s}({\un r}-{\un v}\,t)\left[a_{\uparrow}\,\left({1\atop 0}\right)+a_{\downarrow}\,\left({0\atop 1}\right)\right]
$$
where ${\un v}$ denotes the velocity of the atom, and the coefficients $a_{\uparrow},a_{\downarrow}$ have the property $|a_{\uparrow}|^2+|a_{\downarrow}|^2=1$. The unit spinors are referenced to the direction of the field gradient $\frac{\partial B_z}{\partial z}\,{\un e}_z$. Hence, the expectation value of the force acting on the atom in the SG-magnet is given by Eq.(\ref{eqn:ExpForce}) if we neglect the induction derived term and assume electrostatic forces being absent
$$
\langle {\un F}_{atom} \rangle=\mu_B\,\int {\ul \psi}^{\dagger}_{atom}({\un r},t)\,\frac{\partial B_z}{\partial z}\,{\ul \psi}_{atom}({\un r},t)\,d^3r\,{\un e}_z\,.
$$
For simplicity we equate the field gradient to a constant so that $\langle {\un F}_{atom} \rangle$ reduces to
$$
\langle {\un F}_{atom} \rangle=\mu_B\,\frac{\partial B_z}{\partial z}\,[|a_{\uparrow}|^2-|a_{\downarrow}|^2]\,{\un e}_z\,.
$$
It can obviously attain any value between $-\mu_B\,\frac{\partial B_z}{\partial z}\,{\un e}_z$ and $+\mu_B\,\frac{\partial B_z}{\partial z}\,{\un e}_z$ depending on the value of the coefficients when the atom enters the magnet. Therefore a splitting into two well separated beams cannot possibly occur as long as there is no particular mechanism which inhibits a random distribution. In the following we shall outline such a possible mechanism.\\ 
We assume that the reader is sufficiently familiar with the essential features of the experimental setup. To simplify the line of argument we content ourselves with considering the experiment by Wrede \cite{Wrede} who used a primary beam of hydrogen atoms in a setup that was practically identical with that of Stern and Gerlach. Hydrogen offers the advantage of reducing the spin-orientation problem to that of a single electron. The standpoint we take here is akin to that of Mott and Massey \cite{Mott} who remark: {\it ``From these arguments we must conclude that it is meaningless to assign to the free electron a magnetic moment. It is a property of the electron that when it is bound in an S state in an atom, the atom has a magnetic moment.''}\footnote{However, we want to modify this debatable statement by saying that also free electrons display a magnetic moment when they are exposed to a magnetic field where their motion perpendicular to the field becomes confined to a circular area of a certain diameter.} \\
The hydrogen atoms effuse from some source where they are (almost unavoidably) exposed to the terrestrial magnetic field or at least to the weak fringe field of the SG-magnet. That field causes a weak Zeeman-splitting of the spin up and spin down level of the electronic 1s-state. Because of the weakness of the splitting the two Zeeman-levels are at the temperature of the source equally occupied, that is, 50\% of the effusing atoms have their electronic spins oriented parallel to the weak external field, the spins of the remaining 50\% atoms are anti-parallel. As the atoms approach the SG-magnet they feel in a co-moving coordinate system a magnetic field whose field strength increases continuously and will in general change its direction. We assume for simplicity that the spin orientation is transverse and that the atom moves along the $x$-axis of a laboratory-fixed coordinate system so that changes of the spin orientation will only take place in the $y/z$-plane parallel to the respective plane of the co-moving coordinate system. As soon as the field direction in the co-moving coordinate system departs by a small angle $\delta\theta$ from the original direction of $\vec{B}=B_z\,{\un e}_z$ at the onset of the atom's trajectory, a small $y$-component $\vec{B_y}=B_z\,\sin \delta\theta\,{\un e}_y$ of the field appears as a consequence of which the magnetic moment of the atom experiences a torque $-\mu_B\,B_z\,\sin\delta\theta\,{\un e}_{\varphi}$, where ${\un e}_{\varphi}$ denotes the unit vector in the direction of increasing azimuth angle $\varphi$ in the $x/y$-plane. This torque causes a change $\dot{{\un L}}$ of the spin angular momentum
$$
\dot{{\un L}}=-\frac{\hbar}{2}\,\sin \delta\theta\,\omega_L\,{\un
e}_{\varphi}\,,
$$
where we have used $2\mu_B\,B_z=\hbar\,\omega_L$ (Eq.(\ref{eqn:DefinitionOmega})). Hence, the spin momentum starts precessing about the new  direction of the magnetic field. We ignore the slight tilt of the co-moving new $x/y$-plane perpendicular the new field direction.\\
We envisage a short time span for which we assume the changes of $\theta$ to be small so that 
\begin{eqnarray}
\label{eqn:Neigungswinkel}
\sin\delta\theta\approx\delta\theta=\dot{\theta}\,t\,,
\end{eqnarray}
where $t=0$ coincides with the beginning of the rotation of the field. The following considerations exploit the typical experimental condition that the precession frequency $\omega_L$ is some orders of magnitude larger than the speed of the field rotation. (In the terrestrial magnetic field of magnitude $\approx$ 5$\cdot$10$^{-5}\,T$ the precession frequency of the electronic spin is about 10$^6\,s^{-1}$. At an atomic speed of 10$^5$cm$\,$s$^{-1}$, a distance of about 10$\,$cm and a maximum rotation angle of $\pi/2$ one has $\dot{\theta}\approx$10$^4\,$s$^{-1}$.)  As will become apparent from the following calculations we may limit ourselves to a short time span comprising only few precession periods during which the magnetic field rotates only by a small angle ($\theta\ll 2\pi$) so that one is justified in assuming $\dot{\theta}$ to be constant:
$$
\dot{\theta}=const.
$$
The unit vector ${\un e}_{\varphi}$ may be decomposed
\begin{eqnarray}
\label{eqn:WinkeleinhVektor} {\un e}_{\varphi}=-{\un e}_{x}\,\sin \varphi+{\un
e}_{y}\,\cos \varphi\,.
\end{eqnarray}
At $t=0$ we have  $\varphi(t=0) =-\frac{\pi}{2}$, that is ${\un e}_{\varphi}={\un e}_x$. Thus, it is advisable to replace $\varphi$ with 
$\varphi+\frac{\pi}{2}$, but we omit denoting the new azimuth angle differently. Hence we have $\varphi=0$ for $t=0$, and we obtain instead of Eq.(\ref{eqn:WinkeleinhVektor})
$$
{\un e}_{\varphi}={\un e}_{x}\,\cos \varphi+{\un e}_{y}\,\sin \varphi\,.
$$
The spin precession that now occurs is anti-clockwise
$$
\dot{\varphi}=-\omega_L \quad \mbox{that is} \quad \varphi=-\omega_L\,t\,.
$$
Thus
$$
\dot{{\un L}}=\frac{\hbar}{2}\,\dot{\theta}\,\omega_L\,\left[{\un e}_x\,t\,\cos
\omega_L\,t-{\un e}_y\,t\,\sin \omega_L\,t\right]\,.
$$
This results in a change of the angular momentum after one precession period $T=2\pi/\omega_L$ 
\\[0.2cm]
$\Delta{\un L}=\frac{\hbar}{2}\,\dot{\theta}\,\left[\omega_L\,{\un
e}_x\,\int_{0}^{T}t\,\cos \omega_L\,t\,dt-\right.$
\\[0.2cm]
\hspace*{4.0cm}$\left.{\un e}_y\,\omega_L\,\int_{0}^{T}t\,\sin \omega_L\,t\,dt\right]\,.$
\\[0.2cm]
Hence, using
$$
\int_{0}^{2\pi}\xi\,\sin \xi\,d\xi=-2\pi\quad \mbox{and} \quad
\int_{0}^{2\pi}\xi\,\cos \xi\,d\xi=0\,,
$$
we may $\Delta {\un L}(T)$ cast as 
$$
\Delta {\un L}(T)=\frac{\hbar}{2}\,\underbrace{\dot{\theta}\,T}_{\equiv \Delta
\theta}\,{\un e}_y 
$$
that is in the spirit of our approximation (\ref{eqn:Neigungswinkel})
$$
\Delta {\un L}(T)=\frac{\hbar}{2}\,\sin
\Delta \theta\,{\un e}_y\,.
$$
The $y$-component of the magnetic field which equaled zero at the beginning of the rotation is now given by $\vec{B}_y=B_z\,\sin \Delta \theta\,{\un e}_y$. That means: after one precession period $T$ the magnetic field and the atomic spin angular momentum have turned by the same angle $\Delta \theta$. The spin orientation follows the magnetic field - within the present approximation - without slip, that is adiabatically. (This is similar to the physics of a spinning artillery shell whose spin axis follows the course of the shell's bending trajectory leaving only  a small precession angle.) Thus, the atoms enter the SG-magnet (almost) fully oriented with respect to the SG-magnetic field. This applies to the atoms with anti-parallel spin orientation accordingly. Hence, the two beams leaving the SG-magnet reflect merely the two kinds of atoms associated with the two Zeeman levels before they leave the reservoir.
\\
It is worth mentioning that Leu \cite{Leu} carried out Stern-Gerlach-type experiments using beams of Na-, K-, 
Zn-,\\ Cd- and Tl-atoms instead of Ag-atoms. The Zn- and Cd-atoms possess two s-valence electrons which results in a zero net spin momentum of the atoms and consequently one does not observe a beam splitting in the Stern-Gerlach magnet. On the other hand, Tl-atoms possess a 6p-valence electron that is subjected to spin-orbit coupling. This gives rise to a Land\'e factor $g=\frac{2}{3}$ as a result of which the effective magnetic moment is for $M_j=\frac{1}{2}$ given by
$$
\mu_{eff}=\mu_B\,g\,M_j =\frac{1}{3}\,\mu_B\,.
$$ 
This is, in fact confirmed by the experiments.\\
If one were dealing with atoms that possess a total angular momentum $J=(l\pm \frac{1}{2})\,\hbar$ associated with $2l+2$ different magnetic quantum numbers $M_j$, one would have $2l+2$ different states in the initial weak field and therefore as many different sorts of atoms entering the Stern-Gerlach magnet where they are deflected according to their magmetic moment. That means one would have $2l+2$ different beams instead of 2.\\ 
Our explanation of the SG-experiment is much in the spirit of Stern's conjecture that the spin of an atom responds adiabatically to the directional change of the magnetic field in which it has originally been aligned. In cooperation with Phipps \cite {Stern2} he devised an experiment where one of the beams at the exit of a first SG-magnet was focused into a linear set of three successive magnets whose weaker, essentially homogeneous fields pointed in three different directions perpendicular to the atomic trajectory. The difference between these directions was 120°. If the spin of the selected beam was pointing up after leaving the first SG-magnet and assuming that the spin would adiabatically adjust to the local magnetic field on its passage through the three magnets, it was thus to be expected that it would be finally back to its previous  ``up''-orientation. To test this the beam was sent into a second SG-magnet identically oriented as the first. There was only one  beam coming out of this magnet indicating that the spin was pointing again in the same direction as on entering the three ``turn magnets''. In other words: even after a turn of 360° no slip between spin orientation and the direction of the magnetic field had occurred. We mention here only in passing that our result on the Phipps-Stern experiment agrees with that of Rosen and Zener \cite{Zener} published already in 1932. Different from our more summary analysis these authors attempt to stay close to explicitly solving the time-dependent Pauli equation. \\
Surprisingly, the interpretation of the SG-experiment as demonstrating a coherent splitting of the de Broglie-wave of the incoming atom into two beams has become the most popular view on which a host of considerations on ``measurement'' is based. Papers on the so-called ``Humpty-Dumpty-problem'' (s.~e.~g.~Englert et al. \cite{Englert}) deal explicitly with a possible reconstruction of the original single wave by appropriately merging the two coherent beams at a spot reached later. We believe that such thought experiments are without substance. As we have clearly demonstrated, the SG-magnet does not cause a splitting of the incoming matter wave. The SG-situation is distinctly different from that in neutron spin-flip experiments by Rauch and coworkers \cite{Badurek} where a transverse spin polarized beam of neutrons hits a plate of a Si single crystal such that each matter wave packet splits up into two widely separated beams of packets due to dynamical diffraction within the crystal. This diffraction process is spin-independent. The two beams are coherently merged then by dynamical diffraction at a second Si-plate.     
\\[0.2cm]
Many authors give the impression as if there were not a shadow of doubt that Stern-Gerlach experiments with charged free particles (like electrons) are just as feasible as with spin-carrying neutral atoms. Bohr had very early pointed out (s. Wheeler and Zurek \cite {Wheeler}) that such experiments could not possibly succeed because ``the Lorentz force would inevitably blur any Stern-Gerlach pattern''. Nevertheless, the literature on EPRB- (Einstein-Podolsky-Rosen-Bohm) correlation with pairs of fermions in a singlet state (s. e.$\,$g. Einstein et al. \cite {Einstein1}, Bohm \cite{Bohm}) abounds with allusions to ``measuring separately the $x$/$y$/$z$-spin components'' of the particles by means of Stern-Gerlach magnets. (S. e.~g. Wigner \cite{Wigner}.) Even when one were dealing with neutral fermions what kind of mechanism should yield such information on those spin {\bf components}? How would the time evolution of the respective solution to the time-dependent Pauli equation look like in this case?

\section{The time-dependent Dirac equation}\label{kapitel20.0}

In trying to extend the theory to relativistic systems we retain the following two fundamental assumptions that characterize the non-relativistic quantum mechanics we have been dealing with so far:
\begin{enumerate}
\item The universal existence of stochastic forces that necessitate an ensemble description of the one-particle system under study. The fundamental constituents of this approach are: 
$\rho({\un r},t)$ for the occurrence of the particle at ${\un r}$ and time $t$ and 
${\un p}({\un r},t)$ for the associated ensemble average of the particle momentum
\item Lumping together the two real-valued functions $\rho({\un r},t)$ und ${\un p}({\un r},t)$ in the form of a complex-valued function  $\psi({\un r},t)$\\
\begin{eqnarray}
\label{eqn:2}
\psi({\un r},t)=\sqrt{\rho({\un r},t)}\,\,e^{i\,\varphi({\un r},t)}
\end{eqnarray}
where
\begin{eqnarray}
\label{eqn:2}
{\un p}({\un r},t)=\hbar\,\nabla \varphi({\un r},t) \,.
\end{eqnarray}
From $\psi({\un r},t)=|\psi({\un r},t)|\,e^{i\,\varphi({\un r},t)}$ one then obtains the momentum current density
\\[0.2cm]
\hspace*{2.0cm}${\un j}_p({\un r},t)=\rho({\un r},t)\,{\un p}({\un r},t)=$
\\[0.2cm]
$\frac{1}{2}\,[\psi^{*}({\un r},t)\,\hat{{\un p}}\,\psi({\un r},t)-
\psi({\un r},t)\,\hat{{\un p}}\,\psi^{*}({\un r},t)]$
\\[-0.8cm]
\begin{eqnarray}
\label{eqn:3}
\qquad
\end{eqnarray}
where
\\[0.1cm]
\hspace*{2.5cm}$\hat{{\un p}}\stackrel{\mbox{{\tiny def}}}{=} -i\,\hbar\,\nabla$. 
\end{enumerate}
Eq.(\ref{eqn:2}) implies that ${\un p}({\un r},t)$ is curl-free, that is, the stochastic forces do not cause friction.\\
From Eq.(\ref{eqn:3}) follows for the expectation value of the particle momentum
\begin{eqnarray}
\label{eqn:4}
<{\un p}(t)>=\int{\un j}_p({\un r},t)\,d^3r=\int\psi^{*}({\un r},t)\,\hat{{\un p}}\;\psi({\un r},t)\,d^3r\,.
\end{eqnarray}
If one replaces  $\psi({\un r},t)$ with its Fourier integral
$$
\psi({\un r},t)=(2\pi)^{-\frac{3}{2}}\int C({\un k},t)\,e^{i\,{\un k}\cdot{\un r}}\,d^3k\,,
$$
one obtains on insertion in Eq.(\ref{eqn:4})
\\[0.2cm]
$<{\un p}(t)>=\int\psi^{*}({\un r},t)\,\hat{{\un p}}\;\psi({\un r},t)\,d^3r=$
\\[0.2cm]
\hspace*{4.0cm}$\int C^{*}({\un k},t)\,\hbar\,{\un k}\;C({\un k},t)\,d^3k\,,$
\\[-0.6cm]
\begin{eqnarray}
\label{eqn:5}
\qquad
\end{eqnarray}
and analogously
\\[0.2cm]
$\int \psi^{*}({\un r},t)\frac{\hat{\un p}^2}{2\,m_0}\,\psi({\un r},t)\,d^3r=$
\\[0.2cm]
\hspace*{4.0cm}$\int C^{*}({\un k},t)\,\frac{\hbar^2\,{\un k}^2}{2\,m_0}\;
C({\un k},t)\,d^3k\,.$
\\[-0.6cm]
\begin{eqnarray}
\label{eqn:6}
\qquad
\end{eqnarray}
Newton's modified second law (\ref{eqn:Bohm_equation2}) which we have derived for the non-relativistic case, contains an additional ``quantum force'' ${\un F}_{QP}=-\nabla V_{QP}$ whose expectation value equals zero. As a result one arrives at Ehrenfest's two theorems.  
\begin{eqnarray}
\label{eqn:7}
<{\un v}>=\frac{d}{dt}\,<{\un r}>=<\nabla_{{\un p}}E({\un p})> 
\end{eqnarray}
and
\begin{eqnarray}
\label{eqn:8}
\frac{d}{dt}<{\un p}>=<{\un F}>=<-\nabla V> \,.
\end{eqnarray}
The salient point here is that these two equations apply to the non-relativistic case and we require them to persist unaffected in the relativistic case if the particle is assumed - as before - to perform a dissipationless motion under stochastic extra forces.
\\[0.2cm]
Conversely, one can derive the time-dependent Schrödinger equation just by starting from Eqs.(\ref{eqn:7}) and(\ref{eqn:8}) and going along the same line of argument used in our derivation of the time-dependent Pauli equation in Section \ref{kapitel17}.
In the following we shall refer to the latter. However, instead of 
$$
E({\un p})=\frac{{\un p}^2}{2m_0}+m_0\,c^2+V({\un r})
$$
we now have
\begin{eqnarray}
\label{eqn:E_of_p_relativ}
E({\un p})=\underbrace{\sqrt{{\un p}^2\,c^2+m_0^2\,c^4}}_{=E_{kin}+m_0\,c^2}+V({\un r})
\end{eqnarray}
with $c$ denoting  the velocity of light in vacuo.\\
Hence $<\nabla_{{\un p}}E({\un p})>=<{\un v}>$ in Eq.(\ref{eqn:7}) has to be dealt with differently in the relativistic case. Following Dirac \cite{Dirac} we construct a Fourier-transform  $\ul{\ul{H}}_0({\un k})$ that corresponds to the thought-for energy-operator $\hat{H}_{rel.}$  just as  $\hbar^2\,k^2$ in Eq.(\ref{eqn:5}) relates to the expression $\frac{\hat{\un p}^2}{2\,m_0}$. If one rewrites $E_{kin}({\un k})+m_0\,c^2$ in Eq.(\ref{eqn:E_of_p_relativ}) in the form
\\[0.2cm]
$E_{kin}({\un k})+m_0\,c^2=\hbar\,c\,\sqrt{\sum_{\mu=0}^3 k^2_{\mu}}\quad \mbox{where} \quad p_{\mu}=\hbar\,k_{\mu}$
\\[0.2cm]
and\hspace*{2.0cm}$\quad k_0=\frac{m_0\,c}{\hbar}$
\\[0.2cm]
and replaces the right-hand side with a 4$\times$4-matrix $\ul{\ul{H}}_0({\un k})$ defined by
\\[0.2cm]
\hspace*{2.0cm}$\ul{\ul{H}}_0({\un k})=\hbar\,c\,\sum_{\mu=0}^3 \ul{\ul{\alpha}}_{\mu}\,k_{\mu}\,,$
\\[0.2cm]
{where} $\ul{\ul{\alpha}}_{\mu}$ denotes constant dimensionless 4$\times$4 matrices,
\\
the Fourier transform $\ul{\ul{H}}_0({\un k})$ must obviously possess the property
\begin{eqnarray*}
\ul{\ul{H}}^2_0({\un k})=\hbar^2\,c^2\,\sum_{\mu=0}^3\,\sum_{\mu'=0}^3 k_{\mu}\,k_{\mu'}\,\delta_{\mu\mu'}\,\ul{\ul{1}}=\\
\frac{\hbar^2\,c^2}{2}\,\sum_{\mu=0}^3\,\sum_{\mu'=0}^3 k_{\mu}\,k_{\mu'}\,[\ul{\ul{\alpha}}_{\mu}\,\ul{\ul{\alpha}}_{\mu'}+\ul{\ul{\alpha}}_{\mu'}\,\ul{\ul{\alpha}}_{\mu}]\,.
\end{eqnarray*}
That means that the matrices $\ul{\ul{\alpha}}_{\mu}$ have to comply with the requirement
$$
\frac{1}{2}\,[\ul{\ul{\alpha}}_{\mu}\,\ul{\ul{\alpha}}_{\mu'}+\ul{\ul{\alpha}}_{\mu'}\,\ul{\ul{\alpha}}_{\mu}]=\delta_{\mu\mu'}\,\ul{\ul{1}}\,.
$$
As can be verified by just performing the multiplications, the  matrices $\ul{\ul{\alpha}}_{\mu}$ meet this requirement if they have the form
\begin{eqnarray*}
\ul{\ul{\alpha}}_0=\left( \begin{array}{c c}
\ul{\ul{1}} & \;\;\ul{\ul{0}} \\
\ul{\ul{0}} & -\ul{\ul{1}}
\end{array} \right)\quad \mbox{and} \quad \ul{\ul{\alpha}}_{\mu}=\left( \begin{array}{c c}
\ul{\ul{0}} & \ul{\ul{\sigma}}_{\mu} \\
\ul{\ul{\sigma}}_{\mu} & \ul{\ul{0}}
\end{array} \right)\; \mbox{for} \; \mu=1,2,3\,.
\end{eqnarray*}
Here $\ul{\ul{\sigma}}_{\mu}$ denotes 2$\times$2-matrices that are identical with the Pauli matrices (\ref{eqn:Pauli_Matrizen}). Similar to  the latter one can lump the 4$\times$4-matrices $\ul{\ul{\alpha}}_{\mu}$ together by forming a vector $\ul{\alpha}$ so that  $\ul{\ul{H}}_0({\un k})$ may be cast as
\begin{eqnarray}
\label{eqn:9}
\ul{\ul{H}}_0({\un k})=c\,\ul{\alpha}\cdot \hbar{\un k}+\ul{\ul{\alpha}}_0\,m_0\,c^2\,.
\end{eqnarray}
The feasibility of the above line of thought requires a consistent extension of the hitherto discussed spinor function to a bispinor function
\begin{eqnarray*}
\ul{\psi}({\un
r},t)=\left( \begin{array}{c}
\psi_{\uparrow}^1({\un r},t)\\
\psi_{\downarrow}^1({\un r},t)\\
\psi_{\uparrow}^2({\un r},t)\\
\psi_{\downarrow}^2({\un r},t)
\end{array} \right)
\end{eqnarray*}
where
$$
\psi^{(j)}_{\uparrow(\downarrow)}({\un r},t)=|\psi^{(j)}_{\uparrow(\downarrow)}({\un r},t)|\,e^{i\varphi^{(j)}_{\uparrow(\downarrow)}({\un r},t)}\,, \quad j=1,2\,.
$$
The associated phases $\varphi^{(j)}_{\uparrow(\downarrow)}({\un r},t)$ represent as in Eq.(\ref{eqn:2}) potentials of ensemble averages of momenta which means
\begin{eqnarray*}
{\un p}({\un r},t)=\sum_{\stackrel{j=1,2}{(\uparrow,\downarrow})}\frac{|\psi^{(j)}_{\uparrow(\downarrow)}({\un r},t)|^2}{\rho({\un r},t)}{\un p}^{(j)}_{\uparrow(\downarrow)}({\un r},t)\\ \qquad \mbox{where} \quad {\un p}^{(j)}_{\uparrow(\downarrow)}
({\un r},t)=\hbar\,\nabla\,\varphi^{(j)}_{\uparrow(\downarrow)}({\un r},t)\,.
\end{eqnarray*}
The quantities ${\un p}^{(j)}_{\uparrow(\downarrow)}$ are now different for ``spin up'' and ``spin down'' if the particle in question moves in a  spatially varying potential. Only in the strictly non-relativistic case the spin generating component of the quivering motion and the orbital motion remain unaffected on superposition. In this case we have $\psi_{\uparrow(\downarrow)}({\un r},t)=|\psi_{\uparrow(\downarrow)}({\un r},t)|\,e^{i\,\varphi({\un r},t)}$. \\
If one performs a Fourier transform one obtains in complete analogy to Eq.(\ref{eqn:5}) also 
in the relativistic case
$$
<{\un p}(t)>=\int\ul{C}^{\dagger}({\un k},t)\,\hbar\,{\un k}\;\ul{C}({\un k},t)\,d^3k\,.\qquad \qquad\qquad
$$
Correspondingly one gets
\begin{eqnarray*}
\int\ul{C}^{\dagger}({\un k},t)\,\ul{\ul{H_0}}({\un k})\;\ul{C}({\un k},t)\,d^3k=\qquad\qquad\qquad\qquad\\
\int\ul{\psi}^{\dagger}({\un r},t)\,\underbrace{[c\,\ul{\alpha}\cdot \hat{{\un p}}+\ul{\ul{\alpha}}_0\,m_0\,c^2]}_{\stackrel{\mbox{{\tiny def}}}{=}\ul{\ul{\hat{H}}}_{Dirac}}\;\ul{\psi}({\un r},t)\,d^3r\,.
\end{eqnarray*}
We now form $<{\un v}>$ according to
\begin{eqnarray*}
<{\un v}>=<\nabla_{{\un p}}E({\un p})>=\qquad\qquad\qquad\qquad\\
\int\ul{C}^{\dagger}({\un k},t)\,[\hbar^{-1}\,\nabla_{{\un k}}\ul{\ul{H_0}}({\un k})]\;\ul{C}({\un k},t)\,d^3k\,.
\end{eqnarray*}
If we substitute $\ul{C}({\un k})$ by its Fourier transform we obtain
$$
<{\un v}>=\int\ul{\psi}^{\dagger}({\un r},t)\,c\,\ul{\alpha}\;\ul{\psi}({\un r},t)\,d^3r\,.
$$
Exploiting the identity
$$
\ul{\ul{\hat{H}}}_{Dirac}\,{\un r}-{\un r}\,\ul{\ul{\hat{H}}}_{Dirac}=-i\,c\,\hbar\,\ul{\alpha}\,,
$$
and going through the same set of arguments as with deriving the Pauli equation, we arrive at the time-dependent Dirac equation
$$
[\ul{\ul{\hat{H}}}_{Dirac}+V({\un r})]\,\ul{\psi}({\un r},t)=i\,\hbar\,\frac{\partial}{\partial t}\,\ul{\psi}({\un r},t)\,.
$$ 
The derivation can be extended by including electromagnetic fields, again in complete analogy to the derivation of the Pauli equation.

\section{Spatial particle correlation beyond the limit of entanglement. Spooky action at a distance}\label{kapitel20.1}

As discussed in Section \ref{kapitel15} the electrons of two hydrogen atoms will respond independently to local perturbations once the inter-atomic distance has become macroscopically large. The electronic wave function factorizes then and becomes the product of two one-particle wave functions. One would therefore expect two free fermions that have moved sufficiently far away in opposite directions with their spins being transverse and anti-parallel, to display the same features. If they were still described by an anti-symmetric wave function the particle properties would remain non-locally intertwined in that each of the particles would appear at distant detectors with only half of the total probability. Therefore a realistic description can only be ensured by a product of two one-particle wave functions, wavepackets moving in opposite directions, one for spin up and the other one for spin down or vice versa, the choice randomly distributed among the pairs generated in succession. Consequently, there will be a complete loss of the ``common-cause''-spin correlation of the particles when they hit differently oriented spin detectors. The latter scatter the incoming fermion depending on the angle which the fermion's spin direction encloses with the scattering plane.  To be as concrete as possible we refer in this section to the fundamental experiment by Lamehi-Rachti and Mittig \cite{Lamehi} who were able to generate pairs of protons of about 8 MeV with spins paired anti-parallel and moving apart such that the proton's velocities in the center of mass system have the same absolute value but opposite directions. The spin orientation was analyzed by letting each of the protons impinge on a device akin to a Mott detector familiar from polarized electron detection. The incoming proton is scattered at some carbon atom of a carbon foil. Each Mott-type detector is associated with two particle detectors whose axes point to the scattering center and enclose an angle $\pm \alpha$ with the flight direction of the incoming proton. Together with that direction these axes form the scattering plane. The differential cross section of the carbon scatterer for a proton with spin up perpendicular to the scattering plane is given by
\begin{eqnarray}
\label{eqn:differential_cross_section}
\sigma(\alpha,\beta)=(|f(\alpha)|^2+|g(\alpha)|^2)\,[1-S(\alpha)\sin\beta]
\end{eqnarray}
where $S(\alpha)$ represents the Sherman function for carbon/proton scattering, $\beta$ stands for the azimuthal angle in the plane perpendicular to the proton flight direction and $f(\alpha)$ and $g(\alpha)$ denote the scattering and spin-flip amplitude. The latter is associated with spin-orbit coupling \footnote{Although spin-orbit coupling represents a constituent of the relativistic Pauli approximation to the Dirac equation, we consider it here  as given.} which determines also the magnitude of $S(\alpha)$. In view of the objective of this article, we wish to emphasize at this point that Eq.(\ref{eqn:differential_cross_section}) is a consequence of solving the relativistic Pauli equation, and the experimentally verifiable results that will be discussed below, are another objective consequence which is definitely not affected by the process of particle detection (the ``measurement'').\\
In accordance with the notation familiar from EPRB-experiments we denote the Mott-type analyzer at the end of the left proton track by A and that at the end of the right track by B. Furthermore, the detector on the right side of the scattering plane will be characterized by a ``+''-sign, that on the left side by a ``-''-sign. The two particle detectors of each Mott-type analyzer are located at $\beta=\mp \pi/2$, and $\alpha$ was set $\approx 50^{°}$. Hence the difference between the respective differential cross sections (the ``left-right asymmetry'') is given by $\Delta \sigma=\sigma^{+}-\sigma^{-}=(|f|^2+|g|^2)\,2S$. If the spin of the incoming proton encloses an angle $\Delta$ with the normal of the scattering plane, the $\sin$-factor in Eq.(\ref{eqn:differential_cross_section}) becomes $\sin(\Delta \mp \pi/2)=\mp \cos \Delta$ with $\beta=\mp \pi/2$ denoting the positions of the two particle detectors as before. Thus one has $\Delta \sigma=(|f|^2+|g|^2)\,2S\,\cos \Delta$. In order to capture the general case, we introduce an orthogonal Cartesian coordinate system whose $x/y$-plane is spanned by the two proton tracks before they enter the Mott-type analyzers. The axis of alignment of the proton spins encloses in general an angle $\varphi$ with the $z$-axis thus introduced. The pertinent orientation angles of the scattering planes with respect to that $z$-axis are denoted by $\theta$ and $\phi$ for the normals of the A and B-plane, respectively. That means: $\Delta_A=\theta-\varphi$ and $\Delta_B=\phi-\varphi$. To make contact to the familiar notation, we define a quantity $P_{A(B)}^{\pm}$ through
\begin{eqnarray}
\label{eqn:P_definition}
\frac{\sigma_{A(B)}^{\pm}}{2\,[|f|^2+|g|^2]}= P_{A(B)}^{\pm}=S\,[{\textstyle \frac{1}{2\,S}}\pm {\textstyle \frac{1}{2}}\,\cos \Delta_{A(B)}]
\end{eqnarray}
which has the property
$$
P_{A(B)}^{+}+P_{A(B)}^{-}=1\,.
$$
Obviously, $P_{A(B)}^{\pm}$ is proportional to the count rate of the respective detector, and  $\frac{1}{S}\,(P_{A(B)}^{+}-P_{A(B)}^{-})=\cos \Delta_{A(B)}$ describes the degree of spin orientation of the incoming proton with respect to the normal of the associated scattering plane. If the spin of the proton impinging on the analyzer at A is parallel to that normal, that is perpendicular to the associated scattering plane, we have $\Delta_A =0$ and hence $\frac{1}{S}\,(P_{A}^{+}-P_{A}^{-})=1$. The joint probability of finding the proton pair with one of the protons at A and orientation angle $\Delta_A=\theta-\varphi$ and the other proton at B with orientation angle $\Delta_B=\phi-\varphi-\pi$ is given by 
\begin{eqnarray}
\label{eqn:joint_count_rate}
P_{joint}=(P_{A}^{+}-P_{A}^{-})\,(P_{B}^{+}-P_{B}^{-})=\qquad\qquad \nonumber\\
\qquad\qquad P^{++}+P^{--}-P^{+-}-P^{-+}
\end{eqnarray}
where $P^{\pm \pm}=P_{A}^{\pm}\,P_{B}^{\pm}$ and $P^{\pm \mp}=P_{A}^{\pm}\,P_{B}^{\mp}$. Because of the definition (\ref{eqn:P_definition}) we have
$$
P^{++}+P^{--}+P^{+-}+P^{-+}=1\,.
$$
To make sure that the count rates refer definitely to proton pairs, the counts associated with $P^{\pm \pm}$ and $P^{\pm \mp}$ are filtered by coincidence electronics.\\
Since for principal reasons one has in general $S<0$ (in the case under study $S\approx 0.7$), $P^{+}-P^{-}=S\,\cos \Delta$ can never become unity even when the particle enters the analyzer with its spin perpendicular to the scattering plane, that is when $\Delta=0$. It is therefore suggestive to introduce an $S$-independent joint count rate $\hat{P}_{joint}=\frac{1}{S^2}\,P_{joint}$ which, on combining Eqs.(\ref{eqn:P_definition}) and (\ref{eqn:joint_count_rate}), takes the form
\begin{eqnarray}
\label{eqn:hatP_joint}
\hat{P}_{joint}(\theta,\phi,\varphi)=-\cos(\theta-\varphi)\,\cos(\phi-\varphi)\,.
\end{eqnarray}
In practice the experiments have been carried out with the scattering plane of the B-analyzer lying in the $x/y$-plane, which means $\phi=0$. Since all proton pairs are prepared 100\% polarized, that is with their spins aligned parallel and antiparallel with respect to the $z$-axis, we have also $\varphi=0$ so that Eq.(\ref{eqn:hatP_joint}) simplifies to
\begin{eqnarray}
\label{eqn:hatP_joint_2}
\hat{P}_{joint}=-\cos\theta\,,
\end{eqnarray}
and this is in agreement with the experimental results. \\
We emphasize again that this equation has been obtained by assuming a factorization of the two-proton wave function which means that the motion of the ``A''-proton is controlled only by the potentials specifying the ``A''-analyzer. There is no influence of the potentials that belong to ``B''. Analogous statements apply to the ``B''-proton. Hence, for each pair of protons there is no correlation between their respective ``A'' and ``B''- scattering processes. However, it has been the objective of the experiments, as the authors expressly state, to demonstrate that there is such a correlation. Yet in order to prove that point, the experiments should have allowed a preparation of proton pairs with an axis of spin alignment that encloses an angle $\varphi$ with the $z$-axis as originally assumed above. According to the established terminology that angle has to be regarded as a ``hidden variable''. The values of $\varphi$ associated with the various pairs should have random character. One can form then a new expression from $\hat{P}_{joint}(\theta,\phi,\varphi)$ by averaging over $\varphi$:
\begin{eqnarray}
\label{eqn:hatP_joint_average}
\hat{P}_{av}(\theta,\phi)=\int_{-\frac{\pi}{2}}^{\frac{\pi}{2}}\rho(\varphi)\,\hat{P}_{joint}(\theta,\phi,\varphi)\,d\varphi\,.
\end{eqnarray}
where $\rho(\varphi)$ denotes a weight function normalized to unity. Clearly, in the experiment the averaging occurs automatically and unavoidably.\\
If one assumes a uniform distribution of $\varphi$ over the interval $\pi$, that is $\rho(\varphi)=\frac{1}{\pi}$, and inserts here  $\hat{P}_{joint}(\theta,\phi,\varphi)$ from Eq.(\ref{eqn:hatP_joint}), one obtains 
\begin{eqnarray}
\label{eqn:hatP_joint_average_1}
\hat{P}_{av}(\theta,\phi)=-{\textstyle \frac{1}{2}}\,\cos(\theta-\phi)\,,
\end{eqnarray}
where 
$$
\cos(\theta-\varphi)\,\cos(\phi-\varphi)={\textstyle \frac{1}{2}}\,\cos(\theta-\phi)+{\textstyle \frac{1}{2}}\,\cos(\theta+\phi-2\varphi)
$$
has been used. Hence, for $\phi=0$ as specified in the experiment, Eq.(\ref{eqn:hatP_joint_average_1}) yields
$$
\hat{P}_{av}(\theta,\phi)=-{\textstyle \frac{1}{2}}\,\cos\theta
$$
which differs from (\ref{eqn:hatP_joint_2}) by a factor of $\frac{1}{2}$.\\
At this point it is instructive to contemplate the change that would occur if there would be a non-local correlation between the two analyzers in the following sense:\\
If the ``B''-proton has been specified by the ``B''-analyzer as polarized perpendicular to the associated scattering plane, that is if  $\varphi=\phi$,  and if this property is by some ``spooky action at a distance'' transferred to the ``A''-proton, $\varphi$ attains the same value for the  ``A''-proton. The measurement on the ``A''-proton would then become ``contextual'': it would depend on the result obtained for the ``B''-proton. Consequently, the detection rate (\ref{eqn:hatP_joint}) would take the form
\begin{eqnarray}
\label{eqn:hatP_joint_3}
\hat{P}_{joint}=-\cos (\theta-\phi)=-\cos (\vec{a},\vec{b})=-\vec{a}\cdot\vec{b}\,.
\end{eqnarray}
where we have introduced the quantities $\vec{a}$ and $\vec{b}$ as normal vectors for the ``A``- and ``B''-scattering plane, respectively, which  enclose angles $\theta$ and $\phi$ with the $z$-axis. For the situation specified by the experiment (viz. $\phi=0$), this result becomes identical with (\ref{eqn:hatP_joint_2}). Thus, a distinction between the two mechanisms is not possible within the given limitations. One might argue that a derivation based on a ``spooky-action-at-a-distance''-hypothesis has to be rejected anyway. But the same hypothesis works perfectly for the analogous experiment with pairs of linearly polarized photons where that particular limitation does not exist. (S. Aspect et al. \cite{Aspect}.) \\
By referring to the expectation value
$$
\langle \Psi|\vec{\sigma}_A\cdot\vec{a}\otimes \vec{\sigma}_B\cdot\vec{b}|\Psi\rangle =-\cos (\vec{a},\vec{b})
$$
where $\Psi$ denotes the anti-symmetric singlet-state two-proton wave function and $\vec{\sigma}_{A/B}$ the spin operators,  
Eq.(\ref{eqn:hatP_joint_3}) is commonly discussed as ``the quantum mechanical prediction'' for the experiment in question. Considering all the details of our analysis it is hard to see how this expectation value can have anything to do with the experiment except that it happens to yield the same $-\cos (\vec{a},\vec{b})$.\\
We shortly return to the idea pursued by Lamehi-Rachti and Mittig in their paper. In order to exclude the possibility that their result might accidentally coincide with the prediction of a hidden parameter model, they resort to Bell's theorem \cite{Bell2}. It refers to quantities of the type $\hat{P}_{av}(\vec{a},\vec{b})$ in Eq.(\ref{eqn:hatP_joint_average}) which - according to Eq.(\ref{eqn:hatP_joint_average_1}) - becomes equal to $-\frac{1}{2}\,\cos (\vec{a},\vec{b})$ if $\varphi$ is uniformly distributed. In general the weight function $\rho(\varphi)$ will be unknown, and hence a complete lack of correlation between the ``A''- and ``B''-scattering processes, as implied by our treatment,  will not show up simply as a numerical correction factor of the ``correlated result''. Bell could show that in performing an EPRB-type experiment one is definitely dealing with a non-classical (i.~e. non-local) particle correlation if - irrespective of the form of the weight function and irrespective of the kind of hidden variable -  the following inequality is violated:
$$
|\hat{P}_{av}(\vec{a},\vec{b})-\hat{P}_{av}(\vec{a},\vec{b}')|\leq 2\,|\hat{P}_{av}(\vec{a}',\vec{b}')+\hat{P}_{av}(\vec{a}',\vec{b})|\,,
$$
where $\vec{a},\vec{a}',\vec{b},\vec{b}'$ denote different analyzer settings.  In fact, the authors succeeded in verifying this violation, but it appears to us, because of the limitations discussed above, that this result is not absolutely convincing.
 
\section{Concluding remarks} \label{kapitel21}

In summarizing the essence of quantum mechanics Wigner states in a fundamental article \cite{Wigner} under the headline ``What is the state vector?'': {\it ``We recognize ....that the state vector is only a shorthand expression of that part of information concerning the past of the system which is relevant for predicting (as far as possible) the future behavior thereof.''}  \\
In our view the most impressive success of quantum mechanics in understanding the stability, composition and properties of the building blocks of nature consists in predicting the systematic order in the periodic table, the phenomenon of chemical valency and the ground-state properties of molecules and solids. The state vector of these systems, the ground-state wave function $\Psi({\un r}_1,{\un r}_2, \ldots {\un r}_N)$,  is a function of the particle coordinates ${\un r}_1,{\un r}_2, \ldots {\un r}_N$ in terms of which their Coulomb interaction enters the calculation of the system's total energy. But for every experimentalist there is no doubt that these coordinates are fundamentally inaccessible to measurement, and hence cannot possibly be regarded as ``information gained from measurements''. As is amply demonstrated by modern ab initio-calculations, the wave function allows one to determine the total energy as a function of nuclear positions, bonding angles, vibrational frequencies, lattice constants, elastic moduli, phonon spectra, saturation magnetizations, electric conductivities etc.. These quantities are in the spirit of the common-sense notion true observables whereas the particle coordinates remain definitely hidden parameters. As we have repeatedly explained, this applies to the eigenvalues of hermitian operators as well, thus putting a serious question mark behind ``Kochen-Specker''-type \cite{Kochen} and ``no-go'' theorems (s.~e.~g.\cite{Mermin}) which are all based on the exasperatingly  artificial assumption that ``measurements'' yield eigenvalues or ``probabilities for eigenvalues''. Occasionally a certain awareness of this puzzling inconsistency surfaces as in a statement of Wigner's \cite{Wigner2}:
\\[0.2cm]
{\it``All these are concrete and clearly demonstrated limitations on the measurability of operators. They should not obscure the other, perhaps even more fundamental weakness of the standard theory, that it postulates the measurability of operators but does not give directions as to how the measurement should be carried out.''}
\\[0.2cm]
It is deplorable to notice the impropriety with which certain advocates of the Copenhagen school of thought dismiss supporters of Nelson's attempt on developing a  ``quantum mechanics without observer''  as ``stranded enthusiasts''(s. Streater \cite{Streater}), and ironically base their criticism on the old, actually absurd, arguments how indeterminacy enters the theory through measurement and how commuting ``observables'' correlate with the result of simultaneous measurements. All this has been iterated umpteen times although it is well known to every experimentalist that exactly these ``measurements''  are inexecutable altogether. With the same insensitivity to reality castigators of the Nelson proponents think it fully justified to equate the physics of photon-correlation experiments with analogous, but actually extremely scarce experiments with massive particles. We believe we have presented ample evidence that quantum mechanics is in detail derivable from classical mechanics plus a modified physical vacuum by allowing the latter to undergo energy fluctuations. Their action on massive particles is calibrated by Planck's constant, and despite their presence the conservation of energy (and with free particles: the conservation of particle momentum) is ensured on average.  We hope that the present article can contribute to an unbiased reassessment of present-day quantum mechanics concerning these two questions: 1. Which elements of the old doctrine are obsolete and dispensable? 2. Does ``measurement'' really play a particular role in quantum mechanics or is its alleged importance simply a misunderstanding?

\section{Appendix: Derivation of the Navier-Stokes equation} \label{kapitel22}

Given an ensemble of $N$ similarly prepared one-particle systems we introduce a transition probability $P^{M}({\un r},\vec{\sigma},t, \Delta t)$
which denotes the probability of a particle being in the elementary volume $d^3\sigma$ around a point ${\un r}+\vec{\sigma}$ after a time span $\Delta t$ if it has been with certainty at point  ${\un r}$  at time $t$. This transition probability integrates to unity:
\begin{eqnarray*}
\int P^M({\un r},\vec{\sigma}, t, \Delta
t)\,d^3\sigma=1\,.
\end{eqnarray*}
The superscript ``M'' stands for ``Markov process''.\\
In terms of this transition probability Einstein's law (\ref{eqn:EinsteinLaw}) on the mean square displacement of a particle under the action of stochastic forces may be cast as
\begin{eqnarray*}
\int \sigma_{l}\,\sigma_{k}\,P^M({\un
r},\vec{\sigma}, t, \Delta t)\,d^3\sigma=\delta_{l\,k}\,2\nu\,\Delta t
\end{eqnarray*}
\\[-0.4cm]
where $\vec{\sigma}=(\sigma_1,\sigma_2,\sigma_3)\,.$
\\
Here $\nu=\eta/m_0\,n_0$ denotes the ``kinematic viscosity'' of the embedding system, $\eta$ represents the common ``dynamical viscosity'' and $n_0$ is the particle density of the embedding system.\\
The Smoluchowski equation \cite{Smoluchowski} describes the temporal change of the probability density at ${\un r}$ caused by the motion of the $N$ independent particles under the influence of stochastic forces as a result of which the particles perform transitions from previous positions ${\un r}-\vec{\sigma}$ to ${\un r}$
\\[0.2cm]
$\rho({\un r}, t+\Delta t)=\int \underbrace{\rho({\un
r}-\vec{\sigma}, t)P^M({\un r}-\vec{\sigma},\vec{\sigma}, t, \Delta
t)}_{\equiv G({\un r}-\vec{\sigma}, \vec{\sigma}, t, \Delta t)}d^3\sigma\,.$
\\[-0.6cm]
\begin{eqnarray}
\label{eqn:Smoluchowski}
\qquad
\end{eqnarray} 
One may evaluate the integral on the right-hand side by approximately replacing the integrand $G({\un r}-\vec{\sigma}, \vec{\sigma}, t, \Delta t)$ with a Taylor polynomial of second degree 
\\[0.2cm]
\hspace*{1.0cm}$G({\un r}-\vec{\sigma}, \vec{\sigma}, t, \Delta t)=G({\un r}, \vec{\sigma}, t,\Delta t)-$
\\[0.2cm]
\hspace*{3.5cm}$\sum_{k=1}^3\sigma_{k}\frac{\partial}{\partial x_k}G({\un r},
\vec{\sigma}, t, \Delta t)+$
\\[0.2cm]
\hspace*{3.0cm}$\frac{1}{2}\sum_{l,k}\sigma_l\,\sigma_k\,\frac{\partial^2}{\partial
x_l\,\partial x_k}\,G({\un r}, \vec{\sigma}, t, \Delta t)\,.$
\\[0.2cm]
Inserting this expression under the integral of Eq.(\ref{eqn:Smoluchowski}) yields
$\rho({\un r}, t+\Delta t)=\rho({\un r}, t)
-\sum_{k=1}^3\frac{\partial}{\partial x_k}[\rho({\un r},t)$
\\[0.2cm]
\hspace*{4.0cm}$\times\underbrace{\int \sigma_k
\,P^M({\un r},\vec{\sigma}, t, \Delta t)\,d^3\sigma}_{=v_{ck}({\un r},\bar{t})\,\Delta t}] + $
\\[0.2cm]
$ +\frac{1}{2}\sum_{l, k}\frac{\partial^2}{\partial
x_l\,\partial x_k}[\rho({\un r}, t)\underbrace{\int \sigma_l\,\sigma_k\,P^M({\un
r},\vec{\sigma}, t, \Delta t)\,d^3\sigma}_{=\delta_{lk}\,2\nu\,\Delta t}]\,.$
\\
This may be recast as
\begin{eqnarray}
\label{eqn:Fokker_PLanck_1} \frac{\partial\,\rho}{\partial\,t}+\nabla\cdot {\un
j}_c-\nu \Delta \rho=0 \qquad \mbox{where} \quad {\un j}_c=\rho\,{\un v}_c\,.
\end{eqnarray}
\hspace*{1.2cm}\textbf{``Fokker-Planck Equation''}
\\[0.2cm]
Apart from obeying the Fokker-Planck equation the system of $N$ particles must satisfy the equation of continuity as well 
$$\frac{\partial\,\rho}{\partial\,t}+\nabla
\cdot{\un j}=0 \quad \mbox{where}\quad {\un j}=\rho\,{\un v}\,.\qquad \qquad \qquad \qquad
$$
On forming the difference of these two equations one obtains
$$
\nabla\cdot ({\un j_d+\nu\,\nabla\rho})=0 \quad \mbox{where} \quad {\un j}_d\equiv {\un j}-{\un j}_c\,.\qquad
\qquad \qquad$$
This is equivalent to
\begin{eqnarray*}
\label{eqn:Ficksches Gesetz}
{\bf j}_d=-\nu\,\nabla \rho \quad \mbox{\textbf{``Fick's Law''}}\,.
\end{eqnarray*}
If one writes the diffusion current density ${\un j}_d$ in the form:
\\[0.3cm]
\hspace*{1.5cm}${\un j}_d=\rho\,{\un u}$ $\quad ({\un u}$=\textbf{``osmotic velocity''}), 
\\[0.3cm]
one may recast ${\un u}$ as in Eq.(\ref{eqn:osmotic_velocity})
$$
{\un u}({\un r},t)=-\nu\,\frac{1}{\rho({\un r},t)}\,\nabla \rho({\un r},t)\,.
$$
Thus we have $\;{\un v}={\un v}_c+{\un u}$ which is just Eq.(\ref{eqn:convectivevelocity}) used in advance in Section {\ref{kapitel3}.
\\[0.3cm]
Replacing $\rho({\un r},t)$ in Eq.(\ref{eqn:Smoluchowski}) with $\rho\,v_{c\,k}$,
one obtains 
\begin{eqnarray*}
\label{eqn:AenderungImpulsstromdichte1} \frac{\partial \rho\,
v_{c\,k}}{\partial t}|_{scatter}=-\nabla\cdot(\rho\, v_{c\,k}\,{\un
v_c})+\nu\,\Delta(\rho\,v_{c\,k})\,,
\end{eqnarray*}
There is an additional (local) change in time of the momentum current density effected by the external force 
\begin{eqnarray*}
\label{eqn:AenderungImpulsstromdichte2} \frac{\partial \rho\,
v_{c\,k}}{\partial t}|_{force}=\hat{f}_{k}({\un r})\equiv\frac{1}{m_0}\,\rho({\un
r})\,F_{k}({\un r})\,,\qquad
\end{eqnarray*}
Invoking the Fokker-Planck equation one can write the sum $\frac{\partial \rho\,
v_{c\,k}}{\partial t}|_{force}+\frac{\partial \rho\,
v_{c\,k}}{\partial t}|_{scatter}$ in the form
\begin{eqnarray}
\label{eqn:Fokker_Planck_2} \frac{\partial{\un v}_{c}}{\partial t}+({\un
v}_c+2\,{\un u})\cdot\nabla {\un v}_c-\nu\,\Delta {\un v}_c=\frac{1}{m_0}\,{\un
F}({\un r})\,.
\end{eqnarray}
Substituting here ${\un v}_c$ by ${\un v}-{\un u}$ we arrive at our Eq.(\ref{eqn:Navier_Stokes}):
\begin{eqnarray*}
\frac{\partial}{\partial t}\,({\un v}-{\un
u})+\left[({\un v}+{\un u})\nabla ({\un v}- {\un u})\right]-\nu\,\Delta
({\un v}-{\un u})={\textstyle\frac{1}{m_0}}\,{\un F}({\un r})\,.
\end{eqnarray*}
In hydrodynamics $|{\un u}|$ is usually neglected compared to $|{\un v}|$ and $\frac{1}{m_0}\,{\un F}({\un r})$ is given by the internal mass-referenced force $-\frac{1}{\hat{\rho}({\un r},t)}\,\nabla p\,({\un r},t)$ with $\hat{\rho}({\un r},\,t)=m_0\,\rho({\un r},t)$ denoting the massive density  and $p\,({\un r},t)$ the pressure. Hence Eq.(\ref{eqn:Navier_Stokes}) takes the familiar Navier-Stokes-form:
\\[0.2cm]
$\hat{\rho}({\un r},t) \,\underbrace{\left(\frac{\partial\,{\un v}}{\partial t}\,
+{\un v}\cdot\nabla {\un v}\right)}_{\frac{d{\un v}({\un r},t)}{dt}}-\mu\,\Delta {\un v}+\nabla p\,({\un r},t)=0$
\\
\hspace*{3.0cm}where $\mu=\nu\,\hat{\rho}\,.$
\\[0.1cm]
The derivation of this equation is due to Gebelein \cite{Gebelein}.


\newpage
\begin{thebibliography}{99}
\bibitem{Wick}
D.~Wick, {\it The Infamous Boundary} (Birkhäuser, Boston, 1995)
\bibitem{DavidB}
D.~Bohm, {\it Wholeness and the Implicate Order} (Routledge Kegan Paul, London, 1980), p.~84 
\bibitem{Bohm_1}
D.~Bohm and J.~P.~Vigier, Phys.~Rev.~{\bf 96}, 208 (1954)
\bibitem{Weizel}
W.~Weizel, Z.~Physik {\bf 134}, 264 (1953), {\bf 135}, 270 (1953), {\bf 136}, 582 (1954)
\bibitem{Nelson_1}
E.~Nelson, J.~Math.~Phys.~{\bf 5}, 332 (1964); Phys.~Rev.~{\bf 150}, 1079 (1966)
\bibitem{Nelson_2}
E.~Nelson, {\it Quantum Fluctuations}$\,$, (Princeton University Press, 1985)
\bibitem{Guerra}
F.~Guerra and L.~M.~Morato, Phys.~Rev.~D {\bf 27}, 1774 (1983)
\bibitem{Baublitz}
M.~Baublitz, Progr.~Theor.~Phys.~{\bf 80}, 232 (1988)
\bibitem{Pena}
L.~de la Pe\~na and A.~H.~Cetto, {\it The quantum dice - An introduction to stochastic electrodynamics}$\,$, (Kluwer, Dordrecht 1996)
\bibitem{Petroni}
N.~C.~Petroni and L.~M.~Morato, J.~Phys.~A: Math.~Gen. {\bf 33}, 5833 (2000)
\bibitem{Wallstrom}
T.~C.~Wallstrom, Phys.~Rev.~A {\bf 49}, 1613 (1994)
\bibitem{Streater}
R.~F.~Streater, {\it Lost Causes in and beyond Physics}$\,$, (Springer-Verlag, Berlin, Heidelberg, 2007)
\bibitem{Namsrai}
K.~Namsrai, {\it Nonlocal Quantum Field Theory and Stochastic Quantum Mechanics}, Reidel Publishing, Dordrecht (1986)
\bibitem{FritscheHaugk}
L.~Fritsche and M.~Haugk, Ann. Phys. (Leipzig) {\bf 12}, No.6, 371 (2003)
\bibitem{Groessing_1}
G.~Grössing, Phys.~Lett.~A {\bf 372}, 4556 (2008)
\bibitem{Groessing_2}
G.~Grössing, Physica~A {\bf 388}, 811 (2009)
\bibitem{Bell_1}
J.~S.~Bell, in {\it Sixty-Two Years of Uncertainty}$\,$, edited by A.~I.~Miller, (Plenum, New York, 1989), p.~17
\bibitem{Bess}
L.~Bess, Prog.~Theor.~Phys.~{\bf 49}, 1889 (1973)
\bibitem{Puthoff1}
H.~E.~Puthoff, Phys.~Rev.~D {\bf 35}, 3266 (1987)
\bibitem{Puthoff2}
H.~E.~Puthoff, Phys.~Rev.~A {\bf 40}, 4857 (1989)
\bibitem{Boyer}
H.~E.~Boyer, Scientific American (August), 70 (1985)
\bibitem{Calogero}
F.~Calogero, Phys.~Lett. A{\bf 228}, 335 (1997)
\bibitem{Carati}
A.~Carati and L.~Galgani, Nuovo Cimento B{\bf 114}, 489 (1999)
\bibitem{Ballentine}
L.~E.~Ballentine, Rev.~Mod.~Phys.~{\bf 42}, 358 (1970)
\bibitem{Gebelein}
H.~Gebelein, {\it{Turbulenz}}, p.~75, Springer, Berlin (1935)
\bibitem{Pauli_1}
W.~Pauli,, {\it Handbuch der Physik}, Bd.XXIV/1, 2. Aufl., Springer, Berlin (1933), p. 126
\bibitem{BornJordan}
M.~Born and P.~Jordan, {\it Elementare Quantenmechanik}, p.32, Springer, Berlin (1930)
\bibitem{Bohm1}
D.~Bohm, Phys.~Rev.~{\bf 85}, 166 (1952); {\bf 85}, 180 (1952)
\bibitem{Einstein1}
A.~Einstein, Ann.~d.~Physik {\bf 17}, 549 (1905), Ann.~Phys.~{\bf 19}, 371 (1906)
\bibitem{Broglie}
L.~de Broglie, {\it Une Tentative d'Interpr\'etation Causale et Non-Lin\'eaire de la M\'ecanique Ondulatoire}, Gauthier-Villars, Paris (1956) 
\bibitem{Mielnik}
B.~Mielnik and G.~Tengstrand, Intern.~J.~Theor.~Phys.~{\bf 19}, 239 (1980)
\bibitem{Badurek}
G.~Badurek, H.~Rauch, and D.~Tuppinger, Phys.~Rev.~A~{\bf 34}, 2600 (1986)
\bibitem{Thomson}
W.~Thomson, Proceedings of the Royal Society of Edinburgh, Vol. VI, 1867, pp. 94-105.
\bibitem{Madelung}
E.~Madelung, Z.~Physik {\bf 40}, 322 (1926)
\bibitem{Heisenberg3}
W.~Heisenberg, Z.~Physik {\bf 43}, 172 (1927)
\bibitem{Bell1}
J.~Bell, Physics World, August, p.~33 (1990)
\bibitem{Wigner0}
E.~P.~Wigner in {\it Quantum Theory of measurement}, edited by J.~A.~Wheeler and W.~H.~Zurek, Princeton University Press, Princeton, New Jersey (1983), p.~267
\bibitem{Garbaczewski}
P.~Garbaczewski, arXiv; cond-mat/0703147
\bibitem{Smoluchowski}
M.~V.~Smoluchowski, Ann.~d.~Physik, {\bf 21}, 756 (1906) and {\bf 48}, 1103
(1915); s.~also: {\it Abhandlungen \"uber die Brownische Bewegung und verwandte
Erscheinungen} (Akademische Verlagsgesellschaft, Leipzig (1923))
\bibitem{Wigner1}
E.~P.~Wigner in {\it Quantum Theory of measurement}, edited by J.~A.~Wheeler and W.~H.~Zurek, Princeton University Press, Princeton, New Jersey (1983), p.~313
\bibitem{Feynman}
R.~P.~Feynman, Rev.~Mod.~Phys.~{\bf 20}, 367 (1948)
\bibitem{v_Neumann}
J.~v.~Neumann, {\it Mathematische Grundlagen der Quantenmechanik}, Dover Publications, New York (1943)
\bibitem{Einstein1}
A.~Einstein, B.~Podolsky, and N.~Rosen, Phys. Rev. {\bf 47}, 777 (1935)
\bibitem{Bohm}
D.~Bohm, {\it{Quantum Theory}} (Prentice Hall, Englewood Cliffs, N.J. (1951))
\bibitem{Hartle}
J.~B.~Hartle and M.~Gell-Mann in:{\it Complexity, Entropy and the Physics of Information}, W.~Zurek (ed.) Addison-Wesley, Redwood City, CA (1990)
\bibitem{Kampen}
N.~G.~van~Kampen, (Physica A {\bf 153}, 97 (1988)
\bibitem{Schroedinger1}
E.~Schrödinger, {\it ``The present situation in quantum mechanics''} in: {\it Quantum Theory and Measurement}, J.~A.~Wheeler and W.~H.~Zurek (Eds.), Princeton University Press, Princeton, N.~J. (1983), p.~152
\bibitem{Uhlenbeck}
G.~E.~Uhlenbeck und S.~Goudsmit, Die Naturwissenschaften, {\bf 13}, 953 (1925), Nature {\bf 127}, 264 (1926)
\bibitem{Pauli}
W.~Pauli, Z.~Physik {\bf 31}, 373 (1925)
\bibitem{Van_der_Waerden}
B.~L.~van der Waerden in {\it Theoretical Physics in the Twentieth Century. A Memorial Volume to Wolfgang Pauli}, edited by M.~Fierz and V.~F.~Weisskopf, Interscience Publishers, New York (1960), p.~214 
\bibitem{Schroedinger}
E.~Schr\"odinger, Sitzungsber.~Preuß.~Akad.~Wiss.~Phys.-Math.~Kl. {\bf 24}, 418
(1930)
\bibitem{Weizsaecker}
C.~F.~v.~Weizsäcker in: {\it Zum Weltbild der Physik} S.~Hirzel, Leipzig (1945), p.32)
\bibitem{Dankel}
T.~Dankel, Archiv. Rational Mech. Anal. {\bf 37}, 192 (1971)
\bibitem{Dohrn}
D.~Dohrn, F.~Guerra, and P.~Ruggiero in : {\it Feynman Path Integrals}, edited by S.~Albeverio, Lecture Notes in Physics {\bf 106}, Springer, Heidelberg (1979)
\bibitem{Nelson}
E.~Nelson, {\it Quantum fluctuations}, Princeton University Press, Princeton, New Jersey (1985), p.102
\bibitem{Goldstein}
H.~Goldstein {\it Classical Mechanics}, 2nd edition, Addison-Wesley, Reading, MA (1980), pp.148-158
\bibitem{Slichter}
C.~P.~Slichter, {\it Principles of Magnetic Resonance}, 3rd edition, Springer, Berlin, Heidelberg (1990)
\bibitem{Bloch}
F.~Bloch, Phys.~Rev. {\bf 70}, 460 (1946)
\bibitem{Rauch}
H.~Rauch, Found.~Phys.~ {\bf 23}, 7 (1993)
\bibitem{Werner}
S.~A.~Werner, R.~Colella and A.~W.~Overhauser, and C.~F.~Eagen, Phys.~Rev.~Lett.~{\bf 35}, 1053 (1975)
\bibitem{Bell}
J.~Bell, Journal de Physique, Colloque C2, supplement au n$^{\circ}$3, Tome 42, mare 1981, p.~C~2-45
\bibitem{Stern}
O.~Stern und W.~Gerlach, Z.~Physik {\bf 9}, 349 (1922);\\
W.~Gerlach und O.~Stern, Ann.~d.~Physik {\bf 74}, 673 (1924)
\bibitem{Home}
D.~Home, A.~K.~Pan, Md~M.~Ali and A.~S.~Majundar, J.~Phys.~A: Math.~Theor.~{\bf 40} 3975 (2007)
\bibitem{Wrede}
E.~Wrede, Z.~Physik {\bf 41}, 569 (1927)
\bibitem{Mott}
N.~V.~Mott and H.~S.~W.~Massey in {\it Quantum Theory of measurement}, edited by J.~A.~Wheeler and W.~H.~Zurek, Princeton University Press, Princeton, New Jersey (1983), p.~703
\bibitem{Leu}
A.~Leu, Z.~Physik {\bf 41}, 551 (1927)
\bibitem{Stern2}
T.~E.~Phipps und O.~Stern, Z.~Physik {\bf 73}, 185 (1932)
\bibitem{Zener}
N.~Rosen and C.~Zener, Phys.~Rev.~{\bf 40}, 502 (1932)
\bibitem{Englert}
B.-G.~Englert, J.~Schwinger and M.~O.~Scully, Found.~Phys.~{\bf 18}, 1045 (1988)
\bibitem{Badurek}
G.~Badurek, H.~Rauch, and D.~Tuppinger, Phys.~Rev.~A~{\bf 34}, 2600 (1986)
\bibitem{Wheeler}
J.~A.~Wheeler and W.~H.~Zurek (Eds.) {\it Quantum Theory of measurement}, Princeton University Press, Princeton, New Jersey (1983), p.~699
\bibitem{Dirac}
P.~A.~M.~Dirac, Proc.~Roy.~Soc.~(A), {\bf 117}, 610 (1928)
\bibitem{Lamehi}
M.~Lamehi-Rachti and W.~Mittig, Phys.~Rev. D~{\bf 14}, 2543 (1976)
\bibitem{Aspect}
A.~Aspect, Ph.~Grangier, and G.~Roger, Phys. Rev. Lett. {\bf 49}, 91 (1982)
\bibitem{Bell2}
J.~S.~Bell, Physics (N.Y.), {\bf 1}, 195 (1965)
\bibitem{Wigner}
E.~P.~Wigner in {\it Quantum Theory of measurement}, edited by J.~A.~Wheeler and W.~H.~Zurek, Princeton University Press, Princeton, New Jersey (1983), p.~292
\bibitem{Kochen}
S.~Kochen and E.~P.~Specker, J.~Math.~Mech.~{\bf 17}, 59 (1867)
\bibitem{Mermin}
N.~D.~Mermin, Rev.~Mod.~Phys.~{\bf 65}, 803 (1993)
\bibitem{Wigner2}
E.~P.~Wigner in {\it Quantum Theory of measurement}, edited by J.~A.~Wheeler and W.~H.~Zurek, Princeton University Press, Princeton, New Jersey (1983), p.~313

\end{thebibliography}
\end{document}